\documentclass[review,hidelinks,onefignum,onetabnum]{siamart251104}

\nolinenumbers

\RequirePackage{packages}

\usepackage{lipsum}
\usepackage{amsfonts}
\usepackage{graphicx}
\usepackage{epstopdf}
\usepackage{algorithmic}
\ifpdf
  \DeclareGraphicsExtensions{.eps,.pdf,.png,.jpg}
\else
  \DeclareGraphicsExtensions{.eps}
\fi

\makeatletter
\newcommand{\floral@externalLabel}[2]{%
  \ifnum\pdfmatch{@cref$}{#1}=1
    \unexpanded{#2}%
  \else
    \XR@addURL{#2}%
  \fi}
\patchcmd{\XR@test}{\XR@addURL{#3}}{\floral@externalLabel{#2}{#3}}{}%
  {\PackageError{floral}{Unable to patch external cleveref labels}{Check xr-hyper compatibility.}}
\makeatother

\newsiamremark{remark}{Remark}
\newsiamremark{hypothesis}{Hypothesis}
\crefname{hypothesis}{Hypothesis}{Hypotheses}
\newsiamthm{claim}{Claim}
\newsiamremark{fact}{Fact}
\crefname{fact}{Fact}{Facts}

\headers{Flow matching operator}{S. Bhola and K. Duraisamy}

\title{Residual-Augmented Flow Matching Operators for Probabilistic Partial Differential Equations\thanks{Submitted to the editors DATE.
\funding{This work was supported in part by the OUSD(RE) grant \emph{`Physics-Aware Reduced Order Modeling for Nonequilibrium Plasma Flows'} and in part by the Los Alamos National Laboratory grant {\em `Algorithm/Software/Hardware Co-design for High Energy Density applications'}.}}}

\author{Sahil Bhola\thanks{Department of Aerospace Engineering \& Michigan Institute for Computational Discovery and Engineering,
University of Michigan, Ann Arbor, MI 48109, U.S.A.
  (\email{sbhola@umich.edu}, \email{kdur@umich.edu})}
  \and Karthik Duraisamy\footnotemark[2]
  }

\usepackage{amsopn}

\ifpdf
\hypersetup{
  pdftitle={Flow matching Operators for Residual-Augmented Probabilistic Learning of Stochastic Partial Differential Equations},
  pdfauthor={S. Bhola, and K. Duraisamy}
}
\fi

\nolinenumbers

\begin{document}

\maketitle

\begin{abstract}
Learning surrogate models for physical systems with latent uncertainty remains challenging in data-scarce regimes: deterministic neural operators fail to characterize uncertainty, while generative approaches require large ensembles of high-fidelity solution operator simulations and often sacrifice resolution generalizability.
In this work, we propose a residual-augmented probabilistic operator learning framework that casts flow-matching-based generative modeling in infinite-dimensional function spaces while leveraging inexpensive low-fidelity solution operators as an inductive bias.
Rather than learning the full high-fidelity stochastic solution operator directly, the proposed framework learns probabilistic residual operators that characterize the discrepancy between low- and high-fidelity solutions.
By parameterizing the vector field in flow matching using neural operators conditioned on both the known system input and low-fidelity solution, the framework amortizes probabilistic inference across input conditions while enabling uncertainty-aware and resolution-generalizable predictions across spatial discretizations.
Numerical experiments on stochastic advection, Burgers’, and Darcy flow systems demonstrate that the residual-augmented formulation improves predictive accuracy under the same high-fidelity data budget, while the probabilistic operator learning formulation enables accurate characterization of uncertainty in low-data regimes compared to learning high-fidelity stochastic operators directly from data.
\end{abstract}


\section{Introduction}\label{sec:intro}
Predicting the behavior of complex physical systems governed by partial differential equations (PDEs) is fundamental to science and engineering.
In many such systems, including turbulence~\cite{pope2001turbulent}, weather forecasting~\cite{gneiting2005weather}, and subsurface modeling~\cite{tacher2006geological}, the system state cannot be uniquely determined from the observed inputs alone due to latent forcing, material variability, unresolved physics, and measurement noise.
As a result, the system is more naturally characterized through a \emph{stochastic solution operator} $\mathcal{G}^{\varepsilon}:\mathcal{A}\rightarrow\mathcal{W}$, which maps an input condition $a\in\mathcal{A}$ to a distribution of physically plausible solution fields in the solution space $\mathcal{W}$ induced by an \textit{unobserved latent random variable $\varepsilon$}.
Accurately characterizing this distribution is critical for applications such as uncertainty quantification, predictive control, and design optimization.
However, doing so typically requires generating large ensembles of high-fidelity realizations across varying input conditions, making traditional numerical simulation prohibitively expensive in practice.

Recent advances in data-driven and physics-informed machine learning have enabled learning-based surrogates for PDE systems to be trained using data and physical priors, thereby alleviating the need to repeatedly solve computationally expensive discretized governing equations via traditional numerical methods~\cite{carleo2019machine,brunton2020machine,karniadakis2021physics}.
Notable approaches include (a) finite-dimensional operator learning methods~\cite{guo2016convolutional,zhu2018bayesian,weyn2019can,bhatnagar2019prediction,khoo2021solving,wiewel2019latent,lee2021deep}, (b) neural FEM methods~\cite{dissanayake1994neural,lagaris1998artificial,raissi2019physics,karniadakis2021physics}, and (c) neural operators such as DeepONet~\cite{lu2021learning}, FNO~\cite{li2020fourier}, and GNO~\cite{li2020neural}, which enable resolution-generalizable inference of the solution field across unseen inputs $a$.
However, these methods are predominantly deterministic, that is, given an input $a$ they produce a single solution field rather than characterizing the distribution of physically plausible solution fields.
For a stochastic solution operator $\mathcal{G}^\varepsilon$ and standard mean-squared regression objective, this reduces to learning the conditional mean $\expect[\varepsilon]{\mathcal{G}^\varepsilon[a]}$, thereby failing to capture the uncertainty induced by the latent variable $\varepsilon$.
Consequently, these approaches cannot characterize predictive variability, multi-model behavior, or uncertainty induced by the latent variable.

To characterize the distribution induced by the stochastic solution operator $\mathcal{G}^\varepsilon$, probabilistic generative models have emerged as a powerful alternative to deterministic surrogates.
Approaches such as GANs~\cite{goodfellow2014generative,creswell2018generative}, diffusion models~\cite{song2020denoising,ho2020denoising}, and flow matching~\cite{lipman2022flow,tong2023improving} learn the distribution of solution fields induced by the latent variable $\varepsilon$ for a given input $a$, rather than predicting a single deterministic solution.
This enables sampling of physically plausible solution fields consistent with the observed input and the underlying latent stochasticity.
Such methods have shown significant promise in applications where accurately characterizing the full distribution of solution fields is essential for robust decision-making, including weather forecasting~\cite{li2024generative,fotiadis2024stochastic,chakraborty2026multimodal} and turbulence modeling~\cite{drygala2022generative,li2025generative,liu2025confild}.

Recent efforts have further unified generative modeling with neural operators to enable \textit{probabilistic operator learning in function spaces}~\cite{rahman2022generative,kerrigan2022diffusion,pidstrigach2023infinite,lim2025score,yao2025guided,kerrigan2023functional,shi2025stochastic,hou2025cfo}, allowing resolution-generalizable inference of the solution field while characterizing the uncertainty induced by the stochastic solution operator.
In particular, flow-matching-based operator learning frameworks model continuous-time transport between probability measures on function spaces via vector-field operators, thereby enabling stable training and efficient sampling of solution-field distributions.
However, despite their representational capacity, \textit{existing probabilistic operator learning approaches rely exclusively on large ensembles of high-fidelity solution-field realizations, which are, in practice, prohibitively expensive to obtain.
In scientific applications where high-fidelity simulations of $\mathcal{G}^\varepsilon$ are computationally demanding, this dependence severely limits the practicality of existing probabilistic operator learning methods.}

In this work, we propose a novel framework for probabilistic operator learning that addresses the computational bottleneck of high-fidelity data acquisition.
Our approach leverages an inexpensive and inaccurate low-fidelity approximation of $\mathcal{G}^\varepsilon$ and casts the learning problem as a residual operator learning in function space, where the goal is to characterize the distribution of discrepancy between the low-fidelity approximation and high-fidelity solution, rather than learning the full high-fidelity solution field distribution from scratch.
By casting the flow matching operator learning in a residual-augmented formulation, our framework enables data-efficient, uncertainty-aware, and resolution-generalizable characterization of the stochastic solution operator.
Our key contributions are as follows:

\begin{enumerate}
    \item \textbf{Flow Matching Operators for Conditional Probabilistic Inference in Function Spaces.}
    We introduce a conditional flow matching operator framework for probabilistic operator learning, in which the learned vector-field operator is conditioned on the PDE input and low-fidelity solution operator.
    By amortizing the cost of learning across varying input conditions, the proposed formulation enables probabilistic inference directly in function spaces without retraining for new inputs.

    \item \textbf{Residual-Augmented Probabilistic Operator Learning.}
    We propose a residual-augmented probabilistic operator learning framework that leverages inexpensive low-fidelity solution operators to model the discrepancy between low- and high-fidelity stochastic solution operators.
    By leveraging a low-fidelity prior, the proposed residual-space formulation improves predictive accuracy and uncertainty characterization, particularly in low-data regimes.
    
    \item \textbf{Uncertainty-Aware Resolution-Generalizable Inference.}
    By casting flow matching operators within a conditional operator-learning framework, the proposed framework enables resolution-generalizable probabilistic inference across unseen spatial discretizations and domain locations. 

    \item \textbf{Empirical Validation on Probabilistic PDE Benchmarks.}
    We demonstrate the proposed framework on several probabilistic PDE systems, including stochastic advection and Burgers' equations with stochastic initial conditions, and Darcy flow with stochastic permeability fields. Numerical experiments show that leveraging low-fidelity solution operator priors consistently improves predictive performance and uncertainty characterization across varying data regimes and training resolutions.
\end{enumerate}

We organize the manuscript as follows.
In~\S\ref{sec:background}, we introduce the problem of estimating the state of physical systems with underlying stochasticity and review the necessary background on flow matching and finite-dimensional probabilistic modeling of physical system states.
In~\S\ref{sec:methodology}, we present an operator formulation of flow matching for probabilistic operator learning in function spaces, enabling resolution-generalizable inference of stochastic solution fields. We further introduce our residual-augmented framework, which leverages low-fidelity solution operators to probabilistically model the discrepancy between low- and high-fidelity solutions.
In~\S\ref{sec:experiments}, we evaluate the proposed framework on several synthetic stochastic PDE systems and demonstrate its ability to improve predictive accuracy and uncertainty characterization under limited high-fidelity data.
Finally, in~\S\ref{sec:conclusion}, we provide concluding remarks and discuss potential directions for future work.

\section{Background}\label{sec:background}
Our goal is to solve the \emph{forward problem}: estimate the solution field (or solution function) $w \in \mathcal{W} \define \{w: \Omega_w \subset \real{\kappa_w} \to \real{m_w}\}$ of a physical system given an input $a \in \mathcal{A} \define \{a: \Omega_a \subset \real{\kappa_a} \to \real{m_a}\}$ encoding known conditions such as boundary data, initial conditions, or forcing.
To account for latent stochasticity in the underlying physical system, we model $\varepsilon$ as a random variable with distribution $\prob{\varepsilon}$.
The forward problem can then be expressed as
\begin{align}
    w = \mathcal{G}^\varepsilon[a], \quad \varepsilon \sim \prob{\varepsilon}, \quad a \in \mathcal{A},
\end{align}
where we seek to characterize the conditional distribution of $w$ given an input $a$.
In this section, we first review the flow matching framework for approximating marginal probability distributions.
Next, we first describe how to solve the forward problem in a finite-dimensional setting via flow matching.

\subsection{Flow Matching for Learning Marginal Probability Distributions}
Flow matching seeks a parametric approximation of a target distribution $\prob{y}$ from finite samples by introducing a time $\tau$ dependent flow transformation $\phi_\tau^\zeta:\real{d_y}\to\real{d_y}$ parameterized by $\zeta\in\real{n_\zeta}$ and characterized by the ordinary differential equation (ODE)
\begin{align*}
    \frac{d}{d\tau}\phi_\tau^\zeta(\sample{y}) = u_\tau^\zeta(\phi_\tau^\zeta(\sample{y})),\quad \tau\in[0,1],\\
    \phi_0^\zeta(\sample{y}) = \sample{y} \sim \srcprob{y},
\end{align*}
where $u_\tau^\zeta:\real{d_y}\to\real{d_y}$ is a time-dependent vector field and $\srcprob{y}$ is the prior distribution.
The flow transformation induces the push-forward $\varprob[\tau]{y;\zeta} \define [\phi_\tau^\zeta]_{\#}\srcprob{y}$, where $\varprob[\tau]{y;\zeta}: [0, 1] \times \real{d_y} \ra \real{}_{>0}$ is the time-varying \emph{probability path} induced by the vector field $u_\tau^\zeta$, that is, it satisfies the continuity equation (see Appendix~\ref{sec:continuity_equation}).
To approximate $\prob{y}$, a vector field $v_\tau^\xi:\real{d_y}\to\real{d_y}$ parameterized by $\xi\ireal{n_\xi}$ is regressed onto the true vector field $u_\tau^\zeta$ via the flow matching objective
\begin{align}
    J_{\text{FM}}^\dagger(\zeta, \xi) \define \mathbb{E}_{\tau \sim \mathcal{U}(0,1),\, \sample{y}\sim \varprob[\tau]{\sample{y};\zeta}}\left[\norm[2]{u_\tau^\zeta(\sample{y}) - v_\tau^\xi(\sample{y})}^2\right].
    \label{eq:flow_matching_objective}
\end{align}
In practice, however, the vector field $u_\tau^\zeta$ is intractable. Tong et al.~\cite{tong2023improving} resolve this by introducing a latent random variable $\random{Z}$ distributed as $\varprob{z}$, expressing the vector field $u_\tau^\zeta$ as
\begin{align}
    u_\tau^\zeta \define \expect[\varprob{\sample{z}}]{\frac{\tilde{u}_{\tau,\sample{z}}^\zeta\, \varprob[\tau]{\sample{y}\vert\sample{z};\zeta}}{\varprob[\tau]{\sample{y};\zeta}}},
    \label{eq:latent_vector_field}
\end{align}
introducing a latent vector field $\tilde{u}_{\tau, \vect{z}}^\zeta:\real{d_y}\to\real{d_y}$ that induces the latent probability path $\varprob[\tau]{\sample{y}\vert\sample{z}; \zeta}$.
Thus, obtaining a tractable learning objective for~\cref{eq:flow_matching_objective} as
\begin{align}
    J_{\text{CFM}}^\dagger(\zeta, \xi) \define \mathbb{E}_{\tau \sim \mathcal{U}(0,1),\, \sample{z}\sim\varprob{\sample{z}},\, \sample{y}\sim\varprob[\tau]{\sample{y}\vert\sample{z};\zeta}}\left[\norm[2]{\tilde{u}_{\tau,\sample{z}}^\zeta(\sample{y}) - v_\tau^\xi(\sample{y})}^2\right].
\end{align}
To tractably sample from $\varprob{z}$ and the latent probability path $\varprob[\tau]{\sample{y}\vert\sample{z};\zeta}$, a simple choice is to model $\varprob{z} = \srcprob{y}\prob{y}$ and consider the latent probability path as transport of Gaussian distribution given as
\begin{align*}
    \varprob[\tau]{\sample{y}\vert\sample{z};\zeta} = \mathcal{N}(\sample{y}; \mu_\tau(\sample{z}), \sigma_\tau(\sample{z};\zeta)),\\
    \mu_\tau(\sample{z}) \define \tau \sample{y}_1 + (1-\tau)\sample{y}_0, \quad \sample{y}_0\sim \srcprob{y}, \sample{y}_1\sim \prob{y}, \\
    \sigma_\tau(\sample{z};\zeta) \define \sigma_{\min},
\end{align*}
where $\sigma_{\min}$ is a hyper-parameter.
By defining the latent probability path as a transport of Gaussian distributions, we can evaluate the latent vector field $\tilde{u}_{\tau, \sample{z}}^\zeta$ in closed form as
\begin{align}
    \tilde{u}_{\tau, \sample{z}}^\zeta(\sample{y}) = \frac{\partial \sigma_\tau(\sample{z};\zeta)}{\partial \tau}\frac{(\sample{y} - \mu_\tau(\sample{z}))}{\sigma_\tau(\sample{z};\zeta)} + \frac{\partial \mu_\tau(\sample{z})}{\partial \tau},
    \label{eq:independent_cfm_vector_field}
\end{align}
as shown in~\cite{tong2023improving}.

\subsection{Finite-dimensional view of Solving the Forward Problem via Flow Matching}\label{sec:pde_conditinal_flow_matching_details}
To make the forward problem computationally tractable, we can discretize the domains $\Omega_a$ and $\Omega_w$ using $n_a$ and $n_w$ evaluation points, respectively. This yields finite-dimensional representations $\vect{a} \in \mathbb{R}^{d_a}$ and $\vect{w} \in \mathbb{R}^{d_w}$, where $d_a \define m_a n_a$ and $d_w \define m_w n_w$.
Under this discretization, the stochastic solution operator $\mathcal{G}^\varepsilon$ can be approximated by a stochastic mapping $ \widehat{\mathcal{G}}^\varepsilon : \mathbb{R}^{d_a} \to \mathbb{R}^{d_w}$ between finite-dimensional spaces,
such that $\vect{w} = \widehat{\mathcal{G}}^\varepsilon(\vect{a})$ is a random variable induced by the latent variable $\varepsilon$.
From a probabilistic perspective, the forward problem then amounts to characterizing the conditional distribution $ \prob{\vect{w}\vert \vect{a}} = \int \prob{\vect{w} \vert \vect{a}, \varepsilon}\,\prob{\varepsilon}\, \mathrm{d}\varepsilon$ induced due to the underlying randomness in the physical system.

To approximate $\prob{\vect{w}\vert\vect{a}}$ via flow matching, we can introduce a conditional flow transformation $\psi_{\tau,\vect{a}}^\zeta:\real{d_w}\to\real{d_w}$ characterized by the ODE
\begin{align*}
    \frac{d}{d\tau}\psi_{\tau,\vect{a}}^\zeta(\vect{w}) = f_{\tau,\vect{a}}^\zeta(\psi_{\tau,\vect{a}}^\zeta(\vect{w})), \quad \tau\in [0, 1],\\
    \psi_{0,\vect{a}}^\zeta(\vect{w}) = \vect{w}\sim \srcprob{w}.
\end{align*}
Here, $f_{\tau, \vect{a}}^\zeta: \real{d_w} \ra \real{d_w}$ is a time-dependent conditional vector field parameterized via $\zeta\ireal{n_\zeta}$ that characterizes the conditional flow transformation, and $\srcprob{w}$ denotes the prior (typically modeled as a standard normal distribution).
By conditioning the vector field on $\vect{a}$, the model learns a single amortized representation that generalizes across inputs $\vect{a}$, rather than requiring a separate flow for each instance.
This significantly reduces the learning complexity while enabling the vector field to adapt its dynamics to the input-dependent structure of the target $\prob{w\vert a}$.
The conditional flow transformation induces the push-forward $\varprob[\tau]{w\vert a; \zeta} \define [\psi_{\tau,\vect{a}}^\zeta]_{\#}\srcprob{w}$, where $\varprob[\tau]{w\vert a;\zeta}: [0, 1] \times \real{d_w} \ra \real{}_{>0}$ is the induced \textit{conditional probability path}.
To approximate $\prob{\vect{w}\vert\vect{a}}$, we seek a conditional vector field $f_{\tau,\vect{a}}^\zeta$ that induces a conditional probability path $\varprob[\tau]{w\vert a;\zeta}$ such that $\varprob[1]{w\vert a;\zeta}\approx \prob{w\vert a}$.

Following Lipman et al.~\cite{lipman2022flow}, we can learn a parametric conditional vector field $h_{\tau, \vect{a}}^\xi: \real{d_w} \ra \real{d_w}$ with parameters $\xi\ireal{n_\xi}$ to approximate $f_{\tau, \vect{a}}^\zeta$ by minimizing 
\begin{align}
    J_{\text{FM}}^\ddagger(\zeta, \xi, \vect{a}) \define \mathbb{E}_{\tau \sim \mathcal{U}(0, 1), \vect{w}\sim \varprob[\tau]{w\vert a; \zeta}}\left[\norm[2]{f_{\tau,\vect{a}}^\zeta(\vect{w}) - h_{\tau,\vect{a}}^\xi(\vect{w})}^2\right],
    \label{eq:pde_flow_matching_objective}
\end{align}
such that $h_{\tau,\vect{a}}^{\xi^\ddagger}$ would also induce $\varprob[\tau]{w\vert a; \zeta}$ upon reaching $J_{\text{FM}}^\ddagger=0$ for some $\xi=\xi^\ddagger$.
However, this loss is intractable as the conditional vector field $f_{\tau,\vect{a}}^\zeta$ that results in a conditional probability path $\varprob[\tau]{w\vert a;\zeta}$ such that $\varprob[1]{w\vert a;\zeta}\approx \prob{w\vert a}$ is unknown.
To address this, following Tong et al.~\cite{tong2023improving}, we introduce a latent variable $\random{z}$ with distribution $\varprob{z}$.
We can then introduce a conditional latent vector field $\tilde{f}_{\tau,\vect{a},\vect{z}}^\zeta:\real{d_w}\to\real{d_w}$ that induces the conditional latent probability path $\varprob[\tau]{\vect{y}\vert\vect{a},\vect{z}; \zeta}$, that is, it satisfies the continuity equation.
Using the marginalization trick, we can then write the conditional probability path using the conditional latent probability path as
\begin{align*}
\varprob[\tau]{\vect{y}\vert \vect{a};\zeta} \define \int \varprob[\tau]{\vect{y}\vert\vect{a},\vect{z}; \zeta}\varprob{\vect{z}} \,d\vect{z}.
\end{align*}
As a result, we can express the conditional vector field $f_{\tau,\vect{a}}^\zeta$ as
\begin{align}
    f_{\tau,\vect{a}}^\zeta \define \expect[\varprob{\sample{z}}]{\frac{\tilde{f}_{\tau, \vect{a},\vect{z}}^\zeta \varprob[\tau]{\vect{w}\vert\vect{a},\vect{z}; \zeta}}{\varprob[\tau]{\vect{w}\vert\vect{a};\zeta}}},
    \label{eq:conditional_latent_vector_field}
\end{align}
that extends~\cref{eq:latent_vector_field} to a conditional setting in a straightforward manner.
Using~\cref{eq:conditional_latent_vector_field}, we can then define a tractable unbiased learning objective given as
\begin{align}
    J_{\text{CFM}}^\ddagger(\zeta, \xi, \vect{a}) \define \mathbb{E}_{\tau \sim \mathcal{U}(0, 1), \vect{z} \sim \varprob{\vect{z}}, \vect{w}\sim \varprob[\tau]{\vect{w}\vert \vect{a}, \vect{z}; \zeta}}\left[\left\|\gamma_\tau\left(\tilde{f}_{\tau, \vect{a}, \vect{z}}^\zeta(\vect{w}) - h_{\tau,\vect{a}}^\xi(\vect{w})\right)\right\|_2^2\right],
    \label{eq:pde_conditional_pde_flow_matching}
\end{align}
where $\gamma_\tau$ is a time-dependent scaling.
Minimizing $J_{\text{CFM}}^\ddagger$ is similar to minimizing $J_{\text{FM}}^\ddagger$ since
\begin{align*}
\arg\min_{\xi} J_{\text{FM}}^\ddagger(\zeta, \xi,\vect{a}) = \arg\min_{\xi} J_{\text{CFM}}^\ddagger(\zeta, \xi,\vect{a}),
\end{align*}
as shown in Theorem~\ref{theorem:finite_dim_unbiased_estimator}.
To tractably sample from $\varprob{z}$, we can define it as a joint distribution between the prior and the conditional distribution $\prob{w\vert a}$, that is, $\varprob{z}=\srcprob{w}\prob{w\vert a}$.
We can then model the conditional latent probability path as a transport of a Gaussian distribution
\begin{align*}
    \varprob[\tau]{\sample{w}\vert\sample{a},\sample{z};\zeta} = \mathcal{N}(\sample{w}; \mu_\tau(\sample{z}), \sigma_\tau(\sample{z};\zeta)),\\
    \mu_\tau(\sample{z}) \define \tau \sample{w}_1 + (1-\tau)\sample{w}_0, \quad \sample{w}_0\sim \srcprob{w}, \sample{w}_1\sim \prob{w\vert a}, \\
    \sigma_\tau(\sample{z};\zeta) \define \sigma_{\min} \norm[2]{\sample{w}_1 - \sample{w}_0}^2,
\end{align*}
that results in the conditional vector field $\tilde{f}_{\tau, \sample{a}, \sample{z}}^\zeta = \sample{w}_1 - \sample{w}_0$ using~\cref{eq:independent_cfm_vector_field}.
Here, we introduce a sample-dependent noise $\sigma_\tau$ to ensure a high signal-to-noise ratio when the samples $\sample{w}_0$ and $\sample{w}_1$ are close to each other in $\norm[2]{\cdot}$ sense.
Finally, to learn a vector field $h_{\tau, \sample{a}}^\xi$ that generalizes well across inputs, we can consider the training objective
\begin{align}
J_{\text{CFM}}(\zeta, \xi) \define \expect[\vect{a}\sim\prob{a}]{J_{\text{CFM}}^\ddagger(\zeta, \xi, \vect{a}) },
\end{align}
where $\prob{a}$ characterizes the underlying distribution of the inputs.

\section{Methodology}\label{sec:methodology}
While approximating the stochastic solution operator in finite dimensions yields a tractable formulation of the forward problem under fixed discretizations of the input and solution domains, it restricts inference to fixed domain locations.
Here, we first present a flow matching operator that extends previous work~\cite{kerrigan2023functional,shi2025stochastic} to the forward problem, enabling resolution-generalizable inference of $w$ at arbitrary locations $x_w \in \Omega_w$ given any $a \in \mathcal{A}$, while characterizing the full distribution over $w$ induced by the latent variable $\varepsilon$.
Building on this, we introduce the central contribution of this work: a residual-augmented formulation of the flow matching operator, called $\floral$, that leverages inexpensive low-fidelity approximations of $\mathcal{G}^\varepsilon$ to improve the data-efficiency of characterizing the full solution distribution.

\subsection{Flow Matching Operator for Solving the Forward Problem}\label{sec:operator_view_pde}
\emph{The key idea for an operator view for solving the forward problem via flow matching is to model the vector field as an operator defined on the product space $\mathcal{W} \times \mathcal{A}$}.
Consider a joint probability measure $\nu_1$ on the measurable space $(\mathcal{W} \times \mathcal{A}, \mathcal{B}(\mathcal{W}) \otimes \mathcal{B}(\mathcal{A}))$ that disintegrates as $\nu_1(dw, da) = \nu_1(dw\vert a) \nu_a(da)$, where $\nu_1(\cdot\vert a)$ is a conditional probability measure defined in $\mathcal{B}(\mathcal{W})$ and $\nu_a$ is the measure defined on $\mathcal{B}(\mathcal{A})$ for the input function $a$.
Here, $\mathcal{B}(\cdot)$ denotes the Borel $\sigma$-algebra.
The forward problem in this operator view amounts to characterizing the conditional probability measure $\nu_1(\cdot\vert a)$.

Following Kerrigan et al.~\cite{kerrigan2023functional}, we can approximate $\nu_1(\cdot\vert a)$ by introducing a conditional flow transformation $\varphi_{\tau, a}^\zeta: \mathcal{W}\to\mathcal{W}$ characterized by the ODE
\begin{align*}
    \frac{d}{d\tau}\varphi_{\tau, a}^\zeta[w] = \mathcal{F}_{\tau, a}^\zeta[\varphi_{\tau, a}^\zeta[w]], \quad \tau\in [0, 1],\\
    \varphi_{0, a}^\zeta[w] = w\sim \nu_0.
\end{align*}
Here, $\mathcal{F}_{\tau, a}^\zeta: \mathcal{W} \ra \mathcal{W}$ is a conditional vector field operator with parameter $\zeta\ireal{n_\zeta}$ that maps between function spaces, and $\nu_0$ is the prior measure.
Specifically, we choose $\nu_0$ as a Gaussian measure $\mathcal{GP}(m_0, C_0)$ with mean function $m_0: \Omega_w \ra \real{m_w}$ and trace-class covariance operator $C_0: \mathcal{W} \ra \mathcal{W}$ for which the absolute continuity is well known~\cite{bogachev1998gaussian,shi2025stochastic}.
By conditioning the vector field operator on $a$, the model learns a single amortized representation that generalizes across inputs $a$, rather than requiring a separate flow transformation for each instance.
This significantly reduces the learning complexity while enabling the conditional vector field operator to adapt its dynamics to the input.
The conditional flow $\varphi_{\tau, a}^\zeta$ induces a push-forward of the prior measure given as $\eta_\tau(\cdot\vert a;\zeta) \define [\varphi_{\tau, a}^\zeta]_{\#}\nu_0$ that defines a time-varying conditional probability measure path. 
To approximate the conditional probability measure $\nu_1(\cdot\vert a)$, we seek a conditional vector field operator $\mathcal{F}_{\tau, a}^\zeta$ such that $\eta_1(\cdot\vert a;\zeta)\approx \nu_1(\cdot\vert a)$.
Given we know the conditional vector field operator $\mathcal{F}_{\tau, a}^\zeta$ with $\eta_1(\cdot\vert a;\zeta)\approx \nu_1(\cdot\vert a)$, we can learn its parametric approximation $\mathcal{H}_{\tau, a}^\xi: \mathcal{W} \ra \mathcal{W}$ with learnable parameters $\xi\in\real{n_\xi}$ by minimizing the operator flow matching objective
\begin{align*}
    J_{\text{OFM}}^\dagger(\zeta, \xi, a) \define \mathbb{E}_{\tau \sim \mathcal{U}(0, 1), w\sim \eta_\tau(\cdot\vert a;\zeta)}\left[\norm[\mathcal{W}]{\mathcal{F}_{\tau, a}^\zeta[w] - \mathcal{H}_{\tau, a}^\xi[w]}^2\right],
\end{align*}
as shown by Kerrigan et al.~\cite{kerrigan2023functional}.
However, similar to the finite-dimensional setting the operator $\mathcal{F}_{\tau, a}^\zeta$ that induces $\eta_1(\cdot\vert a;\zeta)\approx \nu_1(\cdot\vert a)$ is unknown.

To obtain a tractable learning objective, we can consider a conditional coupling $\kappa(\cdot\vert a)$ between the prior measure $\nu_0$ and the conditional probability measure $\nu_1(\cdot\vert a)$.
A tractable choice would be model $\kappa(\cdot\vert a)$ as an independent coupling, such that $z\sim \kappa(\cdot | a)$ is given as $z = \{ w_0, w_1 \}$, where $w_0\sim \nu_0$ and $w_1\sim \nu_1(\cdot \vert a)$.
Following~\cite{shi2025stochastic}, we can then define the conditional probability measure path $\eta_\tau(\cdot \vert a;\zeta)$ via the disintegration
\begin{align*}
    \eta_\tau(X\vert a;\zeta) = \int_{\mathcal{W}\times \mathcal{W}} \eta_\tau(X\vert a, z;\zeta)\, \kappa(dz\vert a), \quad \forall X\in\mathcal{B}(\mathcal{W}), a\in\mathcal{A},
\end{align*}
where $\eta_\tau(\cdot\vert a, z;\zeta)$ is a conditional latent probability measure path designed to approximate Dirac measures $\delta_{w_0}$ and $\delta_{w_1}$ for $\tau=0$ and $\tau=1$, respectively.
Therefore, we can express the operator $\mathcal{F}_{\tau, a}^\zeta$ using a conditional latent vector field operator $\tilde{\mathcal{F}}_{\tau, a, z}^\zeta: \mathcal{W}\to\mathcal{W}$ that induces $\eta_\tau(\cdot\vert a,z;\zeta)$ as
\begin{align}
    \mathcal{F}_{\tau, a}^\zeta \define \int_{\mathcal{W}\times \mathcal{W}} \tilde{\mathcal{F}}_{\tau, a, z}^\zeta \frac{d\eta_\tau(\cdot\vert a, z;\zeta)}{d\eta_\tau(\cdot\vert a;\zeta)} \kappa(dz\vert a).
    \label{eq:operator_vector_field}
\end{align}
\Cref{eq:operator_vector_field} extends~\cite[Theorem 1]{kerrigan2023functional} in a straightforward manner to a conditional setting where the conditional flow transformation amortizes the cost of learning separate flows for a given input $a$.
As a result, we can define a learning objective using the conditional latent vector field operator $\tilde{\mathcal{F}}_{\tau, a,z}^\zeta$ as
\begin{align}
J_{\text{COFM}}^\dagger(\zeta, \xi, a)
\define\;
&\mathbb{E}_{\tau \sim \mathcal{U}(0,1), z \sim \kappa(\cdot \mid a),\;
w \sim \eta_\tau(\cdot \mid a, z;\zeta)}
\Bigl[
\bigl\|
\gamma_\tau \bigl(
\tilde{\mathcal{F}}_{\tau,a,z}^\zeta[w]
- \mathcal{H}_{\tau,a}^\xi[w]
\bigr)
\bigr\|_{\mathcal{W}}^2
\Bigr],
\label{eq:pde_conditional_operator_flow_matching}
\end{align}
which results in the same minimizer as the objective $J_{\text{OFM}}^\dagger(\zeta, \xi, a)$.
The equivalence between the learning objectives can be obtained similarly to the finite-dimensional setting (see Theorem~\ref{theorem:finite_dim_unbiased_estimator}) where the conditional vector field operator $\mathcal{F}_{\tau, a}^\zeta$ is defined using~\cref{eq:operator_vector_field}, and inner products are defined in the Hilbert space.

To tractably sample from $\eta_\tau(\cdot\vert a, z;\zeta)$, we can construct it as a Gaussian measure $\mathcal{GP}(\mu_\tau(z), C_\tau(z; \zeta))$ with mean function $\mu_\tau: \Omega_w \ra \real{m_w}$ and trace-class covariance operator $C_\tau: \mathcal{W} \ra \mathcal{W}$ given as
\begin{align*}
    \mu_\tau(z) \define (1-\tau) w_0 + \tau w_1, \quad z\sim\kappa(\cdot\vert a),\\
    C_\tau(z;\zeta) \define \sigma_\tau^2 C_0,
\end{align*}
which remains trace-class given a trace-class covariance operator $C_0$ of the prior measure $\nu_0$ and finite time-dependent scaling $\sigma_{\tau}\ireal{}_{>0}$.
Specifically, we consider $\sigma_\tau = \sigma_{\min} \norm[\mathcal{W}]{w_1 - w_0}$ to ensure high signal-to-noise ratio when the samples $w_0$ and $w_1$ are close to each other in $\norm[\mathcal{W}]{\cdot}$ sense.
Here, $\sigma_{\min}$ is a hyper-parameter.
For such a transformation of Gaussian measure, the conditional latent vector field operator $\tilde{\mathcal{F}}_{\tau, a, z}^\zeta = w_1 - w_0$, as shown by~\cite{kerrigan2023functional}.
Finally, to learn a vector field $\mathcal{H}_{\tau, a}^\xi$ that generalizes well across inputs, we can consider the training objective
\begin{align}
J_{\text{COFM}}(\zeta, \xi) \define \expect[a\sim\nu_{a}]{J_{\text{COFM}}^\dagger(\zeta, \xi, a) }.
\end{align}

Formulating the forward problem in an operator setting using flow matching characterizes the distribution of the solution function $w=\mathcal{G}^\varepsilon[a]$ under the uncertainty induced due to $\varepsilon$.
Moreover, the operator view enables querying the solution function $w$ at arbitrary domain locations $x_w\in\Omega_w$ without model retraining, thereby naturally supporting flexible inference across varying discretizations and resolutions.

\subsection{Solving the Forward Problem in Residual Space}
While flow matching operators enable resolution-generalizable inference, learning $\mathcal{G}^\varepsilon$ directly from high-fidelity data remains prohibitively expensive due to the limited availability of high-fidelity solution realizations.

To improve data-efficiency, we introduce a low-fidelity inductive bias through residual-augmented learning: rather than learning the full distribution of $\mathcal{G}^\varepsilon[a]$ from scratch, we leverage an inexpensive low-fidelity approximation $\mathcal{G}^\varepsilon_\ell[a]$ that captures the dominant large-scale structures of the solution field, and learn the distribution of the residual operator
\begin{align}
    \mathcal{R}^\varepsilon[a] \define \mathcal{G}^\varepsilon[a] - \mathcal{G}_\ell^\varepsilon[a],
\end{align}
which is typically simpler to characterize probabilistically than the full distribution of the solution.
This decomposition is effective under the assumption that $\mathcal{G}_\ell^\varepsilon$ can be obtained at substantially lower computational cost while retaining non-trivial correlation with $\mathcal{G}^\varepsilon$.
The high-fidelity solution is then recovered as $\mathcal{G}^\varepsilon[a] = \mathcal{G}_\ell^\varepsilon[a] + \mathcal{R}^\varepsilon[a]$, where the low-fidelity term provides a cheap structural prior, and the learned residual operator corrects it toward the true solution distribution.
We refer to this framework of probabilistic residual operator learning as \textsc{floral}.

\begin{figure}[ht!]
    \centering
    \includegraphics[width=1.0\textwidth]{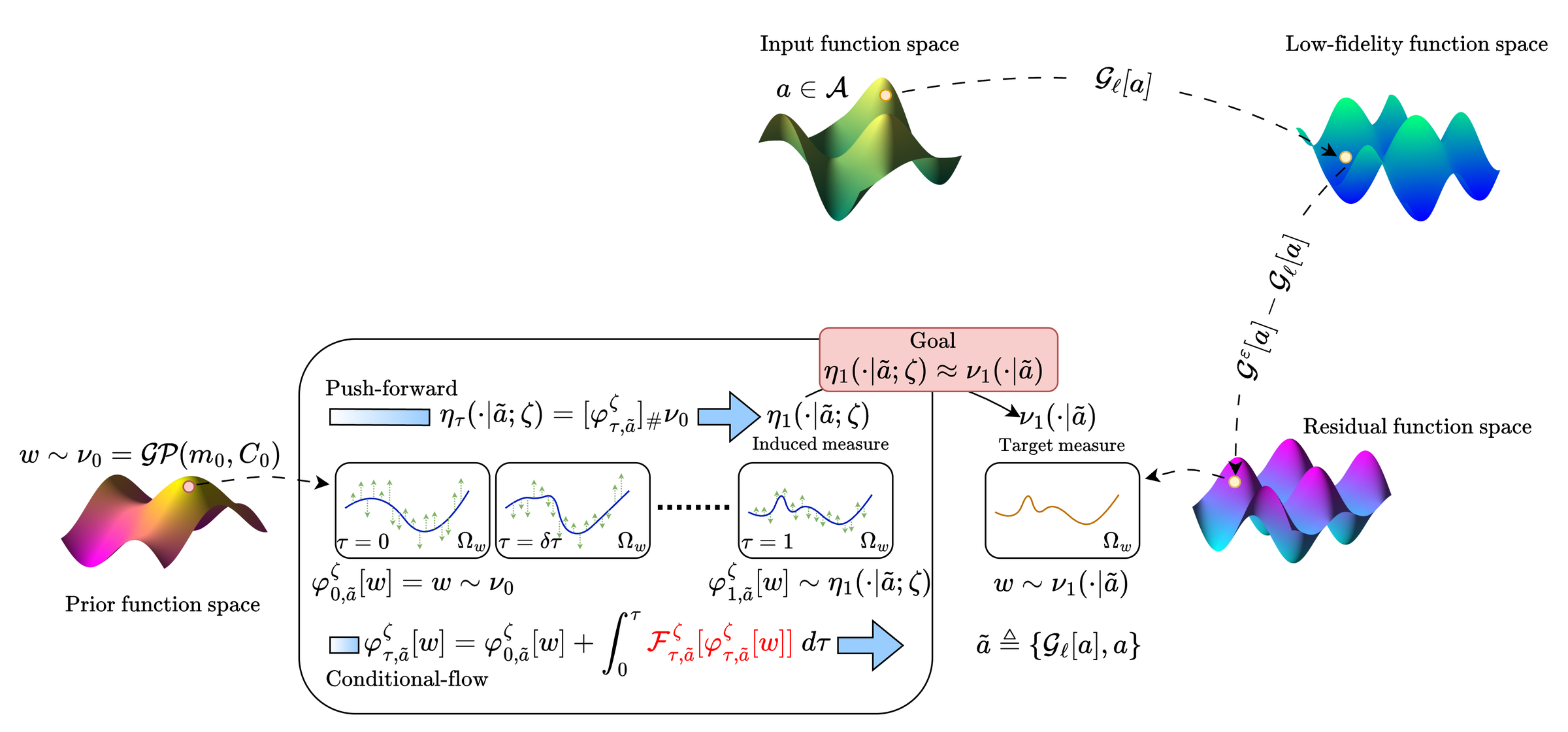}
    \caption{\em Schematic of operator view of flow matching for solving the forward problem in the residual function space by defining an augmented conditioning variable $\tilde{a}=\{\mathcal{G}_\ell[a], a\}$ informed by the low-fidelity solution function.
    During training, the model leverages the available low-fidelity solution to learn the distribution of the residual function $\mathcal{R}^\varepsilon[a] 
= \mathcal{G}^\varepsilon[a] - \mathcal{G}_{\ell}[a]$.
    For inference, the high-fidelity solution function is constructed by generating samples of the residual function given the input $a$ and the low-fidelity approximation $\mathcal{G}_\ell[a]$, followed by performing an additive correction to the low-fidelity approximation.
    }
    \label{fig:methodology/residual_flow}
\end{figure}

Although the low-fidelity solution operator may itself be stochastic, we restrict our analysis to deterministic low-fidelity operators in order to isolate the impact of incorporating low-fidelity information on predictive accuracy, without introducing additional correlations arising from shared randomness.
Under this setting, the residual operator can be expressed as
\begin{align*}
\mathcal{R}^\varepsilon[a] 
= \mathcal{G}^\varepsilon[a] - \mathcal{G}_{\ell}[a],
\end{align*}
where $\mathcal{G}_{\ell}$ denotes a deterministic low-fidelity solution operator.
Importantly, the residual operator depends on the input $a$ and the low-fidelity solution.
Consequently, we define an augmented conditioning variable
\begin{align*}
\tilde{a} = \left\{\mathcal{G}_{\ell}[a], a\right\},
\end{align*}
and seek to characterize the conditional distribution induced by the residual operator using flow matching operators.
An illustration of the proposed framework is shown in~\cref{fig:methodology/residual_flow}.
Note that, compared to directly learning in the full solution function space, learning in the residual space incurs only the additional cost of evaluating the low-fidelity solution function, which is assumed to be available at marginal cost.

\subsubsection{Modeling the Residual Vector Field}
\begin{figure}[ht!]
    \centering
    \includegraphics[width=0.8\textwidth]{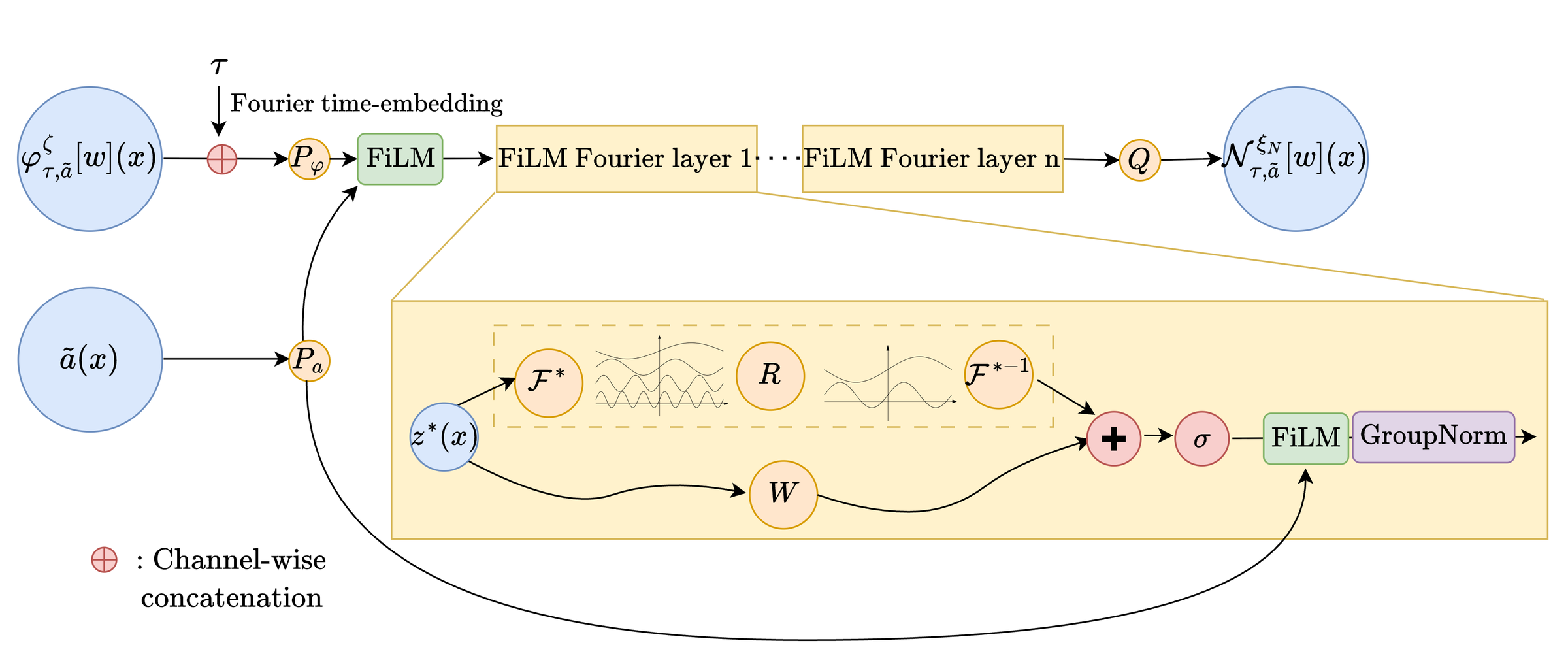}
    \caption{\em Schematic of the proposed FiLMFNO architecture for modeling the nonlinear operator $\mathcal{N}_{\tau, \tilde{a}}^{\xi_N}$.
    The projection operators $P_\varphi$ and $P_a$ lift the inputs into a higher-dimensional channel space.
    A sequence of FiLM Fourier layers: each composed of an integral operator, nonlinear activation, and FiLM-based channel-wise modulation.
    The resulting representation is projected back to the target dimension by the operator $Q$.
    Within each FiLM Fourier layer, the input $z^*$ is mapped to Fourier space via $\mathcal{F}^*$, transformed by a linear operator $R$ acting on the retained low-frequency modes, and returned to physical space through the inverse transform $\mathcal{F}^{*-1}$.
    This output is combined with a skip connection obtained by a local linear map $W$, followed by nonlinear activation and FiLM conditioning.}
    \label{fig:methodology/filmfno}
\end{figure}

To solve the forward problem in a residual-augmented formulation using flow matching operators, the learned conditional vector field operator must be informed by the low-fidelity solution function and the known input function.
Here, we model the conditional vector field operator $\mathcal{H}_{\tau, a}^\xi$ using an FNO~\cite{li2020fourier} conditioned on the augmented input function $\tilde{a}$.
To introduce this input conditioning, we first decompose the model condition vector field operator as 
\begin{align*}
\mathcal{H}^\xi_{\tau, \tilde{a}}[w] = \text{FiLM}(\mathcal{L}_{\tau}^{\xi_L}[w]) + \mathcal{N}_{\tau, \tilde{a}}^{\xi_N}[w],
\end{align*}
where $\mathcal{L}_{\tau}^{\xi_L}$ denotes a linear operator and
$\mathcal{N}_{\tau,\tilde{a}}^{\xi_N}$ denotes a nonlinear operator.
Here $\xi_L\ireal{\xi_L}$ and $\xi_N\ireal{\xi_N}$ denote the learnable parameters of the linear and nonlinear operator.
To condition the linear operator on the augmented inputs, we leverage feature-wise linear modulation (FiLM) layers~\cite{perez2018film}.
The motivation behind this decomposition is to provide a lightweight global conditioning pathway through the linear operator while retaining expressive nonlinear modeling capacity through $\mathcal{N}_{\tau,\tilde{a}}^{\xi_N}$.
In particular, the linear branch introduces an inexpensive skip pathway based on spectral and pointwise linear transformations, enabling efficient propagation of coarse global correlations and input-dependent modulation without substantially increasing the computational complexity of the vector field.
Specifically, following the decomposition introduced in FNO, the linear component is parameterized as
\begin{align*}
\mathcal{L}_{\tau}^{\xi_L}[w]
=
\mathcal{K}_{\tau}^{\xi_L}[w]
+
\mathcal{W}_{\tau}^{\xi_L}[w],
\end{align*}
where $\mathcal{K}_{\tau}^{\xi_L}$ is a spectral convolution operator
that captures global spatial correlations through truncated Fourier modes,
while $\mathcal{W}_{\tau}^{\xi_L}$ is a point-wise $1 \times 1$
convolution operator responsible for local channel mixing.
Since both branches are linear and no activation function is applied, their sum remains a linear operator.
To model the nonlinear component, we propose FiLMFNO that leverages FiLM layers to introduce input-dependent feature modulation while retaining the expressive nonlinear representation capabilities of a traditional FNO. 
An illustration of FiLMFNO is shown in~\cref{fig:methodology/filmfno}.
To form the auxiliary input conditioning, we consider a simple channel-wise concatenation of $a$ and $\mathcal{G}_\ell[a]$.
This implicitly assumes that both quantities are defined on a common domain representation.
We emphasize that this assumption is introduced purely for simplicity and is not fundamental to the proposed framework.
More sophisticated conditioning mechanisms, including cross-attention, latent alignment, or multi-resolution conditioning strategies, could also be considered, but are beyond the scope of the present work.
See Appendix~\ref{sec:training_details} for additional architectural details.

\section{Numerical Experiments}\label{sec:experiments}
In this section, we evaluate $\floral$ on a range of synthetic probabilistic PDEs in regimes where a high-fidelity solution operator is computationally expensive or where a low-fidelity solution operator fails to capture important solution features. 
The goal is to assess whether, under a fixed computational budget, predictive performance can be improved by incorporating an inexpensive low-fidelity inductive bias and learning the residual operator rather than learning the high-fidelity solution operator directly.
As deterministic baselines, we train a standard FNO on high-fidelity data and a residual-learning variant, referred to as Residual FNO, that learns the discrepancy between the low- and high-fidelity solution operators.
As a probabilistic baseline, we train $\flora$, which directly learns the high-fidelity solution operator using the flow matching operator presented in~\S\ref{sec:operator_view_pde}.
These baselines are used to assess the benefit of residual-space learning with low-fidelity inductive bias.

For all experiments, we fix training to $300$ epochs and share the same underlying conditional vector field operator architecture across all baselines to ensure a fair comparison.
See Appendix~\ref{sec:training_details} for additional training details.
Specifically, the conditional vector field operator used for probabilistic operator learning is also employed in the deterministic operator learning variants, without the temporal conditioning variable $\tau$.
This ensures that performance differences reflect the learning formulation rather than model capacity.
To evaluate performance, we report the root-mean-square error (RMSE) and normalized RMSE (nRMSE) following PDEBench~\cite{takamoto2022pdebench}, along with the mean $L_2$ error of the predicted mean and variance fields.
The reported error in variance fields estimates how accurately each model captures the uncertainty structure induced by the latent variable $\varepsilon$.

\subsection{Problem Description}\label{sec:problem_settings}
For all experiments, the low-fidelity solution operator is constructed to have model-form uncertainty introduced through either (a) coarser spatial discretizations or (b) spectral filtering of the initial conditions.
These settings reflect realistic scientific computing scenarios in which inaccurate, low-fidelity priors are available to inform high-fidelity solutions.
In this work, we test the following families of probabilistic PDEs.

\subsubsection{Artificial Benchmark: 1D Correlated Residual (1DCR)}\label{sec:1dcr_setup} 
We consider a low-fidelity solution operator
\begin{align*}
\mathcal{G}_\ell[a](x) = \sin(a(x)) + x - 0.25 a(x),\qquad x\in[0,1],
\end{align*}
where $a(x)=kx - 4$ and $k\sim\mathcal{U}(10,14)$.
We can then construct a high-fidelity stochastic solution operator 
\begin{align*}
    \mathcal{G}^\varepsilon[a](x) = \mathcal{G}_\ell[a](x) + \alpha \mathcal{G}_\ell[a](x) + \beta \cos(2a(x)) + \gamma \varepsilon,
\end{align*}
where $\varepsilon$ is sampled from a zero-mean Gaussian process with a radial basis function covariance kernel with length scale $ 0.2$ and amplitude $0.25$. After sampling the Gaussian random field, we retain the first $10$ modes of its Karhunen--Loève (KL) expansion.
We consider $\alpha=0.1$, $\beta=0.2$, and $\gamma=0.3$.
The high-fidelity stochastic solution operator is constructed such that the residual 
$\mathcal{G}^{\varepsilon}[a]-\mathcal{G}_{\ell}[a]$ remains correlated with the low-fidelity solution operator.
For constructing the high- and low-fidelity datasets, we consider $128$ uniformly spaced locations in $\Omega_w$.


\subsubsection{1D Advection Equation with Stochastic Initial Conditions (Advection)}\label{sec:advection_setup} 
We consider the 1D advection equation that models the transport of a conserved scalar field by a known velocity field given by
\begin{align*}
\frac{\partial u(x, t)}{\partial t} + \beta \frac{\partial u(x, t)}{\partial x} &= 0, \quad x \in [0, 1], \; t \in [0, 1],
\end{align*}
where $\beta$ is the advection velocity and $u(x,0)$ denotes the initial condition given by $u(x,0)=a(x) (1 + \varepsilon)$ with
\begin{align*}
a(x) = \sum_{i=1}^{N} A_i \sin(k_i x + \phi_i),
\label{eq:advection/initial_condition}
\end{align*}
where $k_i = 2\pi n_i$ are the wave numbers with $n_i\sim\mathcal{U}(\{1, \cdots, n_{\max}\})$.
Here, $N$ determines the number of superpositions of sine waves with random amplitudes $A_i\sim \mathcal{U}( 0, 1 )$ and random phase shifts $\phi_i\sim \mathcal{U}(0, 2\pi)$.
Here, we set $n_{\max} = 8$, $N=2$, $\beta=0.05$, and $\varepsilon$ is defined as
\begin{align*}
\varepsilon = \sum_{m=1}^{M}
\alpha_m \cos\left(2\pi m x\right)
+
\beta_m \sin\left(2\pi m x\right)
,
\end{align*}
where $\alpha_m,\beta_m \sim \mathcal{N}(0,\sigma_m^2)$.
We choose $M=8$ and $\sigma_m=0.05$ for all our experiments.

To obtain the high- and low-fidelity solution, we consider $N_x = 128$ spatial grid points and $N_t = 128$ temporal steps.
However, we introduce model-form error in the initial condition for the low-fidelity solution.
Specifically, to construct the low-fidelity initial condition, we apply a spectral filter to the input $a$ to remove high-frequency components and introduce amplitude distortion.
Let $\hat{a}_{\text{h}}(k)$ for $k=0,\cdots K$ be the Fourier transform of $a$, then the low-fidelity spectrum is constructed as $\hat{a}_{\ell}(k) = s_A \hat{a}_{h}(k) r_k e^{- \gamma k / K}$,
where $r_k$ retains only $\max(2, \lfloor f_{\text{keep}} K\rfloor )$ Fourier modes, $e^{ - \gamma k / K}$ introduces a mode-dependent amplitude damping, and $s_A$ scales the amplitude.
We consider $f_{\text{keep}} = 0.4$, $\gamma = 4$, and $s_A = 0.8$ for all the experiments.
The low-fidelity initial condition is then obtained by applying the inverse Fourier transform to $\hat{a}_{\ell}(k)$.
Note, the low-fidelity initial condition does not contain any stochasticity introduced via $\varepsilon$.
To obtain the numerical solution, we solve the advection equation using the first-order upwind finite-difference scheme and the forward Euler method for time integration.

\subsubsection{1D Burgers' Equation with Stochastic Initial Conditions (Burgers')}
We consider the 1D viscous Burgers' equation, a hyperbolic partial differential equation regularized by a diffusive term,
\begin{align*}
\frac{\partial u(x, t)}{\partial t} + u(x, t) \frac{\partial u(x, t)}{\partial x} = \nu \frac{\partial^2 u(x, t)}{\partial x^2}, \quad x\in[0, 1], t\in [0, 0.2],
\end{align*}
which models the nonlinear advection-diffusion process in fluid dynamics.
Here, $\nu$ is the viscosity coefficient, which is assumed to be $0.01$.

To obtain the high- and low-fidelity solution, we consider $N_x = 128$ spatial grid points and $N_t = 128$ temporal steps.
Initial conditions are formed similar to~\S\ref{sec:advection_setup} with $f_{\text{keep}}=0.6$. 
Numerical solution is obtained by using a second-order upwind finite-difference scheme for the advection terms, a second-order central finite-difference scheme for the diffusion term, and a third-order strong stability preserving Runge-Kutta scheme for time integration.

\subsubsection{Flow Through Porous Media with Uncertain Permeability field (Darcy)}
We consider the flow through porous media characterized by Darcy's law that relates pressure $p(\vect{x})$ and permeability $K(\vect{x})$ on $\Omega=[0,1]^2$:
\begin{subequations}
\begin{align*}
\vect{u}(\vect{x}) &= -K(\vect{x}) \nabla p(\vect{x}), \\
\nabla\cdot \vect{u}(\vect{x}) &= f_s(\vect{x}), \\
\vect{u}(\vect{x})\cdot \hat{n}(\vect{x}) &= 0, \quad \vect{x}\in\partial\Omega, \\
\int_\Omega p(\vect{x})\, d\vect{x} &= 0,
\end{align*}
\end{subequations}
where $\hat{n}$ denotes the outward unit normal.
Following~\cite{zhu2018bayesian}, the source term $f_s$ models an injection well at the bottom-left and a production well at the top-right corner, with rate $r=50$ and size $s=0.125$.
Mathematically, the source term is defined as
\begin{align*}
f_s(\vect{x}) =
\begin{cases}
    r, & \text{if } \left| x_i - \frac{1}{2} \right| \le \frac{1}{2} s, \quad i = 1, 2, \\
    -r, & \text{if } \left| x_i - \left( 1 - \frac{1}{2} s \right) \right| \le \frac{1}{2} s, \quad i = 1, 2, \\
    0, & \text{otherwise}.
\end{cases}
\end{align*}

To obtain the permeability field, we first define
$a(\vect{x})=\exp(G(\vect{x}))$, where
$G\sim\mathcal{GP}(\vect{0}, k(\vect{x},\vect{x}'))$ with covariance kernel
$k(\vect{x},\vect{x}')=\exp(-\|\vect{x}-\vect{x}'\|_2/l)$ and $l=0.1$.
We then use a KL expansion of $G$ and retain the first $32$ modes to define the baseline permeability field $a(\vect{x})$.
The stochastic high-fidelity permeability field is constructed as
$K(\vect{x}) = a(\vect{x})(1+\varepsilon(\vect{x}))$.
Here, $\varepsilon$ is sampled from a zero-mean Gaussian process with length scale
$0.2$ and amplitude $1.0$. After sampling $\varepsilon$, we use its KL expansion and retain the first $10$ modes.
For the high-fidelity solution, we solve the governing equations on a uniform $64\times64$ spatial grid.
The low-fidelity solution is obtained by solving the governing equations on a $16\times16$ spatial grid, where the corresponding input permeability field is constructed via strided subsampling of $a(\vect{x})$.
The governing equations are solved using second-order central finite differences, enforcing boundary conditions via Karush–Kuhn–Tucker conditions.
Compared to the high-fidelity solution, the low-fidelity solution can be obtained at approximately $60\times$ lower computational cost.
Since the low-fidelity operator is defined on a coarser spatial discretization, we first interpolate the low-fidelity solution onto the high-fidelity grid using linear interpolation via \texttt{RegularGridInterpolator} available on \texttt{SciPy} before constructing the residual for operator learning.

\subsection{Advantages of Low-fidelity Inductive Bias}\label{sec:predictive_performance}
\begin{figure}[htbp]
    \centering
    \includegraphics[width=0.99\linewidth]{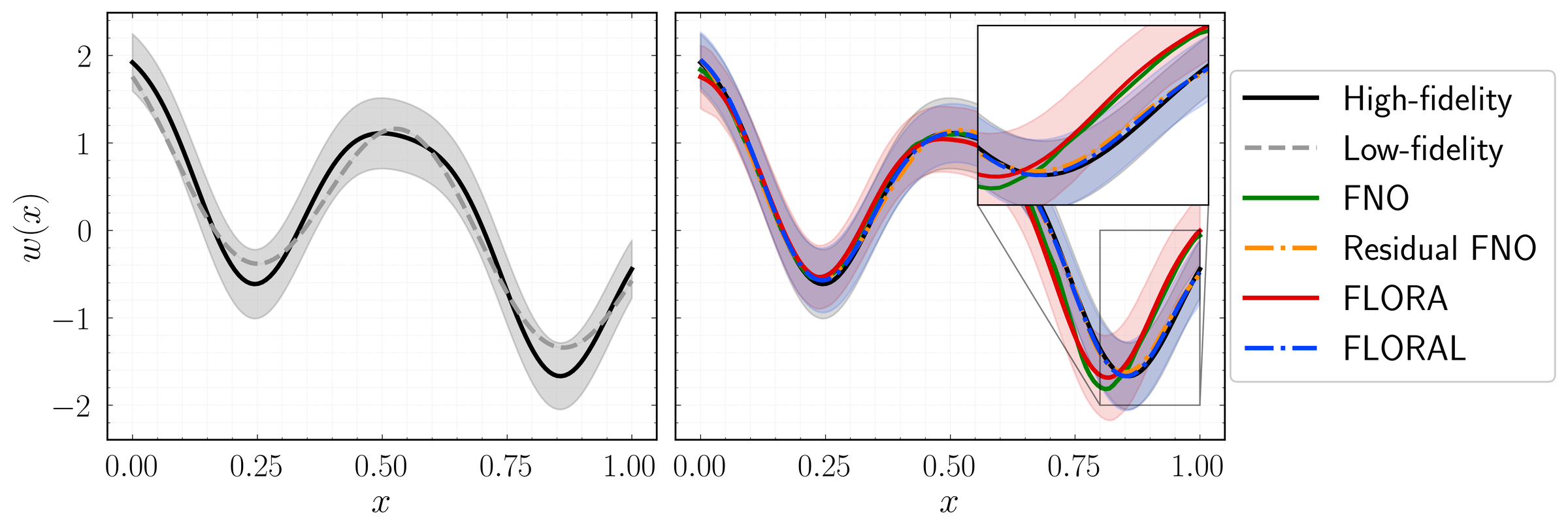}
    \caption{\em Predictive performance for approximating the solution function at an unseen input condition for the 1DCR problem in the low-data regime. The model is trained using a total of $100$ training samples, corresponding to $50$ realizations of the stochastic solution function for each input condition.
    (\textbf{Left}) Comparison between the low- and high-fidelity solution functions. (\textbf{Right}) Predictions obtained using deterministic and probabilistic operator learning methods. Shaded regions denote $\pm 5$ standard deviations about the mean prediction. In the low-data regime, residual-space learning improves the mean predictive accuracy compared to directly learning the high-fidelity solution operator and enables $\floral$ to accurately capture the underlying uncertainty structure of the solution function.}
\label{fig:onedcorr/mean_std_n_train_100_n_val_750_train_res_full}
\end{figure}

\begin{figure}[htbp]
    \centering
    \includegraphics[width=0.99\linewidth]{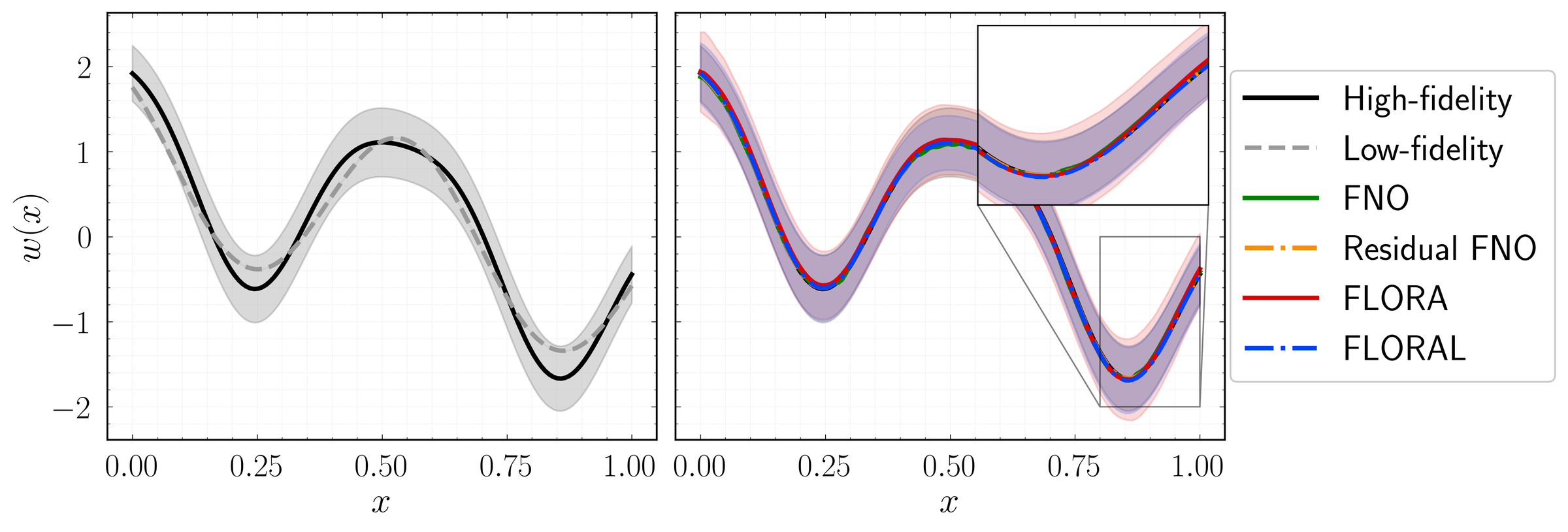}
    \caption{\em Predictive performance for approximating the solution function at an unseen input condition (same as~\cref{fig:onedcorr/mean_std_n_train_100_n_val_750_train_res_full}) for the 1DCR problem in the high-data regime. The model is trained using a total of $500$ training samples, corresponding to $50$ realizations of the stochastic solution function for each input condition.
    (\textbf{Left}) Comparison between the low- and high-fidelity solution functions. (\textbf{Right}) Predictions obtained using deterministic and probabilistic operator learning methods. Shaded regions denote $\pm 5$ standard deviations about the mean prediction.
    }
    \label{fig:onedcorr/mean_std_n_train_500_n_val_750_train_res_full}
\end{figure}

We first qualitatively examine the predictive performance on the toy 1DCR problem.
\Cref{fig:onedcorr/mean_std_n_train_100_n_val_750_train_res_full} illustrates the predictive performance of different methods in approximating the solution function at an unseen input condition in the low-data regime.
As shown, the accuracy of the mean prediction improves by considering the residual-space learning approach for both deterministic and probabilistic methods.
Furthermore, while both $\flora$ and $\floral$ are able to capture the uncertainty induced by the stochastic solution operator, $\floral$ provides improved uncertainty characterization together with more accurate mean predictions compared to directly learning the high-fidelity stochastic solution operator using $\flora$.
The improvement in predictive accuracy becomes marginal in the high-data regime, where directly learning the high-fidelity stochastic solution operator already provides accurate mean predictions and uncertainty estimates.
This is illustrated in~\cref{fig:onedcorr/mean_std_n_train_500_n_val_750_train_res_full}.

Quantitative comparison of predictive accuracy across various probabilistic PDEs for varying numbers of training samples is tabulated in~\Cref{tab:predictive_results_vary_training_size}.
Across various benchmarks, residual-space learning consistently improves predictive accuracy compared to directly learning the high-fidelity stochastic solution operator.
In particular, in the low-data regime, directly learning the high-fidelity stochastic solution operator can perform poorly, in some cases yielding accuracy comparable to, or worse than, that of the low-fidelity model itself.
In such settings, directly learning the high-fidelity operator may not provide sufficient benefit over the low-fidelity approximation despite its significantly higher computational cost.
By contrast, learning in the residual space substantially improves predictive performance for both deterministic and probabilistic methods.
\begin{table}[ht!]
\caption{
\em Quantitative comparison of predictive accuracy across probabilistic PDE benchmarks for varying training sizes.
Training datasets are constructed using $50$ realizations of the stochastic solution function for each input condition, while evaluation statistics are computed over $15$ unseen stochastic solution functions.
For probabilistic methods, $2500$ realizations are drawn per input condition.
Residual-space learning improves predictive accuracy, particularly in low-data regimes where directly learning the high-fidelity operator may underperform the low-fidelity baseline.
}
\label{tab:predictive_results_vary_training_size}
\centering
\small
\setlength{\tabcolsep}{2.5pt}
\renewcommand{\arraystretch}{0.95}
\setlength{\aboverulesep}{0.4pt}
\setlength{\belowrulesep}{0.4pt}
\setlength{\cmidrulesep}{1.0pt}
\begin{tabular}{ll c cccc}
\toprule
\textbf{Method} & \textbf{System} &
\makecell{\textbf{\# Train}\\\textbf{samples}} &
\makecell{\textbf{Mean $L_2$}\\\textbf{Error}$\downarrow$} &
\textbf{RMSE}$\downarrow$ & \textbf{nRMSE}$\downarrow$ &
\makecell{\textbf{Mean $L_2$}\\\textbf{Variance Error}}$\downarrow$\\
\midrule
\multirow{3}{*}{Low-fidelity}
  & Advection & -- & 21.98 & $1.9 \times 10^{-1}$ & $3.7 \times 10^{-1}$ &N/A\\
  & Burgers'  & -- & 13.29 & $1.1 \times 10^{-1}$ & $2.9 \times 10^{-1}$ & N/A\\
  & Darcy     & -- & 4.49  & $7.1 \times 10^{-2}$ & $3.0 \times 10^{-1}$&N/A\\
\midrule
FNO          & Advection & 100  & 29.12 & $2.7 \times 10^{-1}$& $5.2 \times 10^{-1}$ & N/A\\
             &           & 500  & 14.05 & $1.4 \times 10^{-1}$& $2.8 \times 10^{-1}$ & N/A\\
             &           & 2000 & 1.77  & $1.5 \times 10^{-2}$& $3.0 \times 10^{-2}$ & N/A\\
\cmidrule(lr){2-7}
             & Burgers'  & 100  & 23.13 & $2.3 \times 10^{-1}$ & $6.0 \times 10^{-1}$ & N/A\\
             &           & 500  & 18.91 & $2.1 \times 10^{-1}$& $5.3 \times 10^{-1}$& N/A\\
             &           & 2000 & 5.01 & $7.1 \times 10^{-2}$ & $1.8 \times 10^{-1}$&N/A\\
\cmidrule(lr){2-7}
             & Darcy     & 100  & 3.99  & $6.5 \times 10^{-2}$  & $2.7 \times 10^{-1}$ & N/A\\
             &           & 500  & 2.62  & $4.2 \times 10^{-2}$  & $1.7 \times 10^{-1}$& N/A\\
             &           & 2000 & 1.42  & $2.3 \times 10^{-2}$  & $1.0 \times 10^{-1}$ &N/A\\
\midrule
Residual FNO & Advection & 100  & 11.71 & $1.1 \times 10^{-1}$ & $2.2 \times 10^{-1}$&N/A\\
             &           & 500  & 5.03  & $4.8 \times 10^{-2}$& $9.4 \times 10^{-2}$ & N/A\\
             &           & 2000 & 1.35  & $1.1 \times 10^{-2}$& $2.2 \times 10^{-2}$ &N/A\\
\cmidrule(lr){2-7}
             & Burgers'  & 100  & 17.32 & $1.5 \times 10^{-1}$ & $3.8 \times 10^{-1}$&N/A\\
             &           & 500  & 5.49  & $5.2 \times 10^{-2}$& $1.3 \times 10^{-1}$&N/A\\
             &           & 2000 & 3.94  & $4.0 \times 10^{-2}$& $1.0 \times 10^{-1}$&N/A\\
\cmidrule(lr){2-7}
             & Darcy     & 100  &  1.00   & $1.6 \times 10^{-1}$ & $6.9 \times 10^{-2}$ &N/A\\
             &           & 500  &  0.93   & $1.5 \times 10^{-2}$ & $6.4 \times 10^{-2}$&N/A\\
             &           & 2000  & 0.62   & $1.0 \times 10^{-2}$ & $4.4 \times 10^{-2}$&N/A\\
\midrule
FLORA        & Advection & 100  & 31.80 & $2.8 \times 10^{-1}$&  $5.4 \times 10^{-1}$ & 4.42\\
             &           & 500  & 13.88 & $1.3 \times 10^{-1}$&  $2.5 \times 10^{-1}$ & $8.0\times10^{-1}$\\
             &           & 2000 & 4.64  & $3.9 \times 10^{-2}$&  $7.6 \times 10^{-2}$&$8.0\times10^{-1}$ \\
\cmidrule(lr){2-7}
             & Burgers'  & 100  & 26.31 & $2.5 \times 10^{-1}$& $6.5 \times 10^{-1}$ & 3.73\\
             &           & 500  & 21.12 & $2.4 \times 10^{-1}$& $6.2 \times 10^{-1}$ & 2.62\\
             &           & 2000 & 9.46 & $8.2 \times 10^{-2}$& $2.0 \times 10^{-1}$ & 2.62\\
\cmidrule(lr){2-7}
             & Darcy     & 100  &  4.71 &  $7.5 \times 10^{-2}$ & $3.2 \times 10^{-1}$ & $6.4\times10^{-1}$\\
             &           & 500  &  2.58 & $4.2 \times 10^{-2}$  & $1.7 \times 10^{-1}$ &$2.3\times10^{-2}$ \\
             &           & 2000  & 1.30 & $2.1 \times 10^{-2}$  & $9.0 \times 10^{-2}$&$2.4\times10^{-2}$\\
\midrule
FLORAL       & Advection & 100  & 10.92 & $9.9 \times 10^{-2}$ & $1.9 \times 10^{-1}$  & 1.25\\
             &           & 500  & 7.11  & $6.4 \times 10^{-2}$ & $1.2 \times 10^{-1}$ & $7.8\times10^{-1}$\\
             &           & 2000 & 3.86  & $3.2 \times 10^{-2}$& $6.3 \times 10^{-2}$&$7.8\times10^{-1}$\\
\cmidrule(lr){2-7}
             & Burgers'  & 100  & 9.50 & $8.5 \times 10^{-2}$& $2.1 \times 10^{-1}$ & 2.99\\
             &           & 500  & 9.18 & $8.5 \times 10^{-2}$& $2.1 \times 10^{-1}$ & 2.50\\
             &           & 2000 & 10.03 & $8.9 \times 10^{-2}$& $2.2 \times 10^{-1}$ &2.56\\
\cmidrule(lr){2-7}
             & Darcy     & 100  &  1.12    & $1.8 \times 10^{-2}$  & $7.8 \times 10^{-2}$ & $2.6\times10^{-2}$\\
             &           & 500  &  0.93    & $1.5 \times 10^{-2}$  & $6.4 \times 10^{-2}$ &$2.7\times10^{-2}$\\
             &           & 2000  & 0.79    & $1.3 \times 10^{-2}$  & $5.6 \times 10^{-2}$ &$2.5\times10^{-2}$ \\
\bottomrule
\end{tabular}
\end{table}

\begin{figure}
    \centering
    \includegraphics[width=0.99\linewidth]{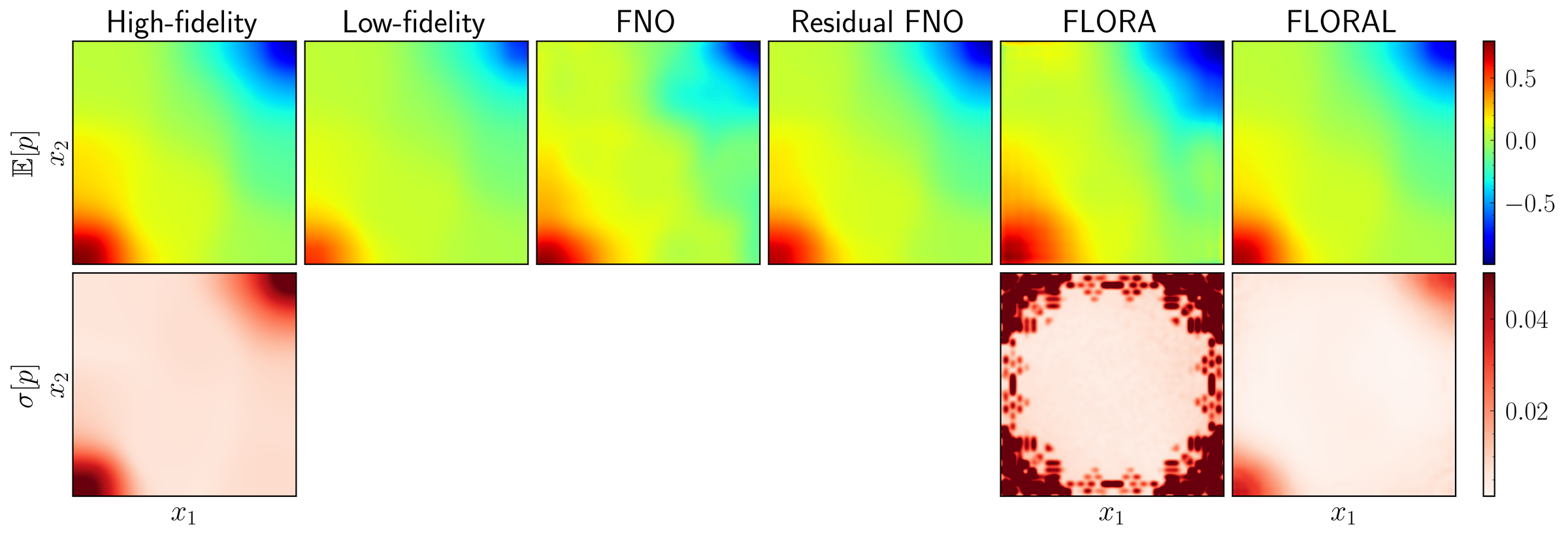}
    \caption{\em Predictive performance for approximating the solution function at an unseen input condition for flow through porous media in the low-data regime.
    The model is trained using a total of $100$ training samples, corresponding to $50$ realizations of the stochastic solution function for each input condition.
    (\textbf{Top}) Mean solution function.
    (\textbf{Bottom}) Standard deviation of the solution function.
    Probabilistic operator learning methods can capture the uncertainty induced by the stochastic solution operator, whereas deterministic methods cannot characterize variability in the solution function.
    Residual-space learning improves mean solution-function prediction and characterizes the uncertainty induced by the solution operator.
    }
    \label{fig:darcy/mean_std_n_train_100_n_val_750_train_res_full}
\end{figure}

While deterministic residual-space learning methods achieve predictive accuracies comparable to those of probabilistic approaches, they cannot characterize the uncertainty induced by the stochastic solution operator.
In contrast, both $\flora$ and $\floral$ are able to model the uncertainty induced by the stochastic solution operator.
Moreover, compared to directly learning the high-fidelity stochastic operator using $\flora$, the residual-space probabilistic formulation $\floral$ provides improved uncertainty characterization, as reflected by the lower variance-field prediction error across several benchmark problems.
An illustration of the predictive performance of different methods for approximating the solution function for the Darcy problem in the low-data regime is presented in~\cref{fig:darcy/mean_std_n_train_100_n_val_750_train_res_full}.
As shown, even with limited training samples, leveraging the low-fidelity solution operator substantially improves the mean predictive accuracy and can provide characterization of uncertainty induced by the high-fidelity stochastic solution operator.
Increasing the number of training samples further improves both the mean predictive accuracy and the characterization of the uncertainty structure, as illustrated in~\cref{fig:darcy/mean_std_n_train_500_n_val_750_train_res_full}.
See Appendices~\ref{sec:appendix/onedcorr}, \ref{sec:appendix/advection}, \ref{sec:appendix/burgers}, and~\ref{sec:appendix/darcy} for additional results.

\begin{figure}
    \centering
    \includegraphics[width=0.99\linewidth]{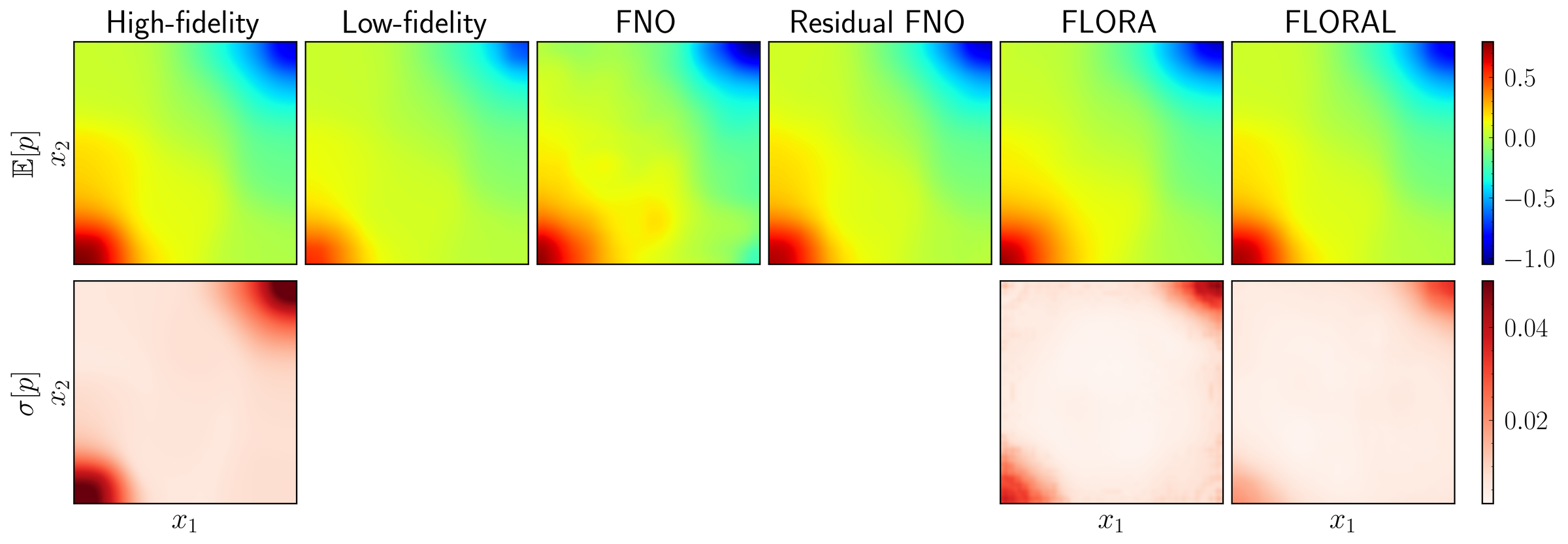}
    \caption{\em Predictive performance for approximating the solution function at an unseen input condition (same as~\cref{fig:darcy/mean_std_n_train_100_n_val_750_train_res_full}) for flow through porous media.
    The model is trained using a total of $500$ training samples, corresponding to $50$ realizations of the stochastic solution function for each input condition.
    (\textbf{Top}) Mean solution function.
    (\textbf{Bottom}) Standard deviation of the solution function.
    } 
    \label{fig:darcy/mean_std_n_train_500_n_val_750_train_res_full}
\end{figure}

\subsection{Computational Cost}\label{sec:computational_complexity}
While deterministic and probabilistic learning approaches achieve comparable mean predictive performance, probabilistic methods additionally enable characterization of the uncertainty induced by the stochastic solution operator.
However, formulating the forward problem in such a probabilistic setting introduces additional computational cost due to both the evaluation of $J_{\text{COFM}}$ and the sampling required during inference for uncertainty characterization.
As shown in \Cref{tab:computational_performance}, probabilistic methods incur nearly an order-of-magnitude increase in training time together with an order-of-magnitude reduction in training throughput compared to deterministic approaches when trained for the same number of epochs.

For inference, directly comparing the inference time per input condition between deterministic and probabilistic methods can be misleading, since probabilistic approaches additionally generate multiple realizations to characterize uncertainty.
Consequently, the reported inference time per input condition for probabilistic methods is nearly three orders of magnitude larger due to the generation of $2500$ realizations per input. A more appropriate comparison is between the deterministic inference time per input and the probabilistic inference time per realization, for which the computational costs are of comparable magnitude.

\begin{table}[ht]
\caption{
\em Quantitative comparison of computational performance across stochastic PDE benchmarks.
Reported metrics include training cost, training throughput, and inference cost.
Training datasets comprised of $500$ samples, corresponding to $50$ realizations of the stochastic solution function for each input condition.
All models were trained for $300$ epochs.
For probabilistic methods, the reported inference time corresponds to generating $2500$ realizations per input condition.
}
\label{tab:computational_performance}
\centering
\small
\setlength{\tabcolsep}{3pt}
\renewcommand{\arraystretch}{0.95}
\setlength{\aboverulesep}{0.4pt}
\setlength{\belowrulesep}{0.4pt}
\setlength{\cmidrulesep}{1.0pt}

\begin{tabular}{llcccc}
\toprule

&
&
\multicolumn{2}{c}{\textbf{Training}} &
\multicolumn{2}{c}{\textbf{Inference}} \\

\cmidrule(lr){3-4}
\cmidrule(lr){5-6}

\textbf{Method} &
\textbf{System} &
\makecell{\textbf{Total Time}\\\textbf{(s)}} &
\makecell{\textbf{Throughput}\\\textbf{(samples/s)}} &
\makecell{\textbf{Time Per}\\\textbf{Input (s)}} &
\makecell{\textbf{Time Per}\\\textbf{Realization (s)}} \\

\midrule

\multirow{3}{*}{FNO}
    & Advection & $4.2\times10^{2}$ & $3.2\times10^{3}$ & $1.4\times10^{-1}$ & -- \\
    & Burgers'  & $3.0\times10^{2}$ & $5.4\times10^{3}$ & $1.2\times10^{-1}$ & -- \\
    & Darcy     & $1.1\times10^{2}$ & $5.0\times10^{3}$ & $9.8\times10^{-2}$ & -- \\
\midrule

\multirow{3}{*}{Residual FNO}
    & Advection & $8.2\times10^{2}$ & $3.2\times10^{3}$ & $1.3\times10^{-1}$ & -- \\
    & Burgers'  & $5.6\times10^{2}$ & $5.3\times10^{3}$ & $1.2\times10^{-1}$ & -- \\
    & Darcy     & $1.3\times10^{2}$ & $4.5\times10^{3}$ & $9.6\times10^{-2}$ & -- \\
\midrule

\multirow{3}{*}{FLORA}
    & Advection & $2.2\times10^{3}$ & $1.5\times10^{2}$ & $4.3\times10^{2}$ & $1.7\times10^{-1}$ \\
    & Burgers'  & $1.6\times10^{3}$ & $1.5\times10^{2}$ & $4.9\times10^{2}$ & $2.0\times10^{-1}$ \\
    & Darcy     & $9.2\times10^{2}$ & $3.7\times10^{2}$ & $8.0\times10^{1}$ & $3.2\times10^{-2}$ \\
\midrule

\multirow{3}{*}{FLORAL}
    & Advection & $1.7\times10^{3}$ & $2.3\times10^{2}$ & $4.7\times10^{2}$ & $1.9\times10^{-1}$ \\
    & Burgers'  & $2.2\times10^{3}$ & $1.5\times10^{2}$ & $4.0\times10^{2}$ & $1.6\times10^{-1}$ \\
    & Darcy     & $9.6\times10^{2}$ & $3.6\times10^{2}$ & $1.0\times10^{2}$ & $4.0\times10^{-2}$ \\

\bottomrule
\end{tabular}
\end{table}
\subsection{Probabilistic Super-resolution}\label{sec:superresolution}
\begin{figure}[htbp]
    \centering
    \includegraphics[width=0.99\linewidth]{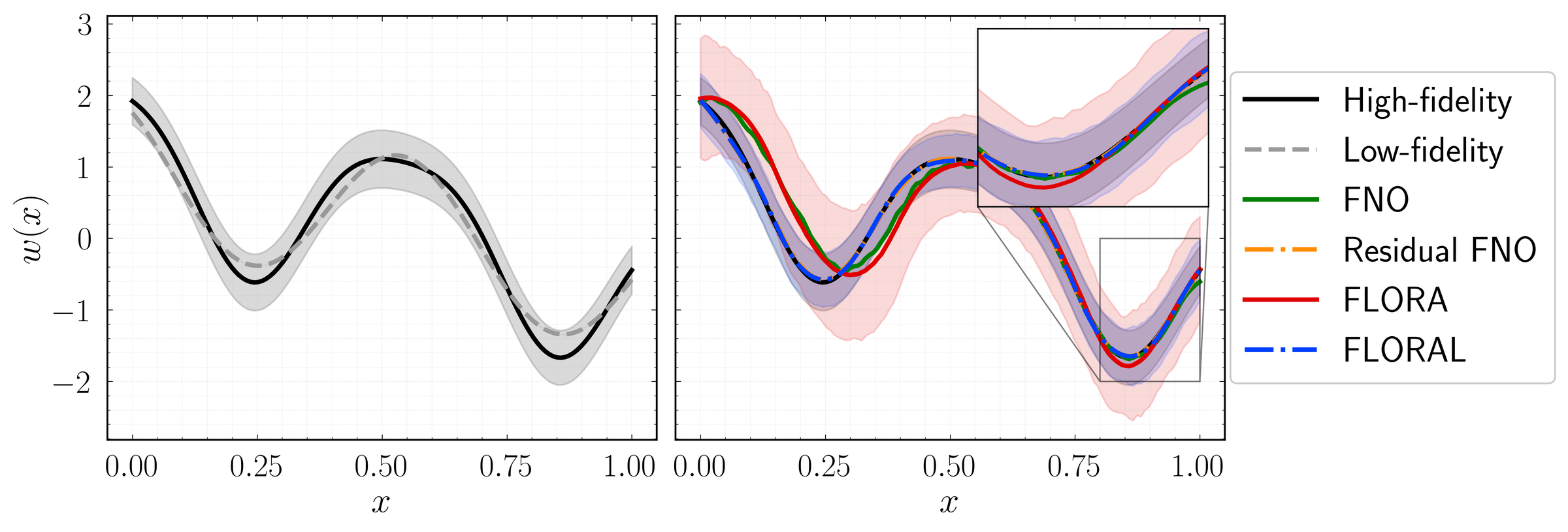}
    \caption{\em Predictive performance for approximating the solution function at an unseen input condition (same as~\cref{fig:onedcorr/mean_std_n_train_500_n_val_750_train_res_full}) for the 1DCR problem. 
    For training, the solution function is evaluated only at $8$ uniformly spaced points in the spatial domain, whereas during inference, the learned operator is evaluated at $128$ spatial locations, corresponding to a $16\times$ super-resolution reconstruction.
    The model is trained using a total of $500$ training samples, corresponding to $50$ realizations of the stochastic solution function for each input condition.
    (\textbf{Left}) Comparison between the low- and high-fidelity solution functions. (\textbf{Right}) Predictions obtained using deterministic and probabilistic operator learning methods. Shaded regions denote $\pm 5$ standard deviations about the mean prediction. 
    Residual-space learning improves super-resolution performance by enhancing both the mean predictive accuracy and the characterization of the uncertainty induced by the stochastic solution operator.
    }
\label{fig:onedcorr/mean_std_n_train_500_n_val_750_train_res_8}
\end{figure}

Formulating the forward problem from an operator-learning perspective enables characterization of the solution as a function over the domain, rather than restricting predictions to fixed discretization points in a finite-dimensional representation.
As a result, once trained, the model can perform resolution-generalizable inference, enabling evaluation of the solution function at unseen spatial discretizations and domain locations.
\Cref{fig:onedcorr/mean_std_n_train_500_n_val_750_train_res_8} illustrates the predictive performance for approximating the same 1DCR solution function shown in~\Cref{fig:onedcorr/mean_std_n_train_500_n_val_750_train_res_full}, but in a super-resolution setting.
During training, the solution function is evaluated only at $8$ uniformly spaced locations in the spatial domain, whereas during inference the learned operator is evaluated at $128$ spatial locations, corresponding to a $16\times$ super-resolution reconstruction.
As shown, incorporating low-fidelity information through residual-space learning significantly improves the quality of the super-resolution reconstruction by enhancing both the mean predictive accuracy and the characterization of the uncertainty induced by the stochastic solution operator.
A quantitative comparison of predictive accuracy across various stochastic PDEs in a super-resolution setting is presented in~\Cref{tab:predictive_results_superresolution}. Similar to the results reported in~\Cref{tab:predictive_results_vary_training_size}, residual-space learning improves the mean predictive accuracy compared to directly learning the high-fidelity stochastic solution operator.
Furthermore, by formulating the problem in a probabilistic operator learning setting, the learned model enables super-resolution reconstruction at unseen spatial discretizations while simultaneously characterizing the uncertainty induced by the stochastic solution operator.
This introduces an important computational tradeoff: although probabilistic operator learning formulations incur higher computational cost during training and inference, they enable uncertainty-informed super-resolution predictions at unseen spatial locations and discretizations by characterizing the distribution induced by the stochastic solution operator.

\begin{figure}[htbp]
    \centering
    \includegraphics[width=0.99\linewidth]{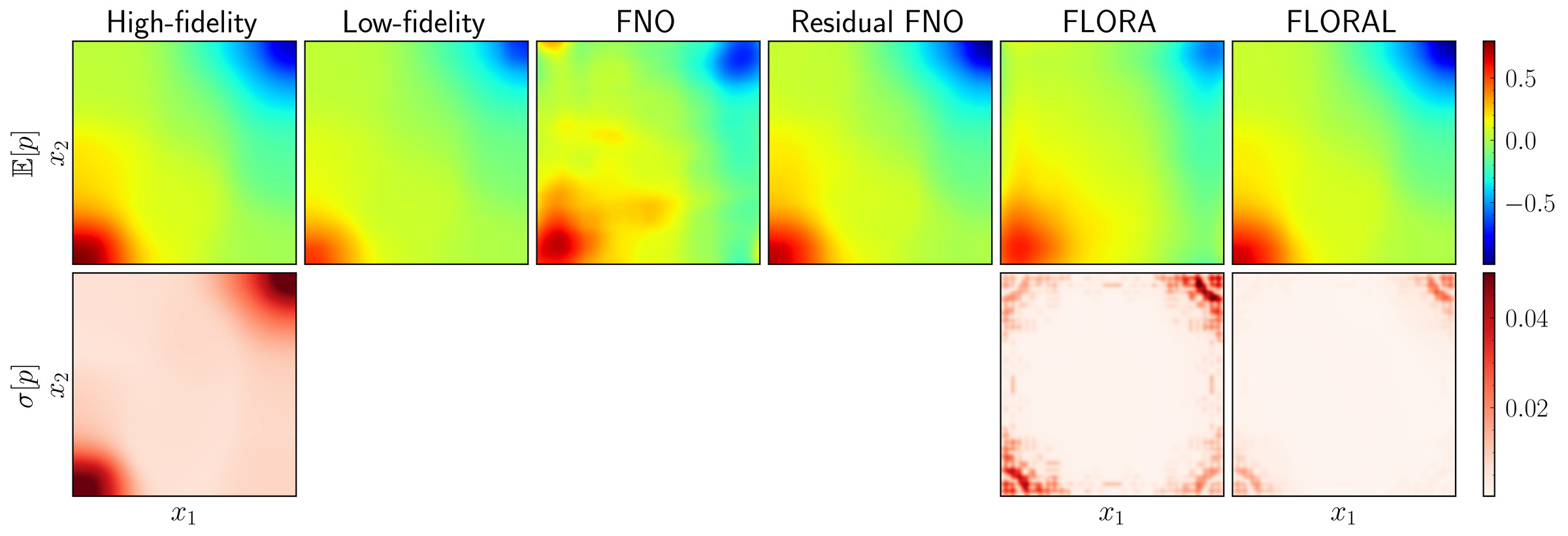}
    \caption{\em Predictive performance for approximating the solution function at an unseen input condition (same as~\cref{fig:darcy/mean_std_n_train_500_n_val_750_train_res_full}) for flow through porous media.
    For training, the solution function is observed only at $8\times8$ uniformly-spaced points in the spatial domain, while during inference the learned operator is evaluated at $64\times 64$ spatial locations, corresponding to a $8\times$ super-resolution reconstruction along each dimension.
    The model is trained using a total of $500$ training samples, corresponding to $50$ realizations of the solution function for each input condition.
    (\textbf{Top}) Mean solution function.
    (\textbf{Bottom}) Standard deviation of the solution function.}
    \label{fig:darcy/mean_std_n_train_500_n_val_750_train_res_8}
\end{figure}

\begin{figure}[htbp]
    \centering
    \includegraphics[width=0.99\linewidth]{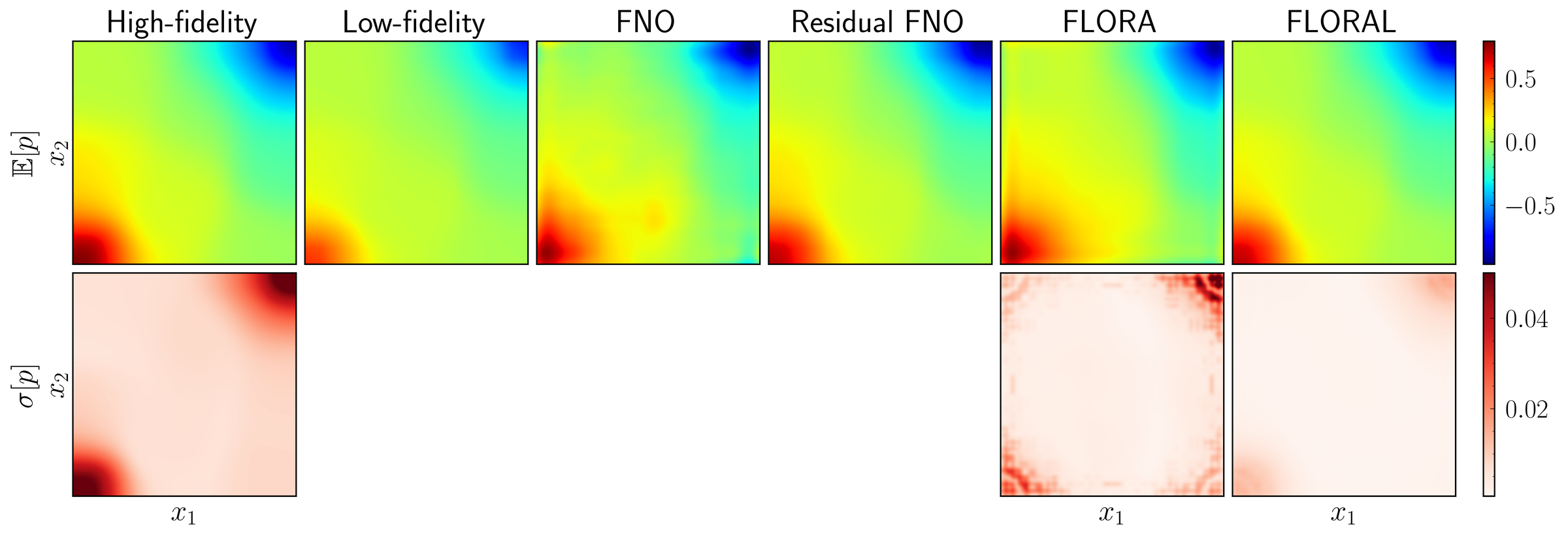}
    \caption{\em Predictive performance for approximating the solution function at an unseen input condition (same as~\cref{fig:darcy/mean_std_n_train_500_n_val_750_train_res_full}) for flow through porous media.
    For training, the solution function is evaluated only at $16\times16$ uniformly spaced points in the spatial domain, whereas during inference, the learned operator is evaluated at $64\times 64$ spatial locations, corresponding to a $4\times$ super-resolution reconstruction along each dimension.
    The model is trained using a total of $500$ training samples, corresponding to $50$ realizations of the solution function for each input condition.
    (\textbf{Top}) Mean solution function.
    (\textbf{Bottom}) Standard deviation of the solution function.}
    \label{fig:darcy/mean_std_n_train_500_n_val_750_train_res_16}
\end{figure}

\Cref{fig:darcy/mean_std_n_train_500_n_val_750_train_res_8,fig:darcy/mean_std_n_train_500_n_val_750_train_res_16} illustrate the predictive performance for approximating the Darcy solution function at an unseen input condition (same as~\Cref{fig:darcy/mean_std_n_train_500_n_val_750_train_res_full}) in a super-resolution setting. During training, the solution function is evaluated only at uniformly spaced $8\times8$ and $16\times16$ spatial locations, whereas during inference, the learned operator is evaluated on a $64\times64$ spatial grid, corresponding to $8\times$ and $4\times$ super-resolution reconstructions along each spatial dimension, respectively.
As shown, incorporating low-fidelity information through residual-space operator learning significantly improves the quality of the super-resolution reconstruction by enhancing both the mean predictive accuracy and the characterization of the uncertainty induced by the stochastic solution operator.
In particular, leveraging the low-fidelity solution operator provides a physically informative inductive bias that improves reconstruction fidelity at unseen spatial discretizations while yielding improved mean predictions and characterization of uncertainty induced by the stochastic solution operator.
See Appendices~\ref{sec:appendix/onedcorr}, \ref{sec:appendix/advection}, \ref{sec:appendix/burgers}, and~\ref{sec:appendix/darcy} for additional results.

\begin{table}[ht]
\caption{
\em Quantitative comparison of mean predictive accuracy across probabilistic PDE benchmarks for varying training resolutions.
Training datasets comprised of $500$ samples, corresponding to $50$ realizations of the stochastic solution function for each input condition.
Inference is performed over $15$ unseen stochastic solution functions.
For the Advection and Burgers' benchmarks, the inference resolution is fixed at $128^2$, while for the Darcy benchmark, the inference resolution is fixed at $64^2$ across all experiments.
For probabilistic methods, $2500$ realizations are drawn per input condition.
}
\label{tab:predictive_results_superresolution}
\centering
\small
\setlength{\tabcolsep}{2.5pt}
\renewcommand{\arraystretch}{0.95}
\setlength{\aboverulesep}{0.4pt}
\setlength{\belowrulesep}{0.4pt}
\setlength{\cmidrulesep}{1.0pt}

\begin{tabular}{ll c cccc}
\toprule
\textbf{Method} & \textbf{System} &
\makecell{\textbf{Train}\\\textbf{Res.}} &
\makecell{\textbf{Mean $L_2$}\\\textbf{Error}$\downarrow$} &
\textbf{RMSE}$\downarrow$ &
\textbf{nRMSE}$\downarrow$  &
\makecell{\textbf{Mean $L_2$}\\\textbf{Variance Error}}$\downarrow$\\
\midrule

\multirow{3}{*}{Low-fidelity}
  & Advection & -- & 21.98 & $1.9 \times 10^{-1}$ & $3.7 \times 10^{-1}$ & N/A\\
  & Burgers'  & -- & 13.29 & $1.1 \times 10^{-1}$ & $2.9 \times 10^{-1}$ & N/A\\
  & Darcy     & -- & 4.49  & $7.1 \times 10^{-2}$ & $3.0 \times 10^{-1}$ & N/A\\
\midrule

FNO
& Advection & $8^2$   & 47.56 & $4.0\times 10^{-1}$ & $7.9\times10^{-1}$ & N/A\\
&            & $16^2$  & 23.71 & $2.3\times10^{-1}$ & $4.5\times10^{-1}$ & N/A\\
\cmidrule(lr){2-7}

& Burgers'   & $8^2$   & $29.88$ & $2.9\times 10^{-1}$ & $7.4\times10^{-1}$ & N/A\\
&            & $16^2$  & $20.77$ & $2.2\times10^{-1}$ & $5.6\times10^{-1}$ & N/A\\
\cmidrule(lr){2-7}

& Darcy      & $8^2$   & $4.56$ & $7.3\times 10^{-2}$& $3.0\times 10^{-1}$ & N/A\\
&            & $16^2$  & $2.91$ & $4.7\times 10^{-2}$ & $1.9\times 10^{-1}$ & N/A\\
\midrule

Residual FNO
& Advection & $8^2$   & 8.47 & $7.9\times10^{-2}$ & $1.5\times10^{-1}$ & N/A\\
&            & $16^2$  & 6.22 & $6.4\times 10^{-2}$ & $1.2 \times 10^{-1}$ & N/A\\
\cmidrule(lr){2-7}

& Burgers'   & $8^2$   & $8.62$ & $7.9\times10^{-2}$ & $2.0\times10^{-1}$ & N/A\\
&            & $16^2$  & $6.15$ & $5.8\times10^{-2}$ & $1.4\times10^{-1}$ & N/A\\
\cmidrule(lr){2-7}

& Darcy      & $8^2$   & $1.10$ & $1.7\times 10^{-2}$& $7.5\times 10^{-2}$ & N/A\\
&            & $16^2$  & $1.0$ & $1.6\times 10^{-2}$ & $6.8\times10^{-2}$ & N/A\\
\midrule

FLORA
& Advection & $8^2$   & 48.27 & $4.0\times 10^{-1}$ & $7.8\times10^{-1}$ & $1.76$\\
&            & $16^2$  & 16.34 & $1.6\times 10^{-1}$ & $3.2\times10^{-1}$ & $0.85$\\
\cmidrule(lr){2-7}

& Burgers'   & $8^2$   & $27.93$ & $2.8\times10^{-1}$ & $7.3\times10^{-1}$ & $2.80$\\
&            & $16^2$  & $23.23$ & $2.5\times10^{-1}$ & $6.5\times10^{-1}$ & $2.81$ \\
\cmidrule(lr){2-7}

& Darcy      & $8^2$   & $5.53$ & $8.9\times 10 ^{-2}$ & $3.7\times 10^{-1}$ & $2.8\times 10^{-2}$\\
&            & $16^2$  & $3.35$ & $5.4\times10^{-2}$ & $2.2\times10^{-1}$ & $2.8\times 10^{-2}$\\
\midrule

FLORAL
& Advection & $8^2$   & 7.30 & $7.9\times 10^{-2}$ & $1.5\times10^{-1}$ & $0.85$\\
&            & $16^2$  & 5.32 & $5.5\times 10^{-2}$ & $1.0\times 10^{-1}$ & $0.86$\\
\cmidrule(lr){2-7}

& Burgers'   & $8^2$   & $8.09$ & $7.4\times 10^{-2}$ & $1.8\times10^{-1}$ & $2.79$ \\
&            & $16^2$  & $8.61$ & $9.1\times10^{-2}$ & $2.3\times10^{-1}$ & $2.78$ \\
\cmidrule(lr){2-7}

& Darcy      & $8^2$   & $1.26$ & $2.0\times 10^{-2}$ & $8.7\times10^{-2}$ & $3.3\times 10^{-2}$\\
&            & $16^2$  & $1.19$ & $2.0\times10^{-2}$ & $8.6\times10^{-2}$ & $3.3\times 10^{-2}$ \\
\bottomrule

\end{tabular}
\end{table}

\section{Concluding remarks}\label{sec:conclusion}
In this work, we introduce a residual-augmented probabilistic operator learning framework, called $\floral$, for modeling stochastic solution operators $\mathcal{G}^\varepsilon$ arising in physical systems with latent stochasticity $\varepsilon$.
By casting flow-matching-based operator learning in a residual learning setting, \textsc{floral}\ learns the distribution of the discrepancy $\mathcal{R}^\varepsilon[a] = \mathcal{G}^\varepsilon[a] - \mathcal{G}_\ell[a]$ between inexpensive low-fidelity approximation $\mathcal{G}^\ell$ and the true high-fidelity stochastic solution operator, substantially improving data-efficiency and predictive accuracy in low-data regimes.
To support resolution-generalizable inference across varying input conditions, we introduce a conditional neural operator architecture for the flow matching vector field, enabling evaluation of $w$ at arbitrary locations $x_w \in \Omega_w$ without retraining for different inputs.
Numerical experiments on stochastic benchmark systems, including the 1D stochastic advection and Burgers' equations and 2D Darcy flow, demonstrate that \textsc{floral}\ accurately characterizes the solution distribution induced by $\varepsilon$ while outperforming deterministic and probabilistic baselines in both predictive accuracy and uncertainty estimation across varying data budgets.
The key technical insights from this work are as follows:

\begin{itemize}

\item \emph{Necessity of probabilistic operator learning.} Deterministic operator approximations are fundamentally insufficient for stochastic physical systems, as they recover only the conditional mean $\mathbb{E}_\varepsilon[\mathcal{G}^\varepsilon[a] \vert a]$ and discard all variability induced by $\varepsilon$.
Accurately characterizing the forward problem requires learning the full distribution over physically plausible solution realizations.

\item \emph{Benefits of residual-space learning.} Incorporating low-fidelity information via residual operator learning consistently improves predictive accuracy in both deterministic and probabilistic settings, with the largest gains in low-data regimes.
In the probabilistic setting, learning the residual distribution $\mathcal{R}^\varepsilon[a]$ additionally improves uncertainty characterization by reducing the complexity of the target distribution relative to modeling $\mathcal{G}^\varepsilon[a]$ directly.

\item \emph{Resolution generalization with uncertainty characterization.} Formulating the forward problem within a probabilistic operator learning framework enables uncertainty-aware, resolution-generalizable inference at unseen spatial discretizations and domain locations.
Low-fidelity information further improves both mean reconstruction and uncertainty characterization during super-resolution inference.

\item \emph{Computational tradeoff.} Probabilistic operator learning results in nearly an order-of-magnitude increase in training time relative to deterministic methods, with additional inference cost from ensemble sampling needed to characterize the distribution of $\mathcal{G}^\varepsilon[a]$.
This tradeoff between computational cost and uncertainty-aware prediction is an important practical consideration.

\item \emph{Future directions.} Open directions include investigating transformer- and graph-based operator backbones to improve scalability, and developing theoretical guarantees under which residual-space learning provably reduces sample complexity in multi-fidelity probabilistic operator learning.

\end{itemize}

\section*{Acknowledgements}\label{sec:acknowledgements}
This work was supported in part by the OUSD(RE) grant \emph{`Physics-Aware Reduced Order Modeling for Nonequilibrium Plasma Flows'} and in part by the Los Alamos National Laboratory grant {\em `Algorithm/Software/Hardware Co-design for High Energy Density applications'}.

\bibliographystyle{siamplain}
\bibliography{references}
\clearpage
\appendix
\section{Continuity Equation}\label{sec:continuity_equation}
Any vector field $u_\tau^\zeta: \real{d_y}\ra\real{d_y}$ defined for $\sample{y}\in \real{d_y}$ induces the probability path $\varprob[\tau]{ \sample{y};\zeta }$ if it satisfies the continuity equation
\begin{align*}
    \frac{\partial \varprob[\tau]{\sample{y};\zeta}}{\partial \tau} + \nabla \cdot (\varprob[\tau]{\sample{y};\zeta} u_\tau^\zeta(\sample{y})) = 0,
\end{align*}
where $\nabla(\cdot)$ denotes the divergence operator~\cite{villani2008optimal}.
The continuity equation is necessary and sufficient to ensure that the probability path $\varprob[\tau]{\sample{y};\zeta}$ evolves according to the flow characterized by the vector field $u_\tau^\zeta$.

\section{Finite-dimensional view of Solving the Forward Problem via Flow Matching}\label{supp:sec:pde_conditinal_flow_matching_details}
To solve the forward problem, we seek to characterize the conditional distribution over $w$ given $a$, where $w \in \mathcal{W} \define \{w: \Omega_w \subset \real{\kappa_w} \to \real{m_w}\}$ denotes the state of the physical system and $a \in \mathcal{A} \define \{a: \Omega_a \subset \real{\kappa_a} \to \real{m_a}\}$ denotes the known input.
To make the forward problem computationally tractable, we can discretize the domains $\Omega_a$ and $\Omega_w$ using $n_a$ and $n_w$ evaluation points, respectively. This yields finite-dimensional representations $\vect{a} \in \mathbb{R}^{d_a}$ and $\vect{w} \in \mathbb{R}^{d_w}$, where $d_a \define m_a n_a$ and $d_w \define m_w n_w$.
Under this discretization, the stochastic solution operator $\mathcal{G}^\varepsilon$ can be approximated by a stochastic mapping $ \widehat{\mathcal{G}}^\varepsilon : \mathbb{R}^{d_a} \to \mathbb{R}^{d_w}$ between finite-dimensional spaces,
such that $\vect{w} = \widehat{\mathcal{G}}^\varepsilon(\vect{a})$ is a random variable induced by the latent variable $\varepsilon$.
From a probabilistic perspective, the forward problem then amounts to characterizing the conditional distribution $ \prob{\vect{w}\vert \vect{a}} = \int \prob{\vect{w} \vert \vect{a}, \varepsilon}\,\prob{\varepsilon}\, \mathrm{d}\varepsilon$ induced due to the underlying randomness in the physical system.

To approximate $\prob{\vect{w}\vert\vect{a}}$ via flow matching, we can introduce a conditional flow transformation $\psi_{\tau,\vect{a}}^\zeta:\real{d_w}\to\real{d_w}$ characterized by the ODE
\begin{align*}
    \frac{d}{d\tau}\psi_{\tau,\vect{a}}^\zeta(\vect{w}) = f_{\tau,\vect{a}}^\zeta(\psi_{\tau,\vect{a}}^\zeta(\vect{w})), \quad \tau\in [0, 1],\\
    \psi_{0,\vect{a}}^\zeta(\vect{w}) = \vect{w}\sim \srcprob{w}.
\end{align*}
Here, $f_{\tau, \vect{a}}^\zeta: \real{d_w} \ra \real{d_w}$ is a time-dependent conditional vector field parameterized via $\zeta\ireal{n_\zeta}$ that characterizes the conditional flow transformation, and $\srcprob{w}$ denotes the prior (typically modeled as a standard normal distribution).
By conditioning the vector field on $\vect{a}$, the model learns a single amortized representation that generalizes across inputs $\vect{a}$, rather than requiring a separate flow for each instance.
This significantly reduces the learning complexity while enabling the vector field to adapt its dynamics to the input-dependent structure of the target $\prob{w\vert a}$.
The conditional flow transformation induces the push-forward $\varprob[\tau]{w\vert a; \zeta} \define [\psi_{\tau,\vect{a}}^\zeta]_{\#}\srcprob{w}$, where $\varprob[\tau]{w\vert a;\zeta}: [0, 1] \times \real{d_w} \ra \real{}_{>0}$ is the induced \textit{conditional probability path}.
To approximate $\prob{\vect{w}\vert\vect{a}}$, we seek a conditional vector field $f_{\tau,\vect{a}}^\zeta$ that induces a conditional probability path $\varprob[\tau]{w\vert a;\zeta}$ such that $\varprob[1]{w\vert a;\zeta}\approx \prob{w\vert a}$.

Following Lipman et al.~\cite{lipman2022flow}, we can learn a parametric conditional vector field $h_{\tau, \vect{a}}^\xi: \real{d_w} \ra \real{d_w}$ with parameters $\xi\ireal{n_\xi}$ to approximate $f_{\tau, \vect{a}}^\zeta$ by minimizing 
\begin{align}
    J_{\text{FM}}^\ddagger(\zeta, \xi, \vect{a}) \define \mathbb{E}_{\tau \sim \mathcal{U}(0, 1), \vect{w}\sim \varprob[\tau]{w\vert a; \zeta}}\left[\norm[2]{f_{\tau,\vect{a}}^\zeta(\vect{w}) - h_{\tau,\vect{a}}^\xi(\vect{w})}^2\right],
    \label{supp:eq:pde_flow_matching_objective}
\end{align}
such that $h_{\tau,\vect{a}}^{\xi^\ddagger}$ would also induce $\varprob[\tau]{w\vert a; \zeta}$ upon reaching $J_{\text{FM}}^\ddagger=0$ for some $\xi=\xi^\ddagger$.
However, as discussed in Section~\ref{sec:background}, this loss is intractable as the conditional vector field $f_{\tau,\vect{a}}^\zeta$ that results in a conditional probability path $\varprob[\tau]{w\vert a;\zeta}$ such that $\varprob[1]{w\vert a;\zeta}\approx \prob{w\vert a}$ is unknown.
To address this, similar to Tong et al.~\cite{tong2023improving}, we can introduce a latent variable $\random{z}$ with distribution $\varprob{z}$.
We can then introduce a conditional latent vector field $\tilde{f}_{\tau,\vect{a},\vect{z}}^\zeta:\real{d_w}\to\real{d_w}$ that induces the conditional latent probability path $\varprob[\tau]{\vect{y}\vert\vect{a},\vect{z}; \zeta}$, that is, it satisfies the continuity equation.
Using the marginalization trick, we can then write the conditional probability path using the conditional latent probability path as
\begin{align*}
\varprob[\tau]{\vect{y}\vert \vect{a};\zeta} \define \int \varprob[\tau]{\vect{y}\vert\vect{a},\vect{z}; \zeta}\varprob{\vect{z}} \,d\vect{z}.
\end{align*}
As a result, we can express the conditional vector field $f_{\tau,\vect{a}}^\zeta$ as
\begin{align}
    f_{\tau,\vect{a}}^\zeta \define \expect[\varprob{\sample{z}}]{\frac{\tilde{f}_{\tau, \vect{a},\vect{z}}^\zeta \varprob[\tau]{\vect{w}\vert\vect{a},\vect{z}; \zeta}}{\varprob[\tau]{\vect{w}\vert\vect{a};\zeta}}},
    \label{supp:eq:conditional_latent_vector_field}
\end{align}
that extends~\cref{eq:latent_vector_field} to a conditional setting in a strigh-forward manner.
Using~\cref{supp:eq:conditional_latent_vector_field} we can then define a tractable unbiased learning objective given as
\begin{align}
    J_{\text{CFM}}^\ddagger(\zeta, \xi, \vect{a}) \define \mathbb{E}_{\tau \sim \mathcal{U}(0, 1), \vect{z} \sim \varprob{\vect{z}}, \vect{w}\sim \varprob[\tau]{\vect{w}\vert \vect{a}, \vect{z}; \zeta}}\left[\left\|\gamma_\tau\left(\tilde{f}_{\tau, \vect{a}, \vect{z}}^\zeta(\vect{w}) - h_{\tau,\vect{a}}^\xi(\vect{w})\right)\right\|_2^2\right],
    \label{eq:pde_conditional_pde_flow_matching_detail}
\end{align}
where $\gamma_\tau$ is a time-dependent scaling.
Minimizing $J_{\text{CFM}}^\ddagger$ is similar to minimizing $J_{\text{FM}}^\ddagger$ since
\begin{align*}
\arg\min_{\xi} J_{\text{FM}}^\ddagger(\zeta, \xi,\vect{a}) = \arg\min_{\xi} J_{\text{CFM}}^\ddagger(\zeta, \xi,\vect{a}),
\end{align*}
as show in~\cref{theorem:finite_dim_unbiased_estimator}.
To tractably sample from $\varprob{z}$ we can define it as a joint distribution between the prior and the conditional distribution $\prob{w\vert a}$, that is, $\varprob{z}=\srcprob{w}\prob{w\vert a}$.
We can then model the conditional latent probability path as a transport of Gaussian distribution
\begin{align*}
    \varprob[\tau]{\sample{w}\vert\sample{a},\sample{z};\zeta} = \mathcal{N}(\sample{w}; \mu_\tau(\sample{z}), \sigma_\tau(\sample{z};\zeta)),\\
    \mu_\tau(\sample{z}) \define \tau \sample{w}_1 + (1-\tau)\sample{w}_0, \quad \sample{w}_0\sim \srcprob{w}, \sample{w}_1\sim \prob{w\vert a}, \\
    \sigma_\tau(\sample{z};\zeta) \define \sigma_{\min} \norm[2]{\sample{w}_1 - \sample{w}_0}^2,
\end{align*}
that results in the conditional vector field $\tilde{f}_{\tau, \sample{a}, \sample{z}}^\zeta = \sample{w}_1 - \sample{w}_0$ using~\cref{eq:independent_cfm_vector_field}.
Here, we introduce a sample-dependent noise $\sigma_\tau$ to ensure high signal-to-noise ratio when the samples $\sample{w}_0$ and $\sample{w}_1$ are close to each other in $\norm[2]{\cdot}$ sense.
Finally, to learn a vector field $h_{\tau, \sample{a}}^\xi$ that generalizes well across inputs, we can consider the training objective
\begin{align}
J_{\text{CFM}}(\zeta, \xi) \define \expect[\vect{a}\sim\prob{a}]{J_{\text{CFM}}^\ddagger(\zeta, \xi, \vect{a}) },
\end{align}
where $\prob{a}$ characterizes the underlying distribution of the inputs.

\section{Proofs}\label{sec:proofs}

\begin{remark}\label{remark:irreducible_mse_error}
\em{Let $\mathcal{G}^\varepsilon$ denote a random operator parameterized by a random variable $\varepsilon$, 
and let $\mathcal{G}^\theta$ denote a deterministic approximation parameterized by $\theta$. 
Define the mean operator
\begin{align*}
\mu(a) \define \expect[\varepsilon]{\mathcal{G}^\varepsilon[a]} .
\end{align*}
Then the expected squared error admits the decomposition
\begin{align*}
\expect[a,\varepsilon]{\norm[\mathcal{W}]{\mathcal{G}^\varepsilon[a] - \mathcal{G}^\theta[a]}^2}
= \expect[a,\varepsilon]{\norm[\mathcal{W}]{\mathcal{G}^\varepsilon[a] - \mu(a)}^2}
+ \expect[a]{\norm[\mathcal{W}]{\mu(a) - \mathcal{G}^\theta[a]}^2},
\end{align*}
where the first term is independent of $\theta$ and represents the irreducible error, 
and the second term is minimized to zero by the optimal parameter $\theta^\dagger$ satisfying
\begin{align*}
\mathcal{G}^{\theta^\dagger}[a] = \mu(a) = \expect[\varepsilon]{\mathcal{G}^\varepsilon[a]}, \quad \forall\, a.
\end{align*}
}
\begin{proof}
Define $\mu(a) \define \expect[\varepsilon]{\mathcal{G}^\varepsilon[a]}$ and 
$r(a,\varepsilon) \define \mathcal{G}^\varepsilon[a] - \mu(a)$.
We can then write
\begin{align*}
\mathcal{G}^\varepsilon[a] - \mathcal{G}^\theta[a] = r(a,\varepsilon) + \mu(a) - \mathcal{G}^\theta[a],
\end{align*}
such that
\begin{align*}
\norm[\mathcal{W}]{\mathcal{G}^\varepsilon[a] - \mathcal{G}^\theta[a]}^2 &= \norm[\mathcal{W}]{r(a,\varepsilon) + \mu(a) - \mathcal{G}^\theta[a]}^2,\\
&= \norm[\mathcal{W}]{r(a,\varepsilon)}^2 + 2\langle r(a,\varepsilon),\, \mu(a) - \mathcal{G}^\theta[a]\rangle_{\mathcal{W}}+\norm[\mathcal{W}]{\mu(a) - \mathcal{G}^\theta[a]}^2.
\end{align*}
Taking the conditional expectation with respect to $\varepsilon$ given $a$, we obtain
\begin{align*}
\expect[\varepsilon]{\norm[\mathcal{W}]{\mathcal{G}^\varepsilon[a] - \mathcal{G}^\theta[a]}^2\mid a} 
&= \expect[\varepsilon]{\norm[\mathcal{W}]{r(a,\varepsilon)}^2\mid a} 
+ 2\left\langle \expect[\varepsilon]{r(a,\varepsilon)\mid a},\, \mu(a) - \mathcal{G}^\theta[a]\right\rangle_{\mathcal{W}} \\
&\quad + \norm[\mathcal{W}]{\mu(a) - \mathcal{G}^\theta[a]}^2.
\end{align*}
Since
\begin{align*}
\expect[\varepsilon]{r(a,\varepsilon)\mid a}
= \expect[\varepsilon]{\mathcal{G}^\varepsilon[a] - \mu(a)\mid a}
= \mu(a) - \mu(a) = 0,
\end{align*}
the cross term vanishes, and hence
\begin{align*}
\expect[\varepsilon]{\norm[\mathcal{W}]{\mathcal{G}^\varepsilon[a] - \mathcal{G}^\theta[a]}^2 \mid a}
= \expect[\varepsilon]{\norm[\mathcal{W}]{r(a,\varepsilon)}^2 \mid a}
+ \norm[\mathcal{W}]{\mu(a) - \mathcal{G}^\theta[a]}^2.
\end{align*}
Taking the expectation over $a$ on both sides:
\begin{align*}
\expect[a,\varepsilon]{\norm[\mathcal{W}]{\mathcal{G}^\varepsilon[a] - \mathcal{G}^\theta[a]}^2}
= \underbrace{\expect[a,\varepsilon]{\norm[\mathcal{W}]{r(a,\varepsilon)}^2}}_{\text{(I) independent of } \theta} 
+ \underbrace{\expect[a]{\norm[\mathcal{W}]{\mu(a) - \mathcal{G}^\theta[a]}^2}}_{\text{(II) minimizable}}.
\end{align*}
Since term~(I) is independent of $\theta$, it represents the irreducible error.
Term~(II) is minimized to zero for $\theta = \theta^\dagger$ and 
\begin{align*}
\mathcal{G}^{\theta^\dagger}[a] = \mu(a) = \expect[\varepsilon]{\mathcal{G}^\varepsilon[a]}.
\end{align*}
\end{proof}
\end{remark}
\begin{theorem}[Finite-dimensional Unbiased Estimator]\label{theorem:finite_dim_unbiased_estimator}
    Given the conditional latent vector field $\tilde{f}_{\tau, \vect{a}, \vect{z}}^\zeta$ induces the conditional latent probability path $\varprob[\tau]{\vect{w}\vert \vect{a}, \vect{z};\zeta}$ with $\vect{a}\sim\prob{\vect{a}}$, $\vect{z}\sim\varprob{\vect{z}}$, and $\zeta \ireal{n_\zeta}$,
    then up to a constant that is independent of trainable parameters $\xi\ireal{n_\xi}$ of the vector field $h_{\tau, \vect{a}}^\xi$
\begin{align*}
    \arg\min_{\xi} J_{\text{FM}}^\ddagger(\zeta, \xi,\vect{a}) = \arg\min_{\xi} J_{\text{CFM}}^\ddagger(\zeta, \xi,\vect{a}).
\end{align*}
\begin{proof}
    Similar to~\cite{lipman2022flow,tong2023improving}, we assume that the conditional probability path $\varprob[\tau]{ \vect{w}\vert\vect{a};\zeta } > 0$ for the entire support of $\prob{a}$.
    To ensure the existence of all the involved integrals and changing of order, we assume that $\varprob{\vect{z}}$, $\varprob{\vect{w}\vert \vect{a},\vect{z};\zeta}$ are decreasing to zero at sufficient speed as $\norm{\vect{w}} \ra \infty$ for the entire support of $\prob{a}$.
    We assume that the vector fields $f_{\tau,\vect{a}}^\zeta$, $h_{\tau,\vect{a}}^\xi$, and $\nabla_\xi h_{\tau,\vect{a}}^\xi$ are bounded.
Using the bi-linearity of inner products,
\begin{align*}
\norm[2]{f_{\tau, \vect{a}}^\zeta(\vect{w}) - h_{\tau, \vect{a}}^\xi(\vect{w})}^2
&= \norm[2]{f_{\tau, \vect{a}}^\zeta(\vect{w})}^2
   - 2\bigl\langle f_{\tau, \vect{a}}^\zeta(\vect{w}),
                  h_{\tau, \vect{a}}^\xi(\vect{w}) \bigr\rangle
   + \norm[2]{h_{\tau, \vect{a}}^\xi(\vect{w})}^2,
\end{align*}
and
\begin{align*}
\bigl\| \gamma_\tau\bigl(
\tilde{f}_{\tau, \vect{a}, \vect{z}}^\zeta(\vect{w})
- h_{\tau, \vect{a}}^\xi(\vect{w})
\bigr)\bigr\|_2^2
&= \gamma_\tau^2 \Bigl(
    \norm[2]{\tilde{f}_{\tau, \vect{a}, \vect{z}}^\zeta(\vect{w})}^2 \\
&\qquad
    - 2\bigl\langle
      \tilde{f}_{\tau, \vect{a}, \vect{z}}^\zeta(\vect{w}),
      h_{\tau, \vect{a}}^\xi(\vect{w})
    \bigr\rangle
    + \norm[2]{h_{\tau, \vect{a}}^\xi(\vect{w})}^2
\Bigr).
\end{align*}
Using the law of total expectation we have
\begin{align*}
    \expect[\mathpzc{q}_\tau(\vect{w}\vert\vect{a};\zeta)]{\norm[2]{h_{\tau, \vect{a}}^{\xi}(\vect{w})}^2} = \expect[\mathpzc{q}(\vect{z})]{\expect[\mathpzc{q}_\tau(\vect{w}\vert\vect{a}, \vect{z};\zeta)]{\norm[2]{h_{\tau, \vect{a}}^\xi(\vect{w})}^2}}.
\end{align*}
Now, using Equation~\ref{supp:eq:conditional_latent_vector_field} to define the conditional vector field $f_{\tau, \vect{a}}^\zeta$, we have
\begin{align*}
    \mathbb{E}_{\varprob[\tau]{\vect{w}\vert\vect{a};\zeta}}\left[\left<f_{\tau, \vect{a}}^\zeta(\vect{w}), h_{\tau, \vect{a}}^\xi(\vect{w})\right>\right] &=
    \int\varprob[\tau]{\vect{w}\vert\vect{a};\zeta}
        \begin{aligned}[t]
        \Biggl\langle
            \int \varprob{\vect{z}}{\frac{\tilde{f}_{\tau, \vect{a}, \vect{z}}^\zeta(\vect{w}) \varprob[\tau]{\vect{w}\vert\vect{a}, \vect{z};\zeta}}{\varprob[\tau]{\vect{w}\vert\vect{a};\zeta}}} \, d\vect{z}, \\
            h_{\tau, \vect{a}}^\xi(\vect{w})
        \Biggr\rangle \, d\vect{w},
        \end{aligned}\\
    &= \int{\left<\int \varprob{\vect{z}}{\tilde{f}_{\tau, \vect{a}, \vect{z}}^\zeta(\vect{w}) \varprob[\tau]{\vect{w}\vert\vect{a}, \vect{z};\zeta}} \, d\vect{z}, h_{\tau, \vect{a}}^\xi(\vect{w})\right>} \, d\vect{w},\\
    &= \int{\int \left <\varprob{\vect{z}}{\tilde{f}_{\tau, \vect{a}, \vect{z}}^\zeta(\vect{w}) \varprob[\tau]{\vect{w}\vert\vect{a}, \vect{z};\zeta}},  h_{\tau, \vect{a}}^\xi(\vect{w})\right> \,d\vect{z}} \, d\vect{w},\\
    &= \int{\int \varprob{\vect{z}}\varprob[\tau]{\vect{w}\vert\vect{a}, \vect{z};\zeta}\left <{\tilde{f}_{\tau, \vect{a}, \vect{z}}^\zeta(\vect{w})},  h_{\tau, \vect{a}}^\xi(\vect{w})\right> \,d\vect{z}} \, d\vect{w},\\
    &= \expect[\varprob{\vect{z}}]{\expect[\mathpzc{q}_\tau(\vect{w}\vert\vect{a}, \vect{z};\zeta)]{\left<\tilde{f}_{\tau, \vect{a}, \vect{z}}^\zeta(\vect{w}), h_{\tau, \vect{a}}^\xi(\vect{w})\right>}}.
\end{align*}
Since $\nabla_\xi \norm[2]{f_{\tau, \vect{a}}^\zeta(\vect{w})}^2 = 0$, we have
\begin{align*}
\nabla_\xi \, J_{\text{FM}}^\ddagger(\zeta, \xi, \vect{a})
    &= \mathbb{E}_{\tau\sim\mathcal{U}(0, 1)} \left[ \nabla_\xi\,\mathbb{E}_{\varprob[\tau]{\vect{w}\vert\vect{a};\zeta}} \left[ \norm[2]{ f_{\tau, \vect{a}}^\zeta(\vect{w}) - h_{\tau, \vect{a}}^\xi(\vect{w}) }^2 \right] \right],\\
    &= \begin{aligned}[t]
    \mathbb{E}_{\tau\sim\mathcal{U}( 0, 1 )} \bigl[ \nabla_\xi \, \mathbb{E}_{\varprob{\vect{z}}} \bigl[ \mathbb{E}_{\varprob[\tau]{\vect{w}\vert\vect{a},\vect{z};\zeta}} \bigl[ -2 \langle \tilde{f}_{\tau, \vect{a}, \vect{z}}^\zeta(\vect{w}), h_{\tau, \vect{a}}^\xi(\vect{w}) \rangle + \\
        \norm[2]{ h_{\tau, \vect{a}}^\xi(\vect{w}) }^2 \bigr] \bigr] \bigr],
    \end{aligned}\\
    &= \mathbb{E}_{\tau\sim\mathcal{U}( 0, 1 )} \left[\nabla_\xi \, \mathbb{E}_{\varprob{\vect{z}}, \varprob[\tau]{w\vert a, z;\zeta}} \left[ \left\| \left( \tilde{f}_{\tau, \vect{a}, \vect{z}}^\zeta (\vect{w})  - h_{\tau, \vect{a}}^\xi(\vect{w}) \right)\right\|_2^2\right] \right],\\
    &= \nabla_\xi \, \mathbb{E}_{\tau\sim\mathcal{U}( 0, 1 ),\varprob{\vect{z}}, \varprob[\tau]{w\vert a, z;\zeta} } \left[\left\| \left( \tilde{f}_{\tau, \vect{a}, \vect{z}}^\zeta (\vect{w})  - h_{\tau, \vect{a}}^\xi(\vect{w}) \right)\right\|_2^2\right],\\
    &\define \nabla_\xi \, J^\ast_{\text{FM}}(\zeta, \xi, \vect{a}),
\end{align*}
where $J^\ast_{\text{FM}}(\zeta, \xi, \vect{a})\define  \mathbb{E}_{\tau\sim\mathcal{U}( 0, 1 )} \ell^\ast_\text{FM}$ with
\begin{align*}
\ell^\ast_\text{FM} \define \mathbb{E}_{\varprob{\vect{z}}, \varprob[\tau]{w\vert a, z;\zeta}} \left[\left\| \left( \tilde{f}_{\tau, \vect{a}, \vect{z}}^\zeta (\vect{w})  - h_{\tau, \vect{a}}^\xi(\vect{w}) \right)\right\|_2^2\right].
\end{align*}
As a result, minimizing $J_\text{FM}^\ddagger(\zeta, \xi, \vect{a})$ with respect to $\xi$ is equivalent to minimizing $J^\ast_{\text{FM}}(\zeta, \xi,\vect{a})$ with respect to $\xi$.
Using the definition of $J_{\text{CFM}}^\ddagger(\zeta, \xi,\vect{a})$ from Equation~\ref{eq:pde_conditional_pde_flow_matching}, we have
\begin{align*}
    J_\text{CFM}^\ddagger(\zeta, \xi,\vect{a}) &= \mathbb{E}_{\tau\sim\mathcal{U}( 0, 1 )} [\gamma_\tau^2 \ell^\ast_\text{FM}],
\end{align*}
such that $ \nabla_\xi \, J_\text{CFM}^\ddagger(\zeta, \xi,\vect{a}) = \mathbb{E}_{\tau\sim\mathcal{U}(0, 1)}[\gamma_\tau^2 \nabla_\xi \, \ell^\ast_\text{FM}]$.
As a result,
\begin{align*}
   \arg\min_{\xi}\,J_{\text{CFM}}^\ddagger(\zeta, \xi,\vect{a}) = \arg\min_{\xi}\,J^\ast_{\text{FM}}(\zeta, \xi,\vect{a}) = \arg\min_{\xi}\,J_{\text{FM}}^\ddagger(\zeta, \xi,\vect{a}),
\end{align*}
for $\gamma_\tau > 0$ since it is a time-dependent constant that is independent of $\xi$, and therefore rescales the gradients without altering the location of the optima.
\end{proof}
\end{theorem}

\section{Model Architecture and Training Details}\label{sec:training_details}
For each experiment, we require samples drawn from a Gaussian process measure.
Specifically, we consider a zero-mean Gaussian process (that is, $m_0 = 0$) parameterized by a Matérn covariance kernel with smoothness parameter $\nu = 0.5$ and length scale $10^{-3}$.
All Gaussian process samples are generated using the $\texttt{GPyTorch}$ library~\cite{gardner2018gpytorch}.

To parameterize the conditional vector fields, we employ a conditional Fourier neural operator (FiLMFNO) based on the Fourier neural operator (FNO)~\cite{li2020fourier} augmented with feature-wise linear modulation (FiLM)~\cite{perez2018film}.
The implementation is based on the $\texttt{neuraloperator}$ library~\cite{kossaifi2025librarylearningneuraloperators}, and all FiLMFNO hyper-parameters follow the conventions defined therein.

For all experiments, the FiLMFNO architecture consists of $4$ Fourier layers with $64$ hidden channels, a lifting channel ratio of $4$, and a projection channel ratio of $4$.
Across all spatial resolutions considered, the number of Fourier modes in each spatial dimension is set to one-half of the corresponding resolution.
The FiLM modulation coefficients are computed using fully connected neural networks with $3$ hidden layers, each containing $64$ neurons and SiLU activation functions.

All models are trained using the Adam optimizer with a learning rate of $10^{-3}$ and weight decay of $10^{-4}$.
An exponential learning rate scheduler with decay factor $0.99$ is employed throughout training.
A batch size of $32$ is used for all experiments.

To ensure a comparable computational budget across all methods and experiments, each model is trained for a maximum of $300$ epochs while employing early stopping based on the validation loss.
Specifically, we use an early stopping criterion with patience equal to $0.3 \times 300$ epochs and a minimum improvement threshold of $10^{-4}$.
This strategy enables models to terminate training upon convergence while maintaining a consistent upper bound on the computational cost across all experiments.

For integrating the learned vector fields, we use the $\texttt{dopri5}$ ODE solver provided in $\texttt{torchdiffeq}$~\cite{chen2018neural} with absolute and relative tolerances set to $\texttt{atol} = \texttt{rtol} = 10^{-5}$.

All experiments are performed on a single NVIDIA H100 GPU.
\clearpage
\section{Artificial Benchmark: 1D Correlated Residual (1DCR)}\label{sec:appendix/onedcorr}
\begin{figure}[htbp]
    \centering
    \includegraphics[width=0.7\linewidth]{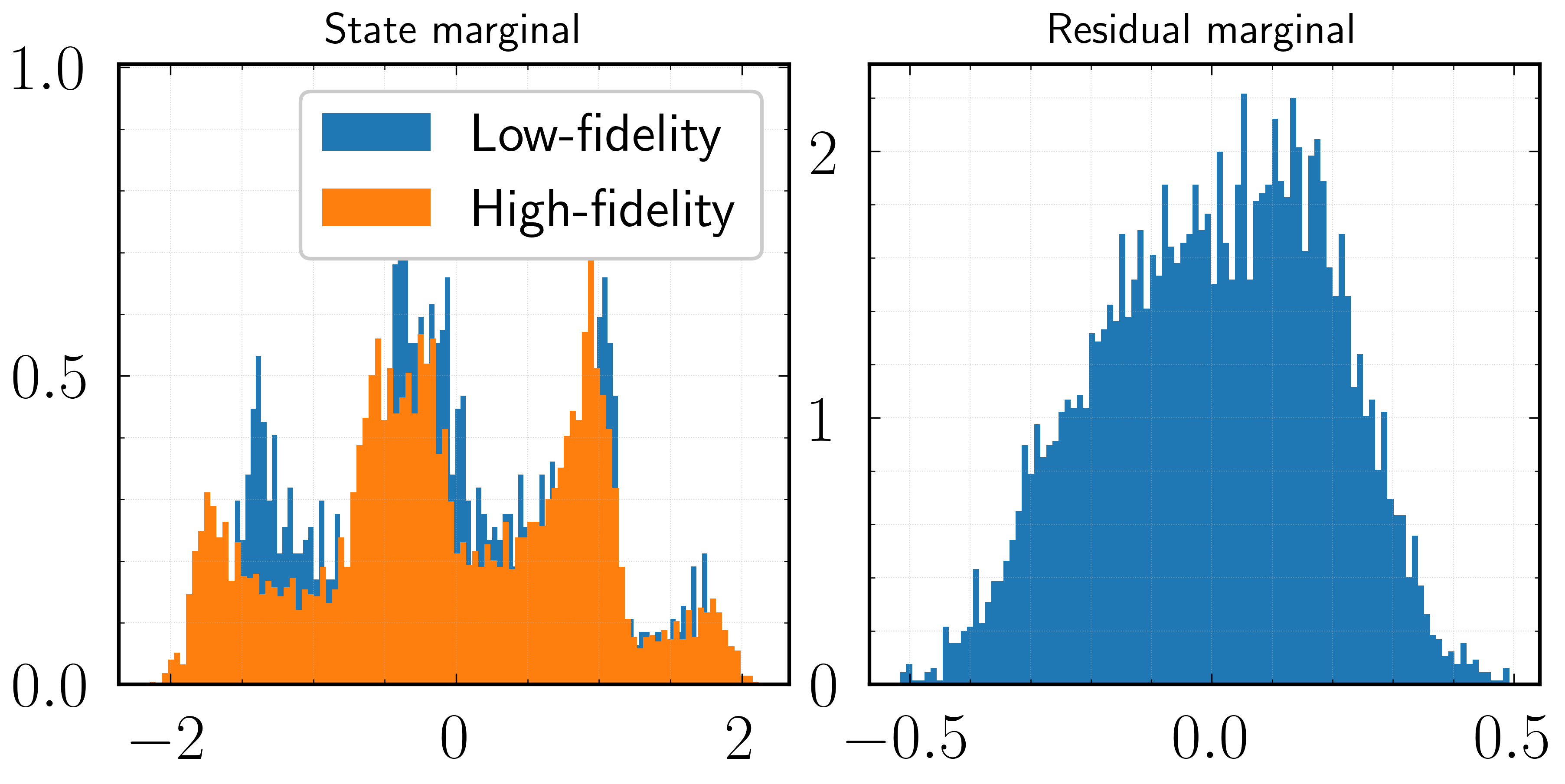}
    \caption{\em Marginal distributions for the 1DCR benchmark. (\textbf{Left}) Distribution of the solution state. (\textbf{Right}) Distribution of the residual between the low- and high-fidelity solution functions.}
    \label{fig:appendix/onedcorr/data_marginal}
\end{figure}
\begin{figure}[htbp]
    \centering
    \includegraphics[width=0.9\linewidth]{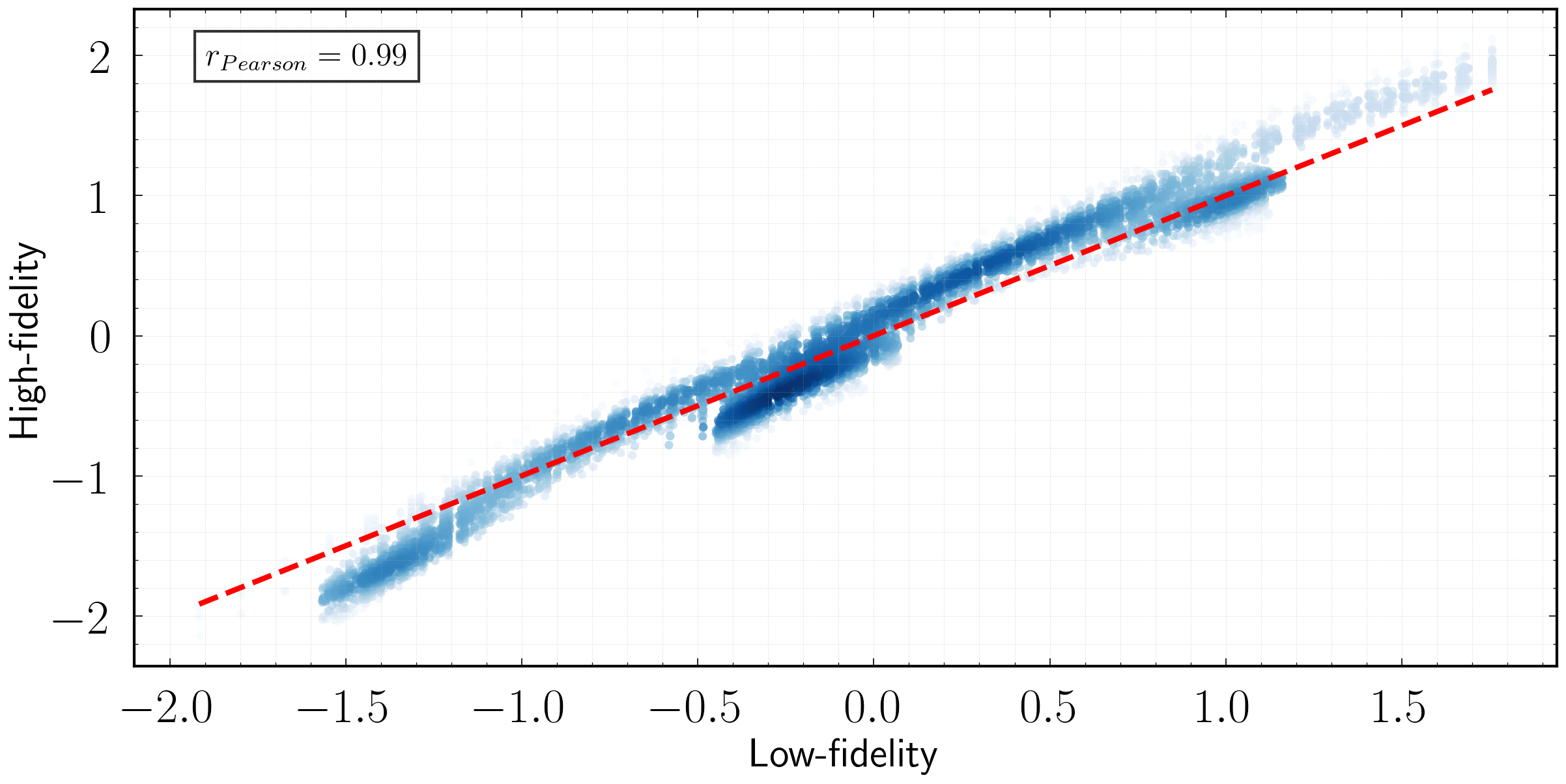}
    \caption{\em Joint distribution between the low- and high-fidelity solution states for the 1DCR benchmark.
    }
    \label{fig:appendix/onedcorr/LF_HF_joint_plot}
\end{figure}
\begin{figure}[htbp]
    \centering
    \includegraphics[width=0.9\linewidth]{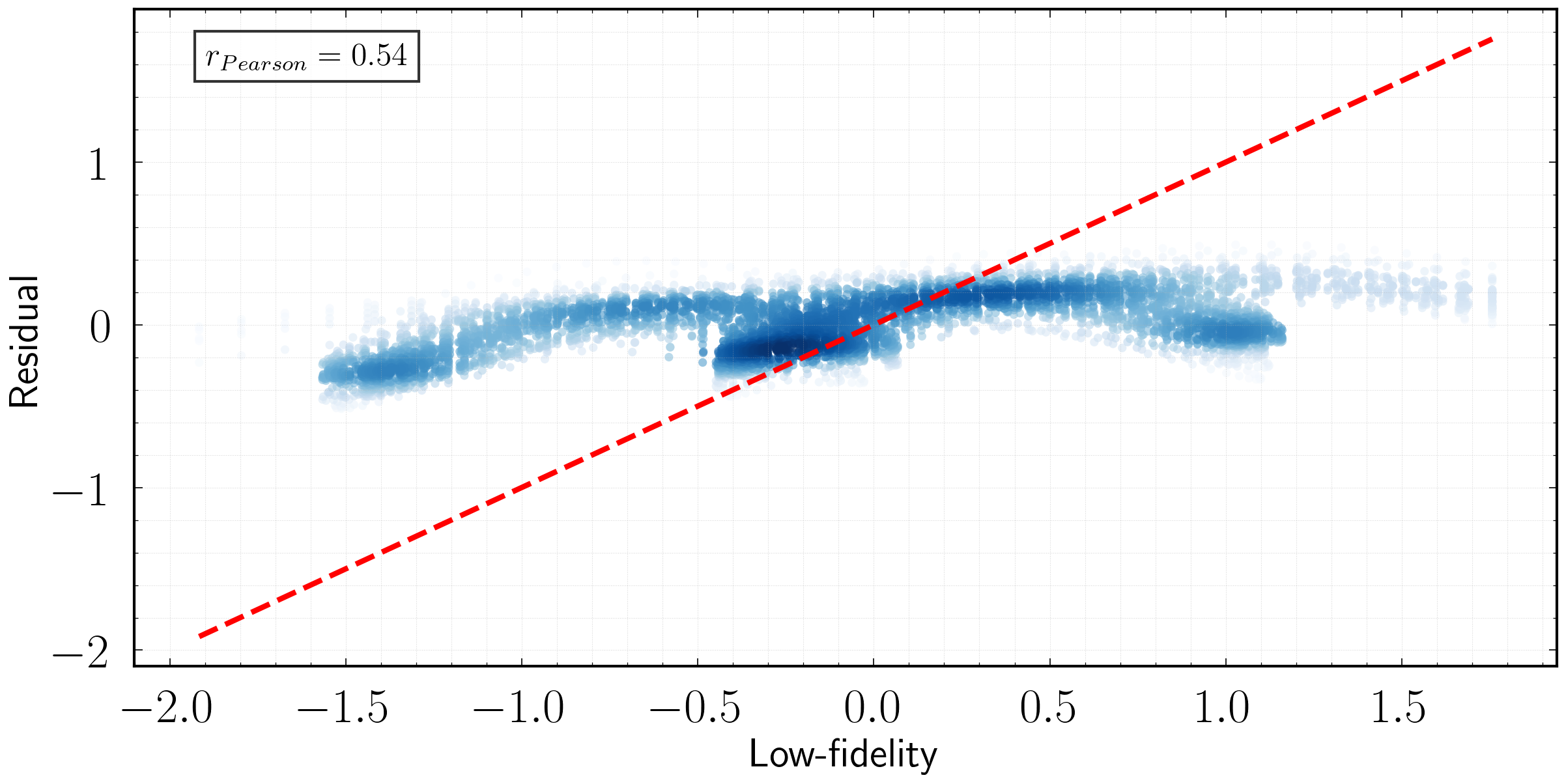}
    \caption{\em Joint distribution between the low-fidelity and residual for the 1DCR benchmark.
    }
    \label{fig:appendix/onedcorr/Residual_joint_plot}
\end{figure}
\begin{figure}[htbp]
    \centering
    \includegraphics[width=0.99\linewidth]{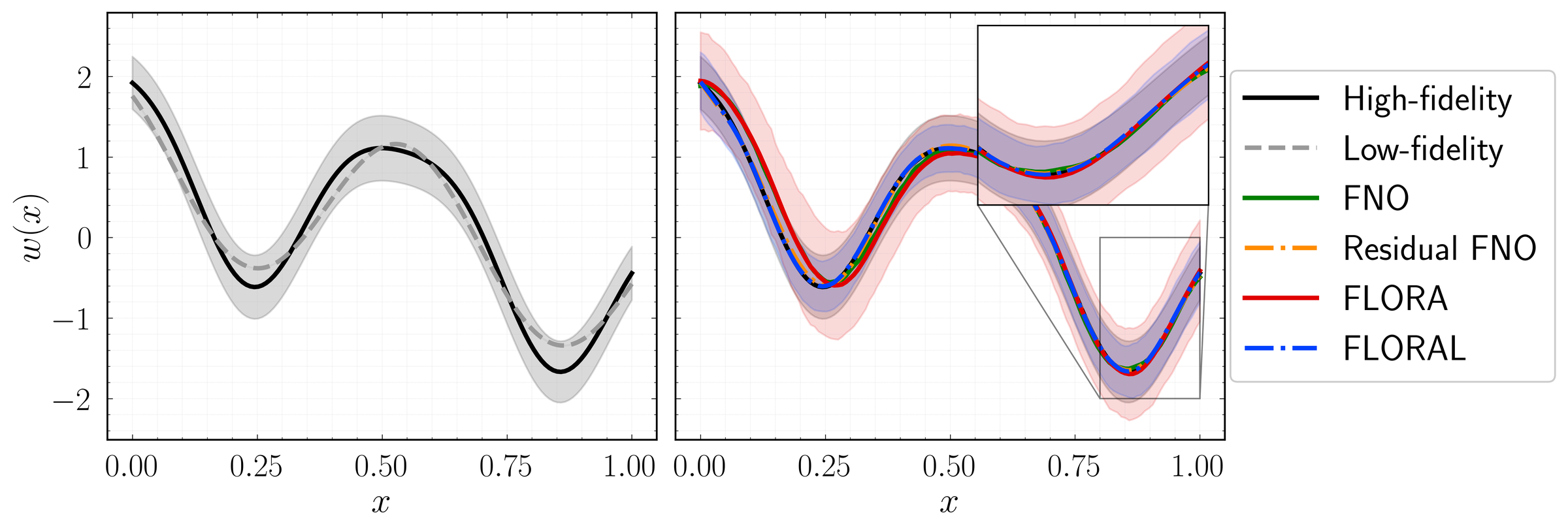}
    \caption{\em Predictive performance for approximating the solution function at an unseen input condition (same as Figure~\ref{fig:onedcorr/mean_std_n_train_500_n_val_750_train_res_full}) for the 1DCR problem. 
    For training, the solution function is evaluated only at $16$ uniformly spaced points in the spatial domain, whereas during inference, the learned operator is evaluated at $128$ spatial locations, corresponding to a $8\times$ super-resolution reconstruction.
    The model is trained using a total of $500$ training samples, corresponding to $50$ realizations of the stochastic solution function for each input condition.
    (\textbf{Left}) Comparison between the low- and high-fidelity solution functions. (\textbf{Right}) Predictions obtained using deterministic and probabilistic operator learning methods. Shaded regions denote $\pm 5$ standard deviations about the mean prediction. 
    Residual-space learning improves super-resolution performance by enhancing both the mean predictive accuracy and the characterization of the uncertainty induced by the stochastic solution operator.
    }
\label{fig:appendix/onedcorr/mean_std_n_train_500_n_val_750_train_res_16}
\end{figure}
\begin{figure}[htbp]
    \centering
    \includegraphics[width=0.99\linewidth]{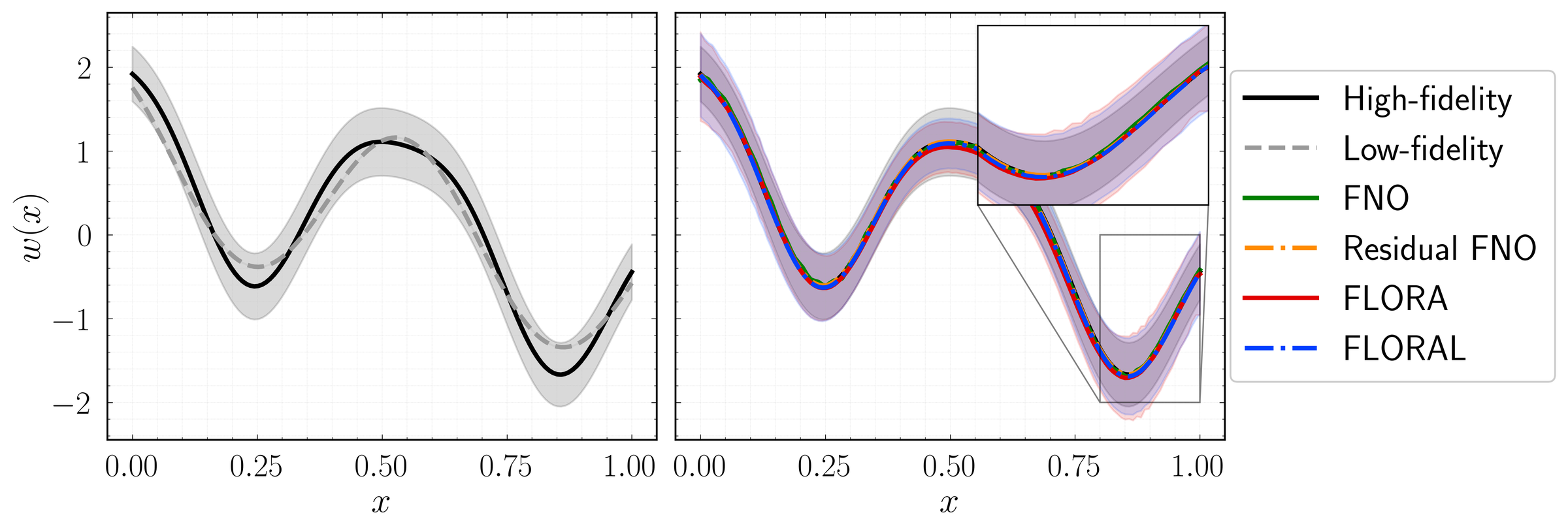}
    \caption{\em Predictive performance for approximating the solution function at an unseen input condition (same as Figure~\ref{fig:onedcorr/mean_std_n_train_500_n_val_750_train_res_full}) for the 1DCR problem. 
    For training, the solution function is evaluated only at $64$ uniformly spaced points in the spatial domain, whereas during inference, the learned operator is evaluated at $128$ spatial locations, corresponding to a $2\times$ super-resolution reconstruction.
    The model is trained using a total of $500$ training samples, corresponding to $50$ realizations of the stochastic solution function for each input condition.
    (\textbf{Left}) Comparison between the low- and high-fidelity solution functions. (\textbf{Right}) Predictions obtained using deterministic and probabilistic operator learning methods. Shaded regions denote $\pm 5$ standard deviations about the mean prediction. 
    Residual-space learning improves super-resolution performance by enhancing both the mean predictive accuracy and the characterization of the uncertainty induced by the stochastic solution operator.
    }
\label{fig:appendix/onedcorr/mean_std_n_train_500_n_val_750_train_res_64}
\end{figure}

\clearpage
\section{1D Advection Equation with Stochastic Initial Conditions (Advection)}\label{sec:appendix/advection}
\begin{figure}[htbp]
    \centering
    \includegraphics[width=0.7\linewidth]{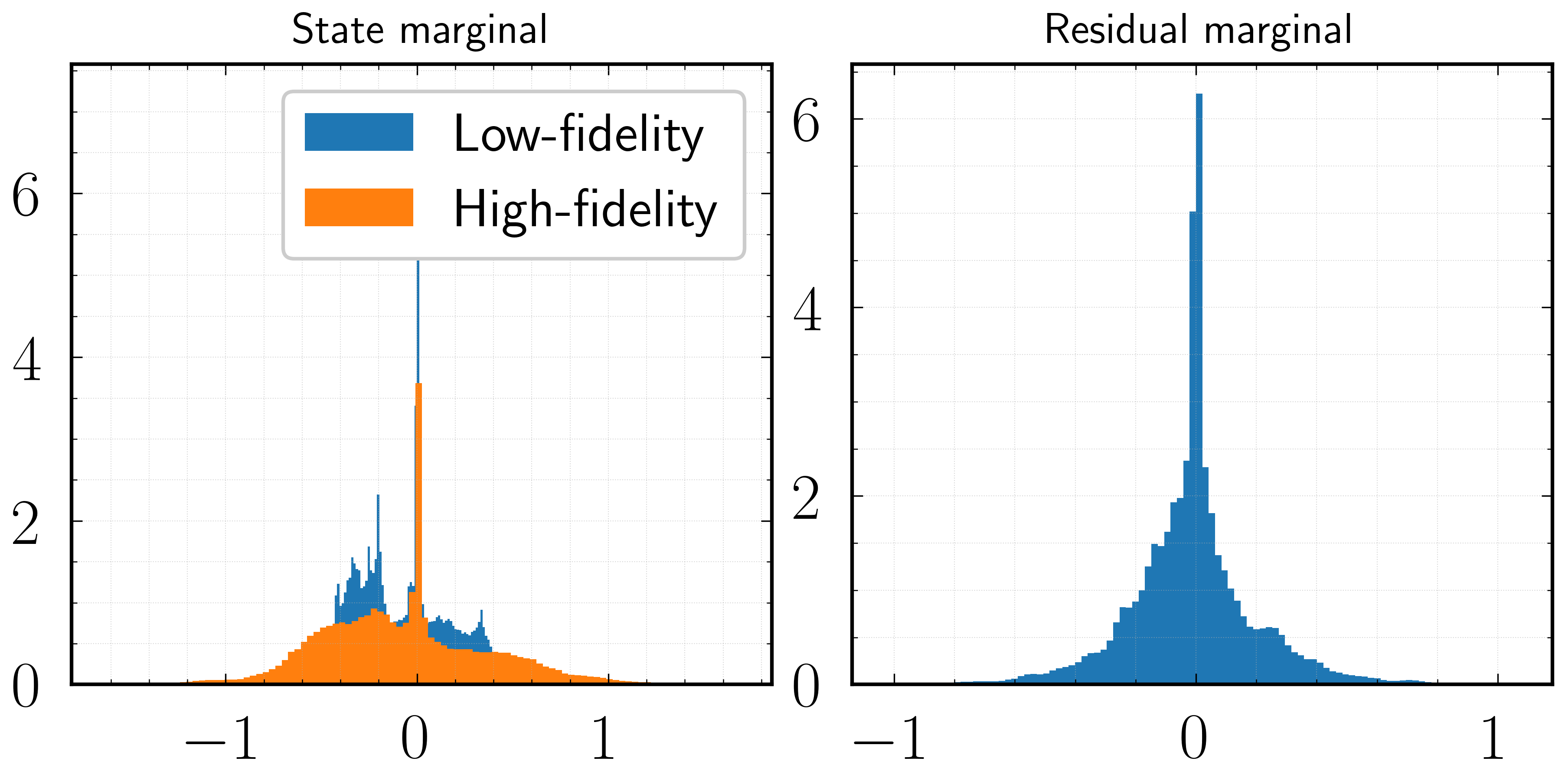}
    \caption{\em Marginal distributions for the 1D Advection benchmark. (\textbf{Left}) Distribution of the solution state. (\textbf{Right}) Distribution of the residual between the low- and high-fidelity solution functions.}
    \label{fig:appendix/advection/data_marginal}
\end{figure}
\begin{figure}[htbp]
    \centering
    \includegraphics[width=0.9\linewidth]{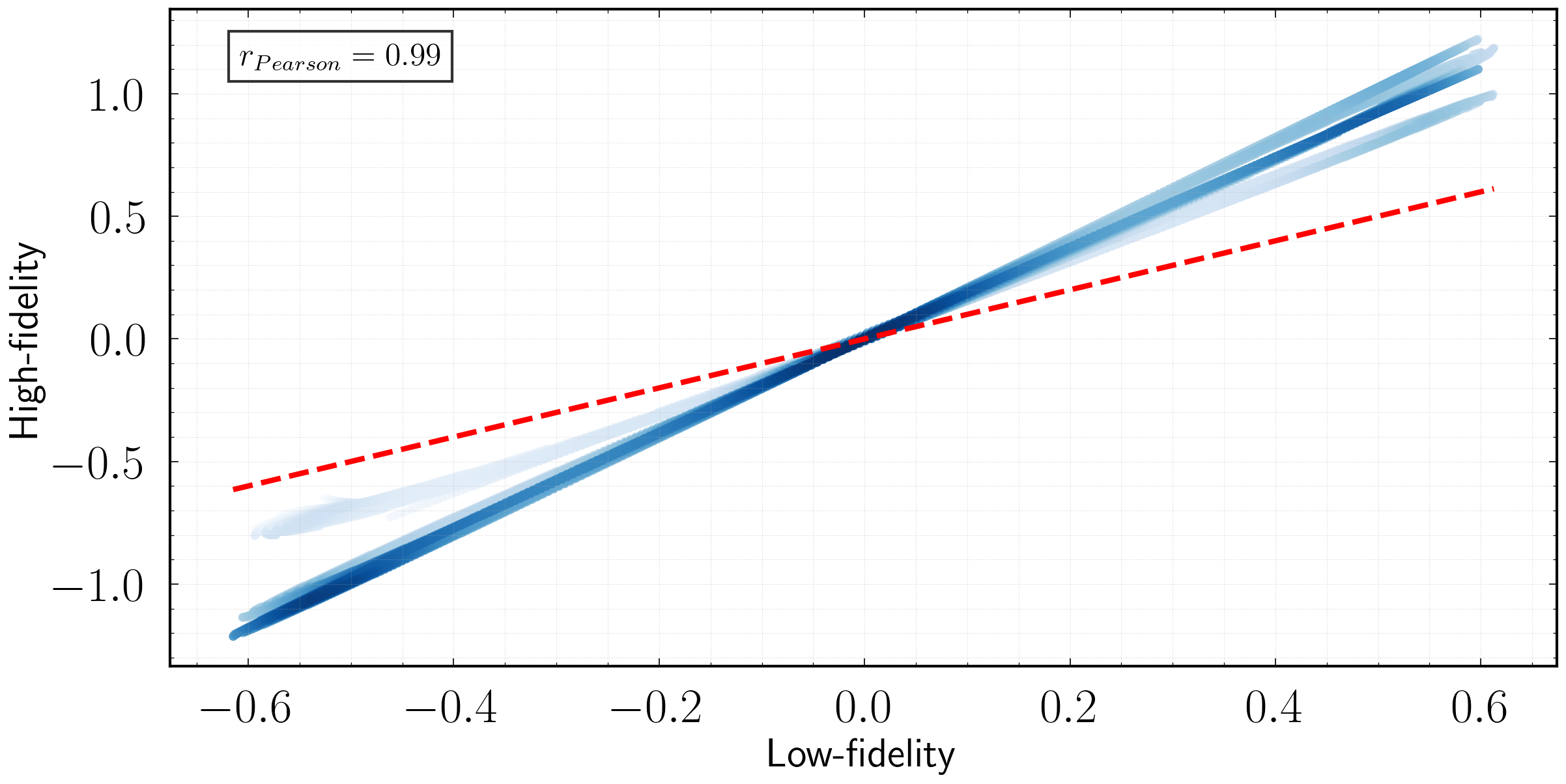}
    \caption{\em Joint distribution between the low- and high-fidelity solution states for the 1D Advection benchmark.
    }
    \label{fig:appendix/advection/LF_HF_joint_plot}
\end{figure}
\begin{figure}[htbp]
    \centering
    \includegraphics[width=0.9\linewidth]{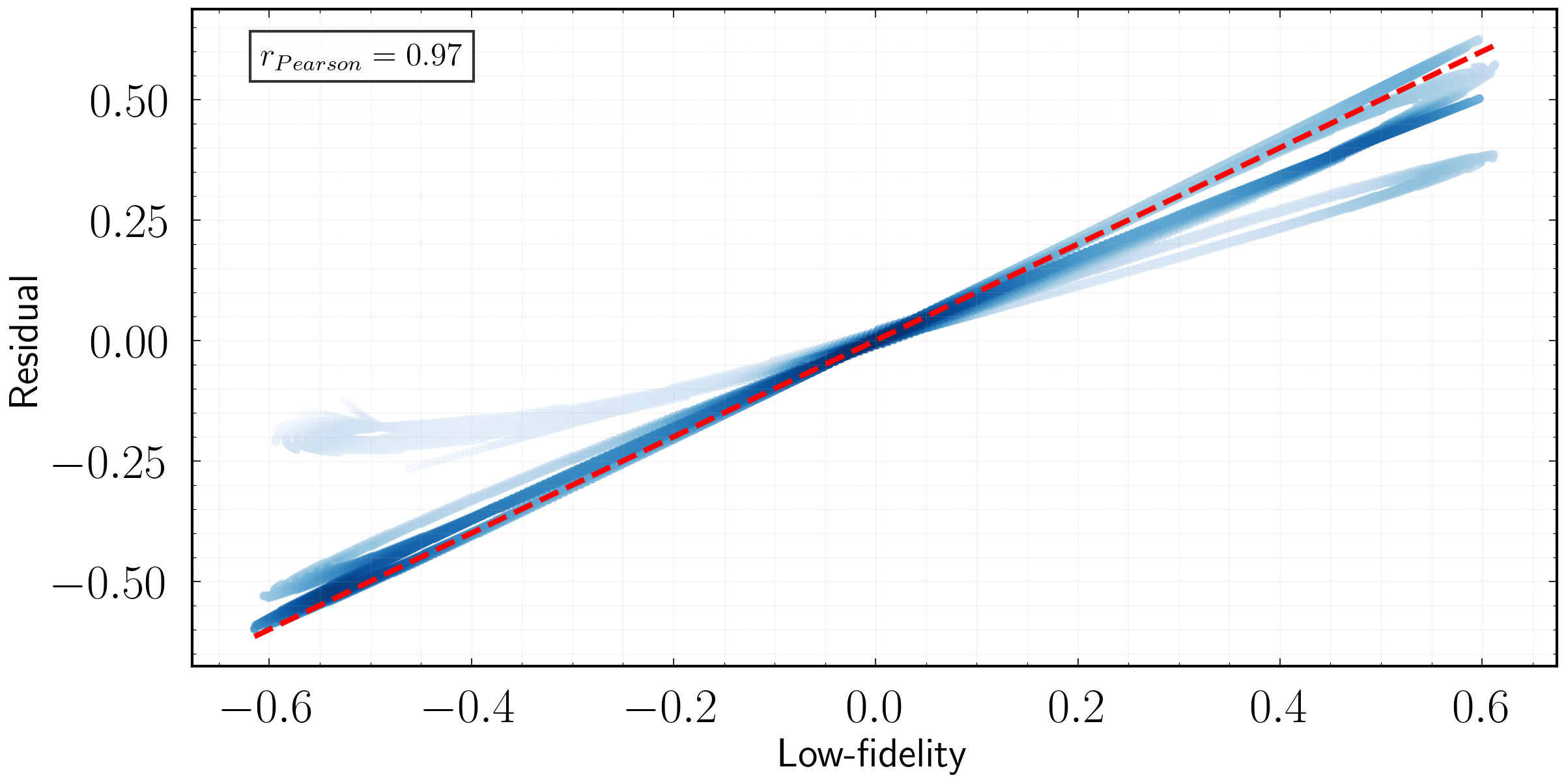}
    \caption{\em Joint distribution between the low-fidelity and residual for the 1D Advection benchmark.
    }
    \label{fig:appendix/advection/Residual_joint_plot}
\end{figure}
%
%
%
\begin{figure}
    \centering
    \includegraphics[width=0.99\linewidth]{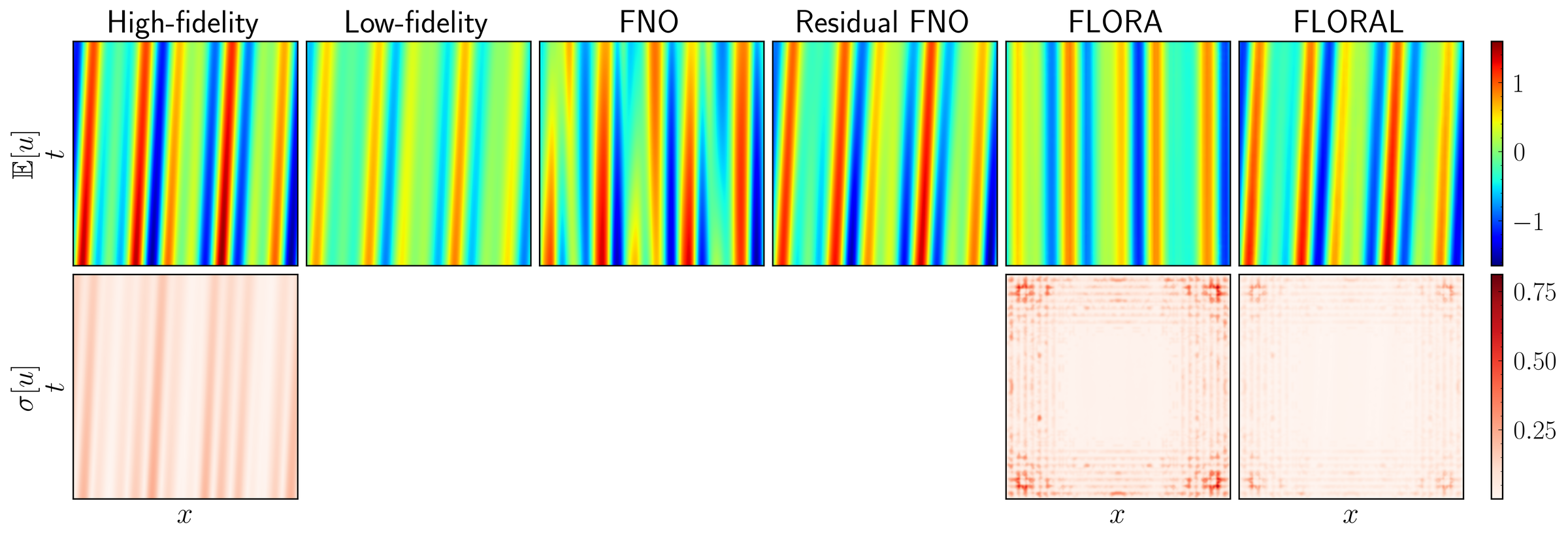}
    \caption{\em Predictive performance for approximating the solution function at an unseen input condition for the 1D Advection benchmark in the low-data regime.
    The model is trained using a total of $100$ training samples, corresponding to $50$ realizations of the stochastic solution function for each input condition.
    (\textbf{Top}) Mean solution function.
    (\textbf{Bottom}) Standard deviation of the solution function.
    }
    \label{fig:appendix/advection/mean_std_n_train_100_n_val_750_train_res_full}
\end{figure}
\begin{figure}
    \centering
    \includegraphics[width=0.99\linewidth]{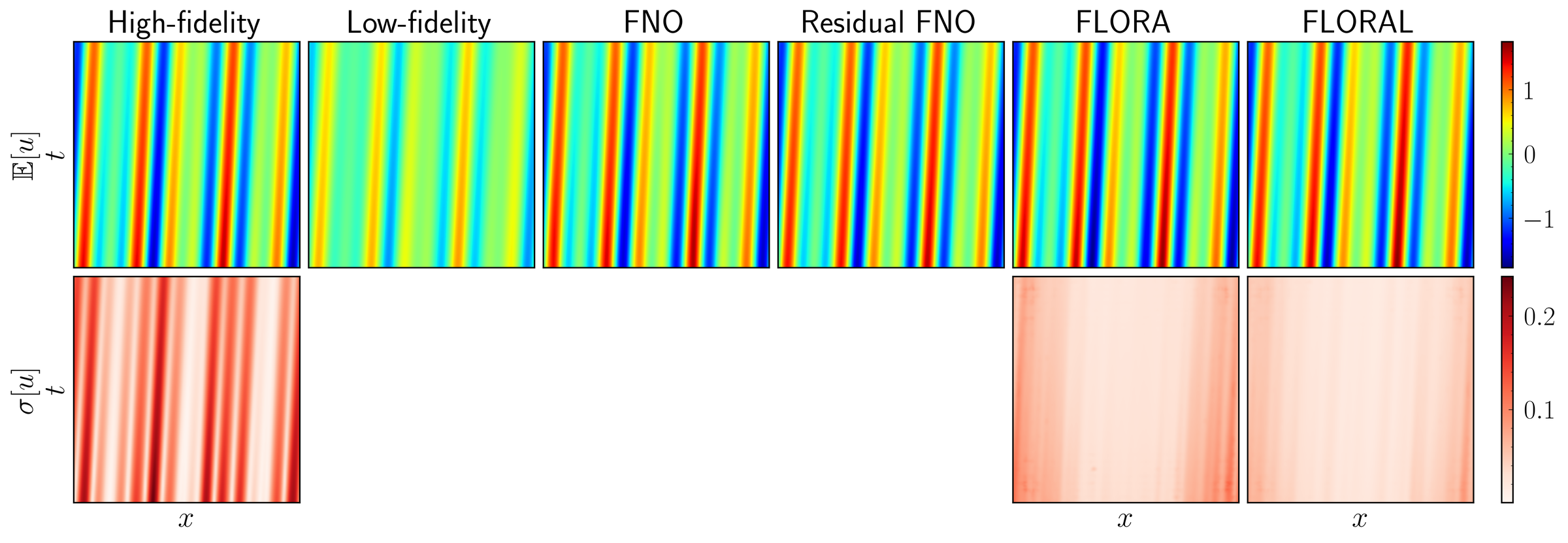}
    \caption{\em Predictive performance for approximating the solution function at an unseen input condition (same as~\cref{fig:appendix/advection/mean_std_n_train_100_n_val_750_train_res_full}) for the 1D Advection benchmark.
    The model is trained using a total of $500$ training samples, corresponding to $50$ realizations of the stochastic solution function for each input condition.
    (\textbf{Top}) Mean solution function.
    (\textbf{Bottom}) Standard deviation of the solution function.
    }
    \label{fig:appendix/advection/mean_std_n_train_500_n_val_750_train_res_full}
\end{figure}
\begin{figure}
    \centering
    \includegraphics[width=0.99\linewidth]{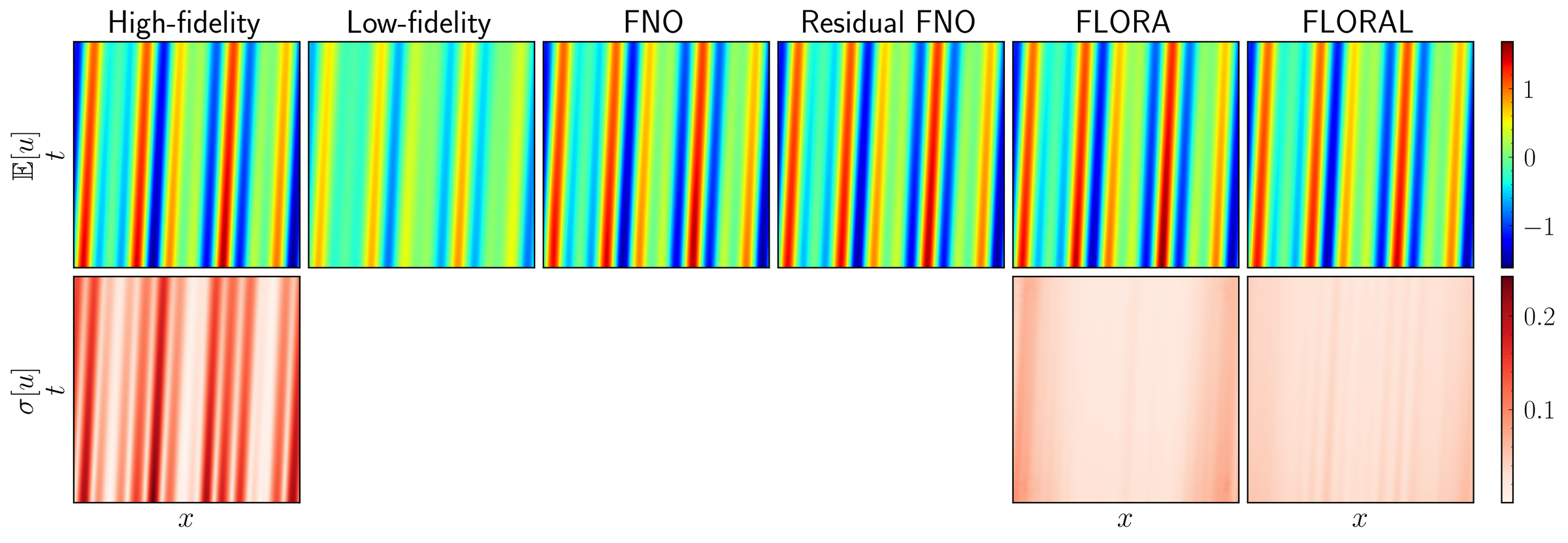}
    \caption{\em Predictive performance for approximating the solution function at an unseen input condition (same as~\cref{fig:appendix/advection/mean_std_n_train_100_n_val_750_train_res_full}) for the 1D Advection benchmark. 
    The model is trained using a total of $2000$ training samples, corresponding to $50$ realizations of the stochastic solution function for each input condition.
    (\textbf{Top}) Mean solution function.
    (\textbf{Bottom}) Standard deviation of the solution function.
    }
    \label{fig:appendix/advection/mean_std_n_train_2000_n_val_750_train_res_full}
\end{figure}
\begin{figure}[htbp]
    \centering
    \includegraphics[width=0.99\linewidth]{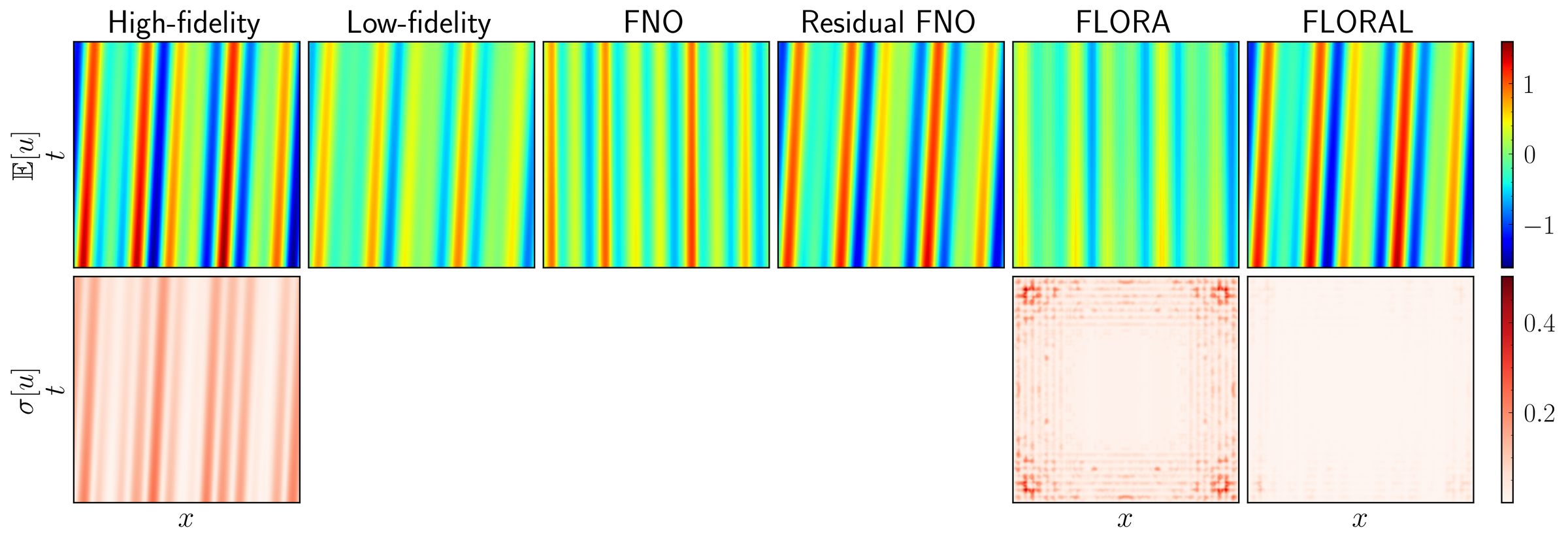}
    \caption{\em Predictive performance for approximating the solution function at an unseen input condition (same as~\cref{fig:appendix/advection/mean_std_n_train_100_n_val_750_train_res_full}) for the 1D Advection benchmark.
    For training, the solution function is observed only at $8\times8$ uniformly-spaced points in the spatial domain, while during inference the learned operator is evaluated at $64\times 64$ spatial locations, corresponding to a $8\times$ super-resolution reconstruction along each dimension.
    The model is trained using a total of $500$ training samples, corresponding to $50$ realizations of the solution function for each input condition.
    (\textbf{Top}) Mean solution function.
    (\textbf{Bottom}) Standard deviation of the solution function.}
    \label{fig:appendix/advection/mean_std_n_train_500_n_val_750_train_res_8}
\end{figure}
\begin{figure}[htbp]
    \centering
    \includegraphics[width=0.99\linewidth]{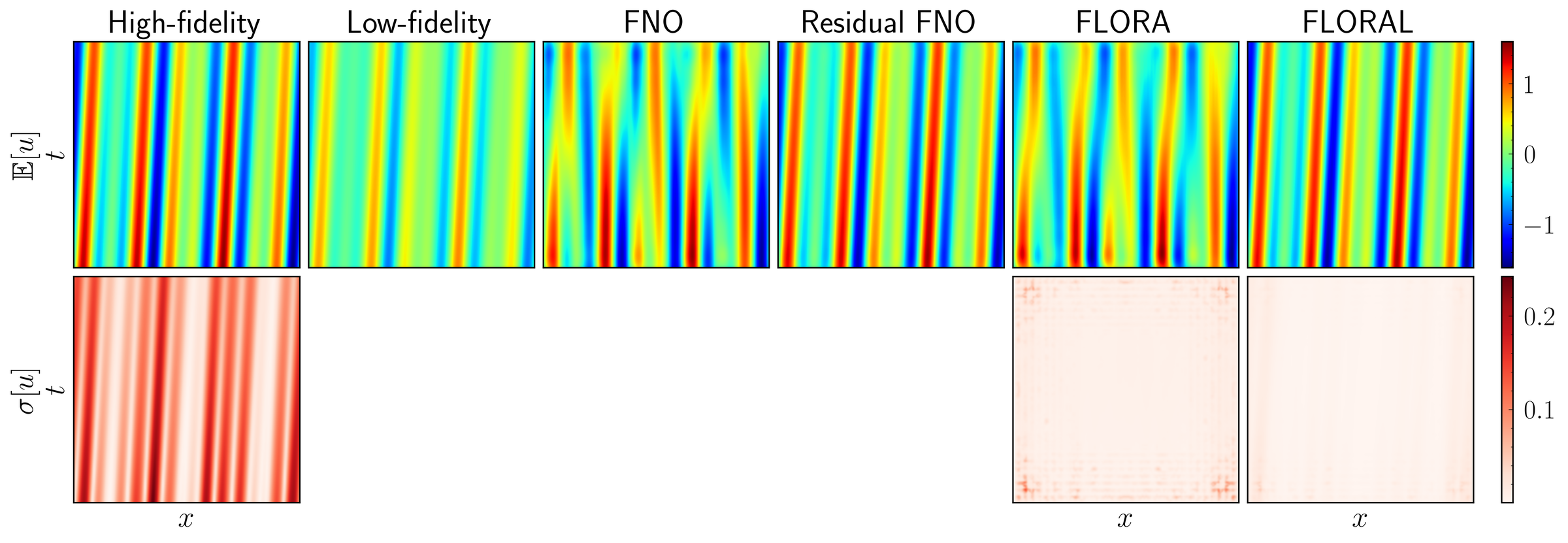}
    \caption{\em Predictive performance for approximating the solution function at an unseen input condition (same as~\cref{fig:appendix/advection/mean_std_n_train_100_n_val_750_train_res_full}) for the 1D Advection benchmark.
    For training, the solution function is evaluated only at $16\times16$ uniformly spaced points in the spatial domain, whereas during inference, the learned operator is evaluated at $64\times 64$ spatial locations, corresponding to a $4\times$ super-resolution reconstruction along each dimension.
    The model is trained using a total of $500$ training samples, corresponding to $50$ realizations of the solution function for each input condition.
    (\textbf{Top}) Mean solution function.
    (\textbf{Bottom}) Standard deviation of the solution function.}
    \label{fig:appendix/advection/mean_std_n_train_500_n_val_750_train_res_16}
\end{figure}
\begin{figure}
    \centering
    \includegraphics[width=0.99\linewidth]{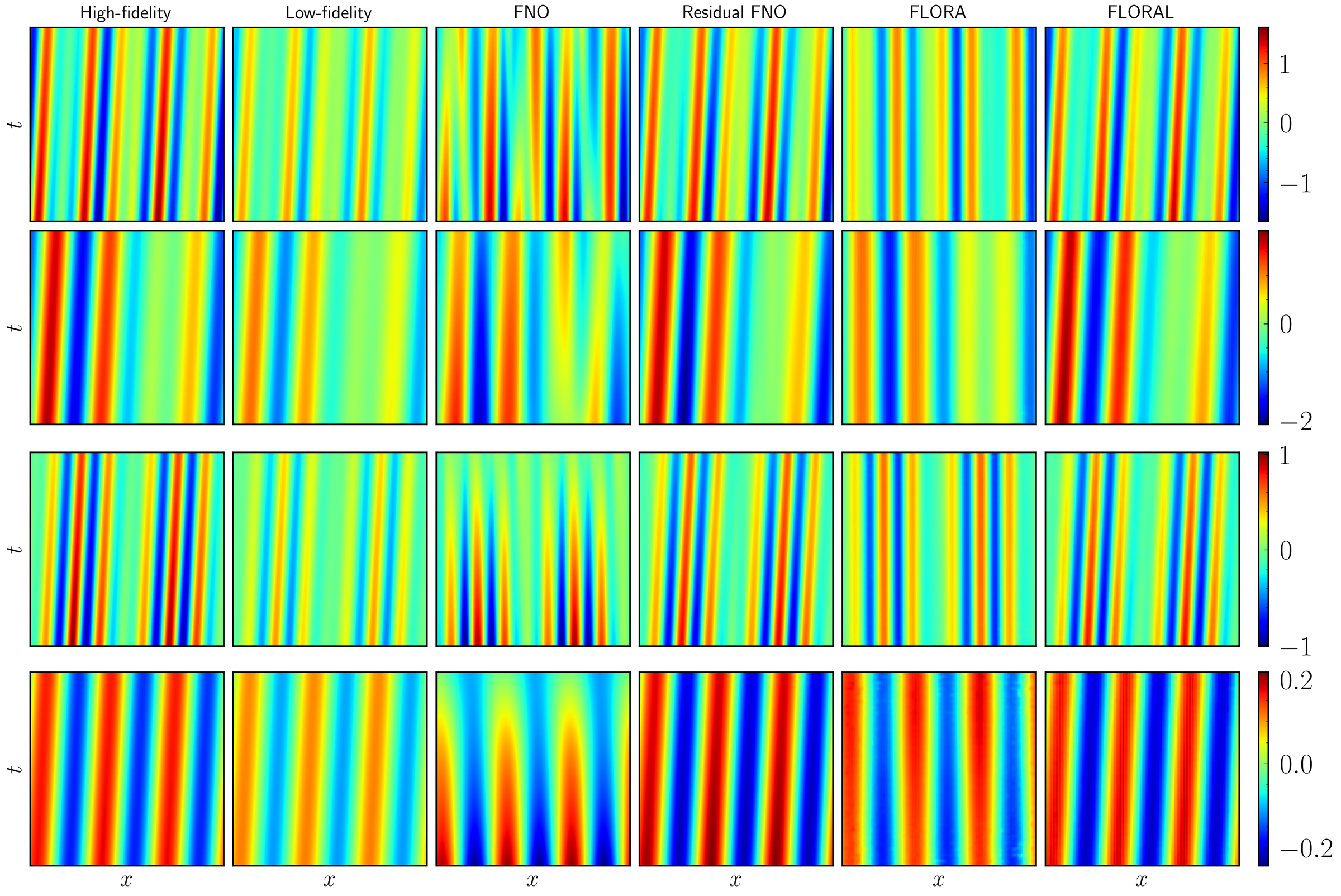}
    \caption{\em Predictive samples of the stochastic solution function for the 1D Advection benchmark at unseen input conditions.
    Each row corresponds to a distinct unseen input condition, with $2500$ realizations generated from the learned probabilistic operator, and mean solution function is shown.
    The model is trained using a total of $100$ training samples, corresponding to $50$ realizations of the solution function for each input condition. }
    \label{fig:appendix/advection/field_samples_n_train_100_n_val_750}
\end{figure}
\begin{figure}
    \centering
    \includegraphics[width=0.99\linewidth]{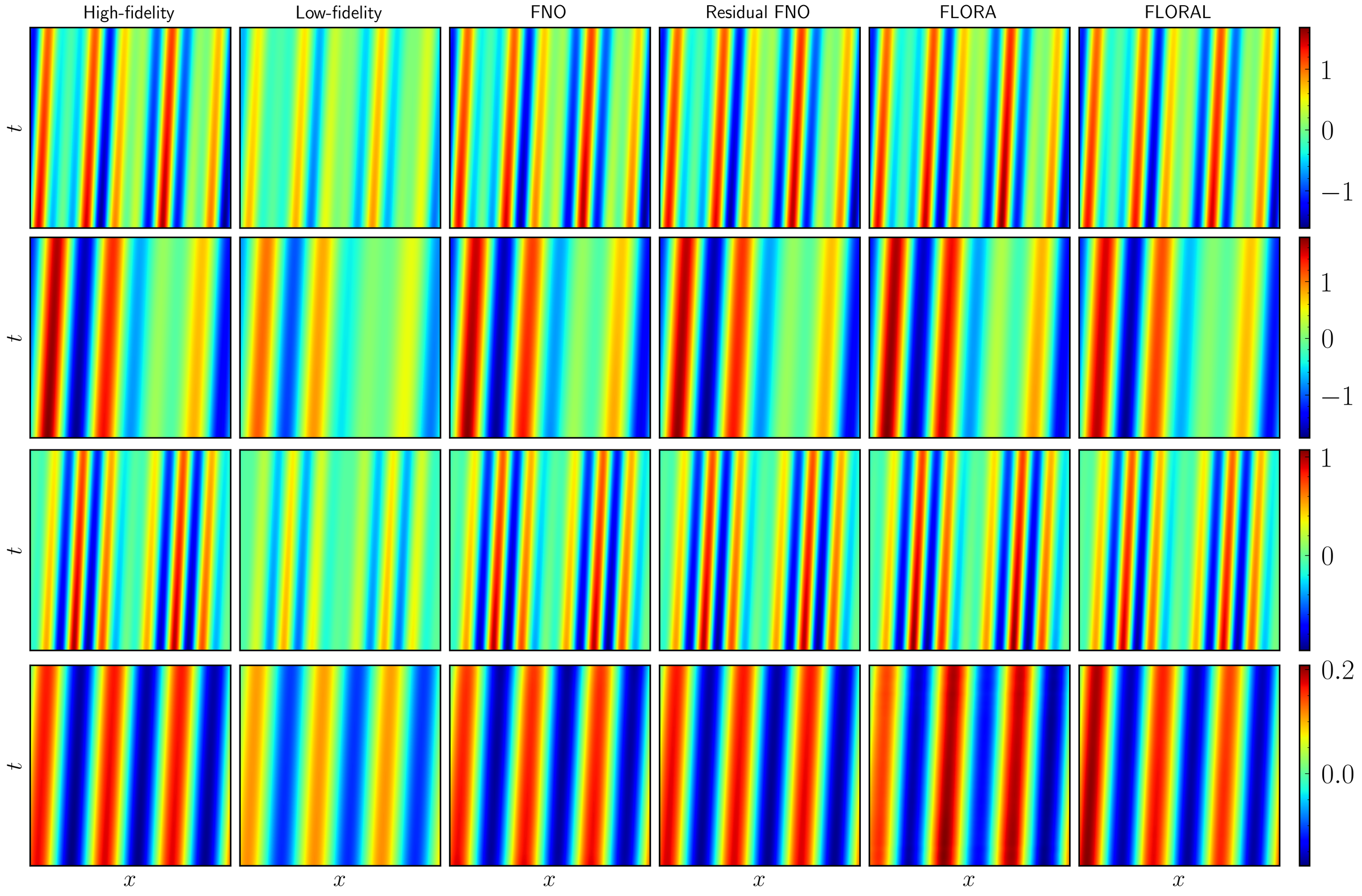}
    \caption{\em Predictive samples of the stochastic solution function for the 1D Advection benchmark at unseen input conditions (same as~\cref{fig:appendix/advection/field_samples_n_train_100_n_val_750}).
    Each row corresponds to a distinct unseen input condition, with $2500$ realizations generated from the learned probabilistic operator, and mean solution function is shown.
    The model is trained using a total of $2000$ training samples, corresponding to $50$ realizations of the solution function for each input condition. }
    \label{fig:appendix/advection/field_samples_n_train_2000_n_val_750}
\end{figure}
\clearpage
\section{1D Burgers' Equation with Stochastic Initial Conditions (Burger's)}\label{sec:appendix/burgers}
\begin{figure}[htbp]
    \centering
    \includegraphics[width=0.7\linewidth]{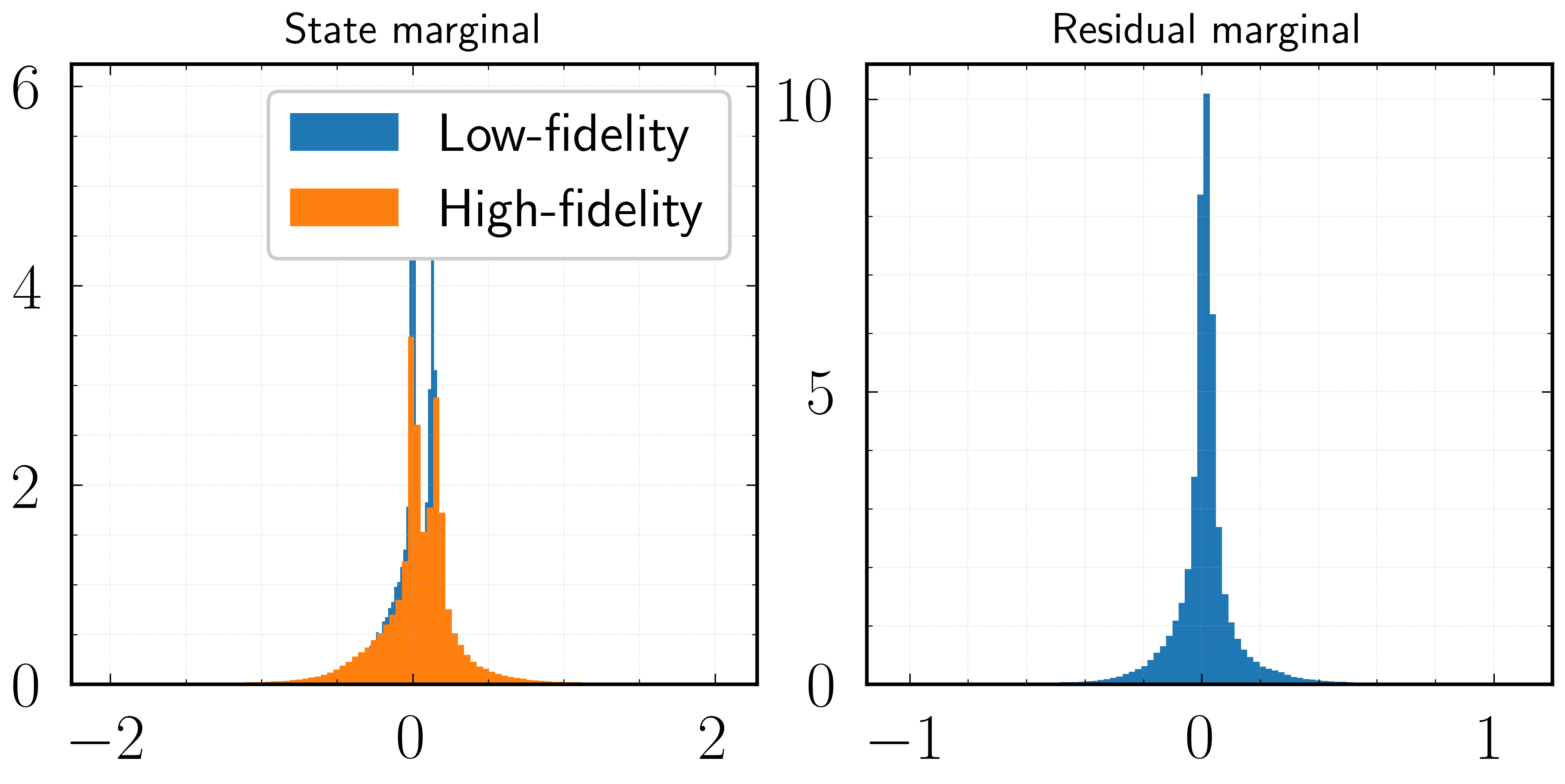}
    \caption{\em Marginal distributions for the 1D Burgers' benchmark. (\textbf{Left}) Distribution of the solution state. (\textbf{Right}) Distribution of the residual between the low- and high-fidelity solution functions.}
    \label{fig:appendix/burgers/data_marginal}
\end{figure}
\begin{figure}[htbp]
    \centering
    \includegraphics[width=0.9\linewidth]{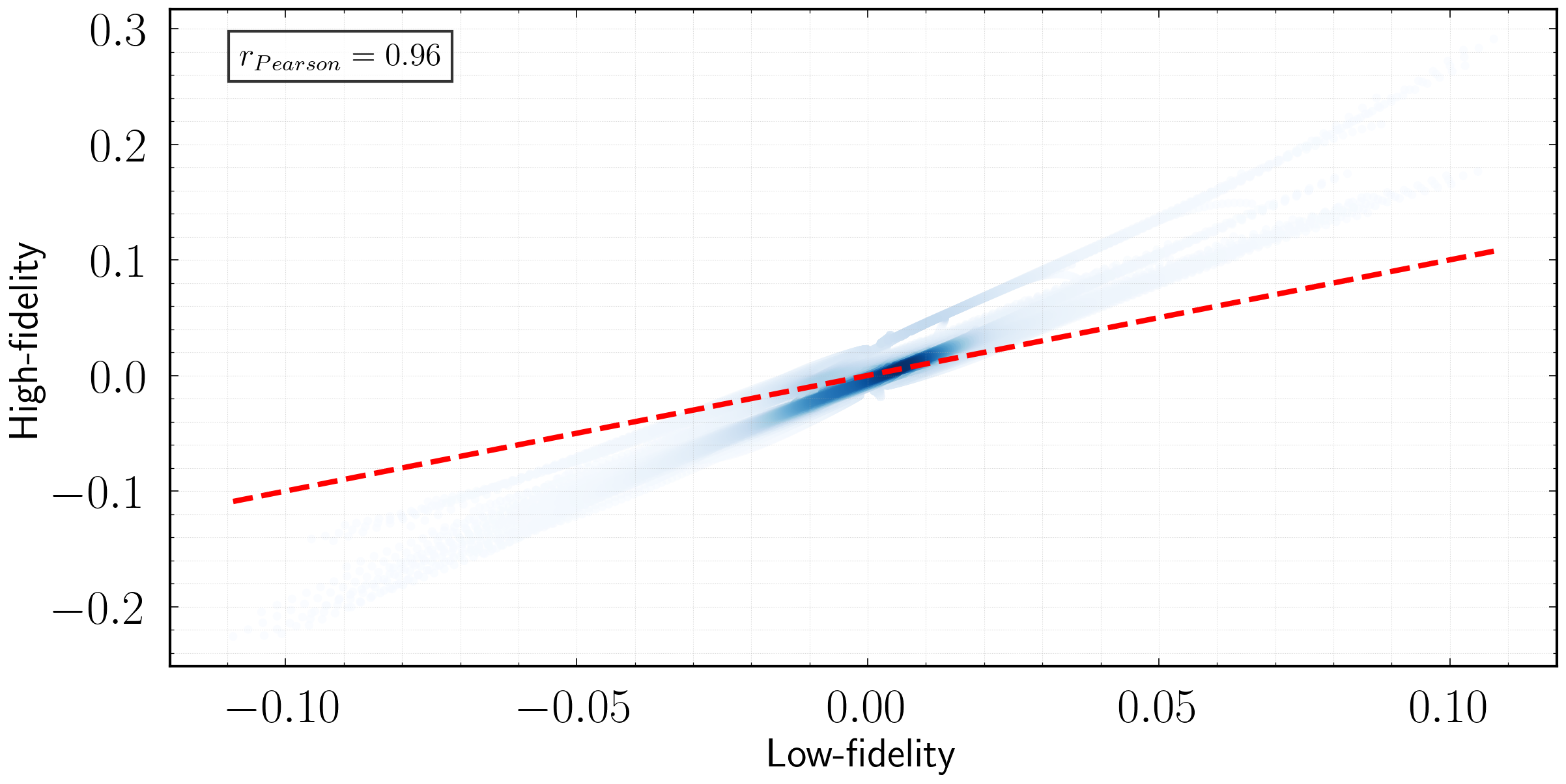}
    \caption{\em Joint distribution between the low- and high-fidelity solution states for the 1D Burgers' benchmark.
    }
    \label{fig:appendix/burgers/LF_HF_joint_plot}
\end{figure}
\begin{figure}[htbp]
    \centering
    \includegraphics[width=0.9\linewidth]{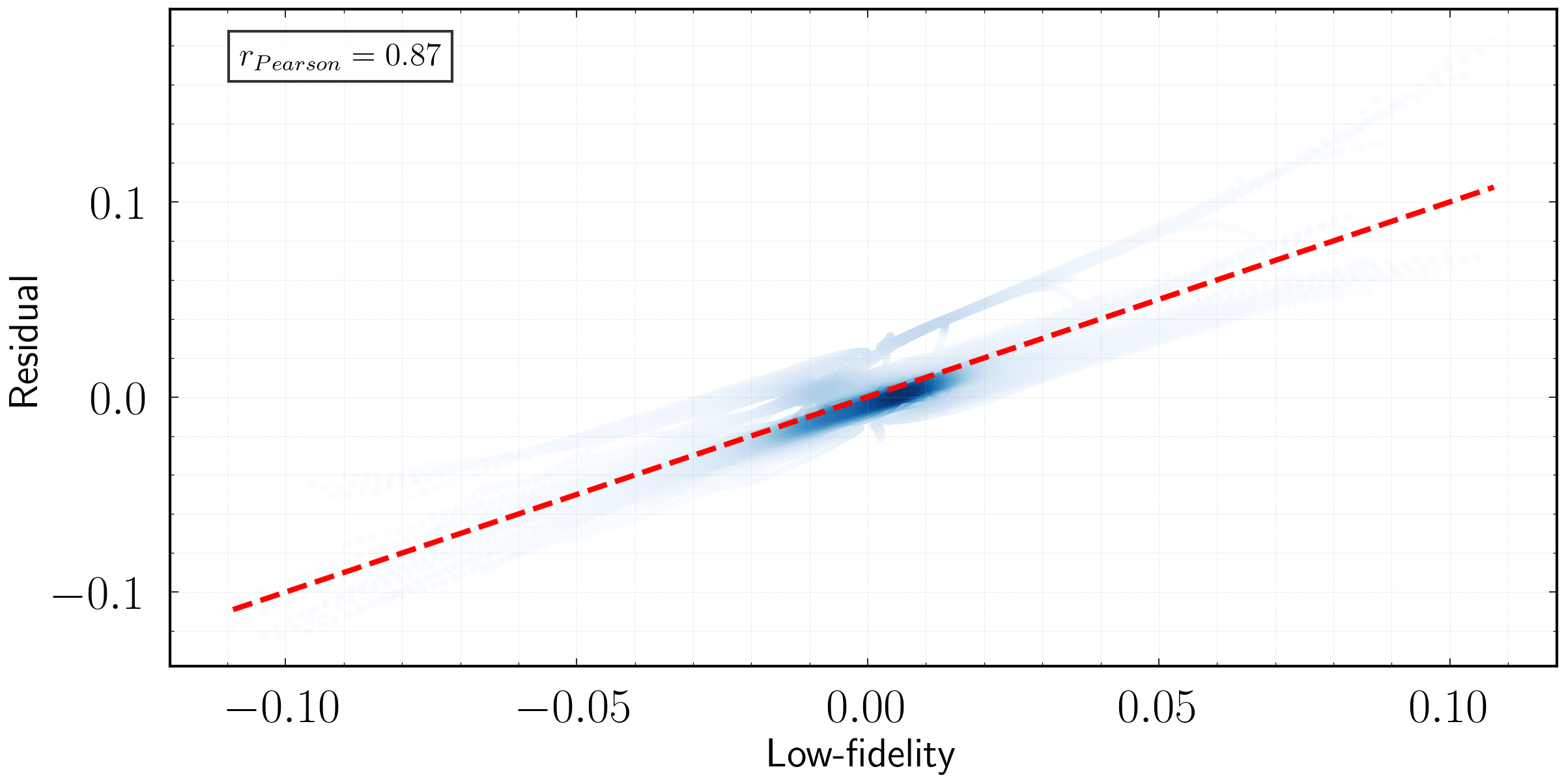}
    \caption{\em Joint distribution between the low-fidelity and residual for the 1D Burgers' benchmark.
    }
    \label{fig:appendix/burgers/Residual_joint_plot}
\end{figure}
%
%
%
\begin{figure}
    \centering
    \includegraphics[width=0.99\linewidth]{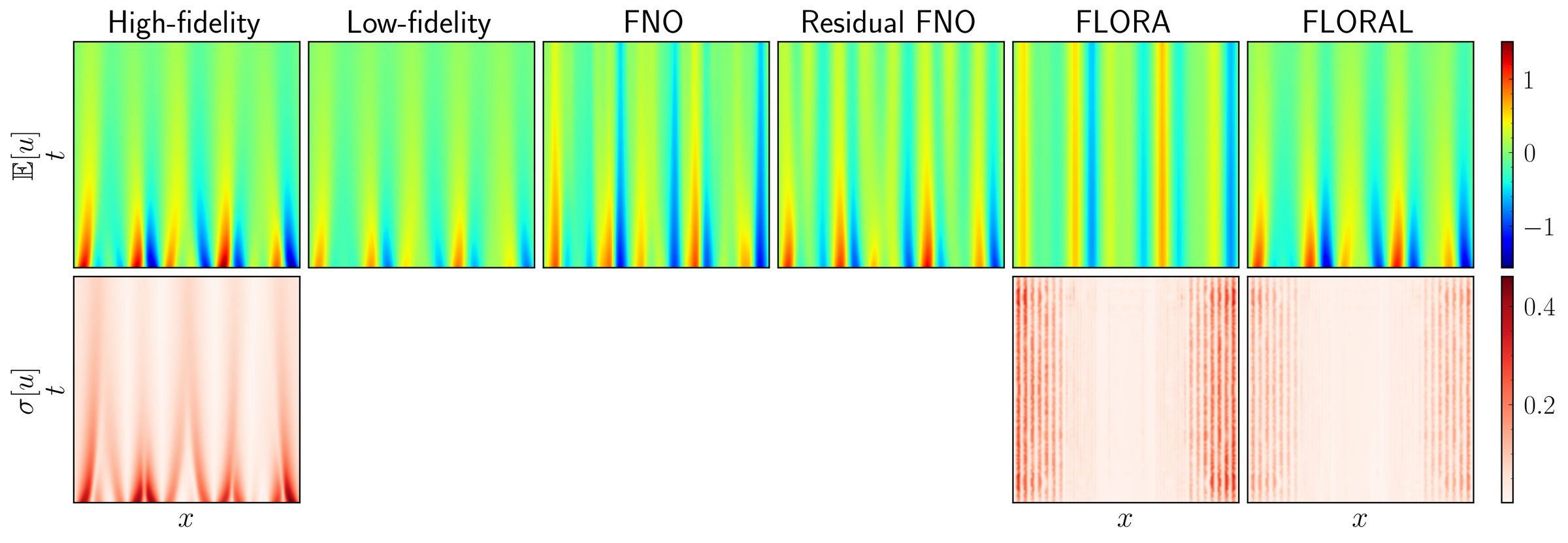}
    \caption{\em Predictive performance for approximating the solution function at an unseen input condition for the 1D Burgers' benchmark in the low-data regime.
    The model is trained using a total of $100$ training samples, corresponding to $50$ realizations of the stochastic solution function for each input condition.
    (\textbf{Top}) Mean solution function.
    (\textbf{Bottom}) Standard deviation of the solution function.
    }
    \label{fig:appendix/burgers/mean_std_n_train_100_n_val_750_train_res_full}
\end{figure}
\begin{figure}
    \centering
    \includegraphics[width=0.99\linewidth]{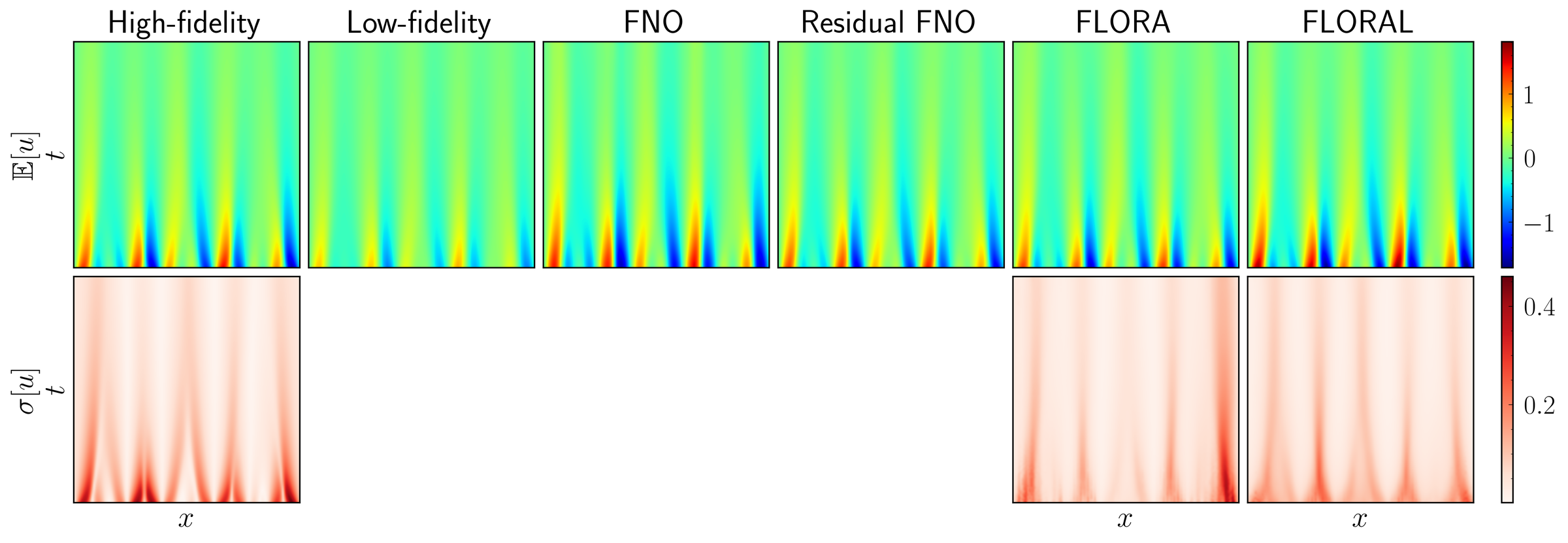}
    \caption{\em Predictive performance for approximating the solution function at an unseen input condition (same as~\cref{fig:appendix/burgers/mean_std_n_train_100_n_val_750_train_res_full}) for the 1D Burgers' benchmark.
    The model is trained using a total of $500$ training samples, corresponding to $50$ realizations of the stochastic solution function for each input condition.
    (\textbf{Top}) Mean solution function.
    (\textbf{Bottom}) Standard deviation of the solution function.
    }
    \label{fig:appendix/burgers/mean_std_n_train_500_n_val_750_train_res_full}
\end{figure}
\begin{figure}
    \centering
    \includegraphics[width=0.99\linewidth]{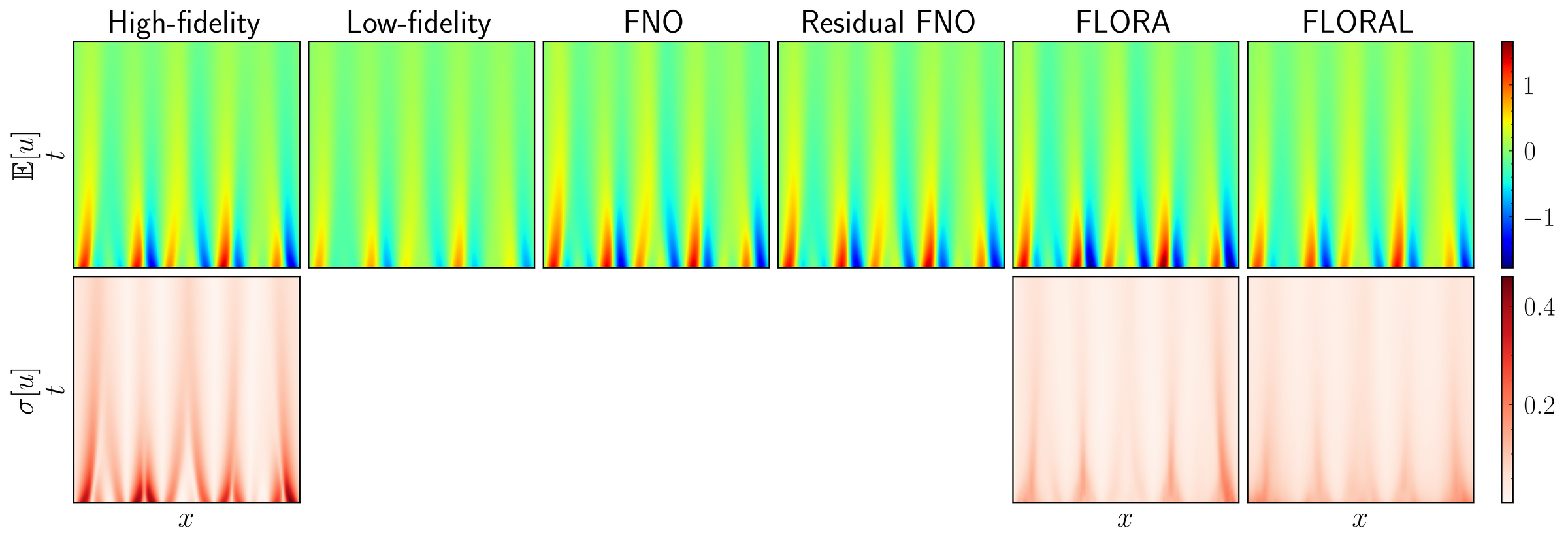}
    \caption{\em Predictive performance for approximating the solution function at an unseen input condition (same as~\cref{fig:appendix/burgers/mean_std_n_train_100_n_val_750_train_res_full}) for the 1D Burgers' benchmark. 
    The model is trained using a total of $2000$ training samples, corresponding to $50$ realizations of the stochastic solution function for each input condition.
    (\textbf{Top}) Mean solution function.
    (\textbf{Bottom}) Standard deviation of the solution function.
    }
    \label{fig:appendix/burgers/mean_std_n_train_2000_n_val_750_train_res_full}
\end{figure}
\begin{figure}[htbp]
    \centering
    \includegraphics[width=0.99\linewidth]{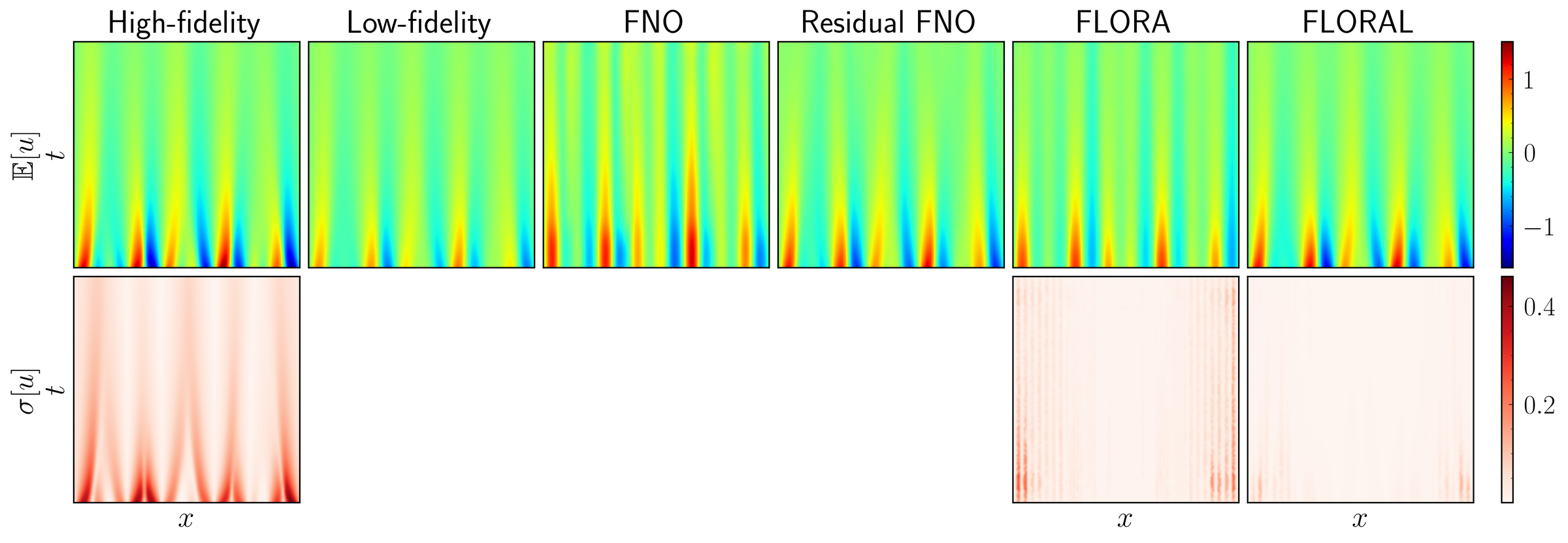}
    \caption{\em Predictive performance for approximating the solution function at an unseen input condition (same as~\cref{fig:appendix/burgers/mean_std_n_train_100_n_val_750_train_res_full}) for the 1D Burgers' benchmark.
    For training, the solution function is observed only at $8\times8$ uniformly-spaced points in the spatial domain, while during inference the learned operator is evaluated at $64\times 64$ spatial locations, corresponding to a $8\times$ super-resolution reconstruction along each dimension.
    The model is trained using a total of $500$ training samples, corresponding to $50$ realizations of the solution function for each input condition.
    (\textbf{Top}) Mean solution function.
    (\textbf{Bottom}) Standard deviation of the solution function.}
    \label{fig:appendix/burgers/mean_std_n_train_500_n_val_750_train_res_8}
\end{figure}
\begin{figure}[htbp]
    \centering
    \includegraphics[width=0.99\linewidth]{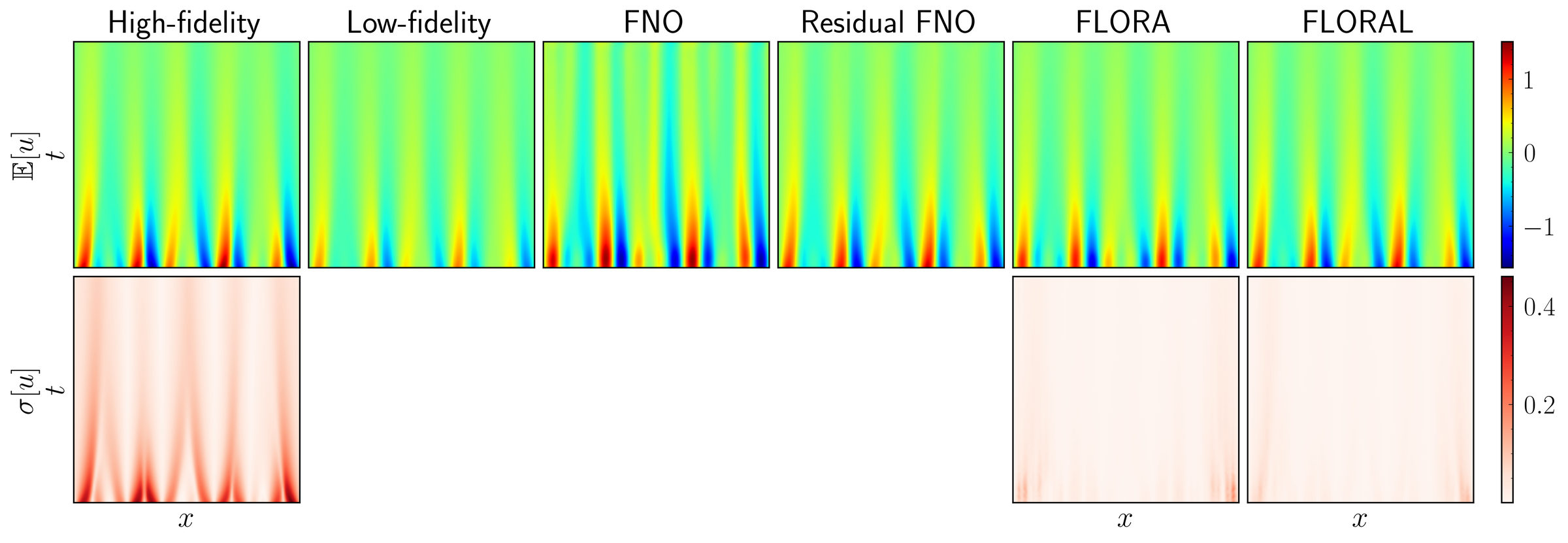}
    \caption{\em Predictive performance for approximating the solution function at an unseen input condition (same as~\cref{fig:appendix/burgers/mean_std_n_train_100_n_val_750_train_res_full}) for the 1D Burgers' benchmark.
    For training, the solution function is evaluated only at $16\times16$ uniformly spaced points in the spatial domain, whereas during inference, the learned operator is evaluated at $64\times 64$ spatial locations, corresponding to a $4\times$ super-resolution reconstruction along each dimension.
    The model is trained using a total of $500$ training samples, corresponding to $50$ realizations of the solution function for each input condition.
    (\textbf{Top}) Mean solution function.
    (\textbf{Bottom}) Standard deviation of the solution function.}
    \label{fig:appendix/burgers/mean_std_n_train_500_n_val_750_train_res_16}
\end{figure}
\begin{figure}
    \centering
    \includegraphics[width=0.99\linewidth]{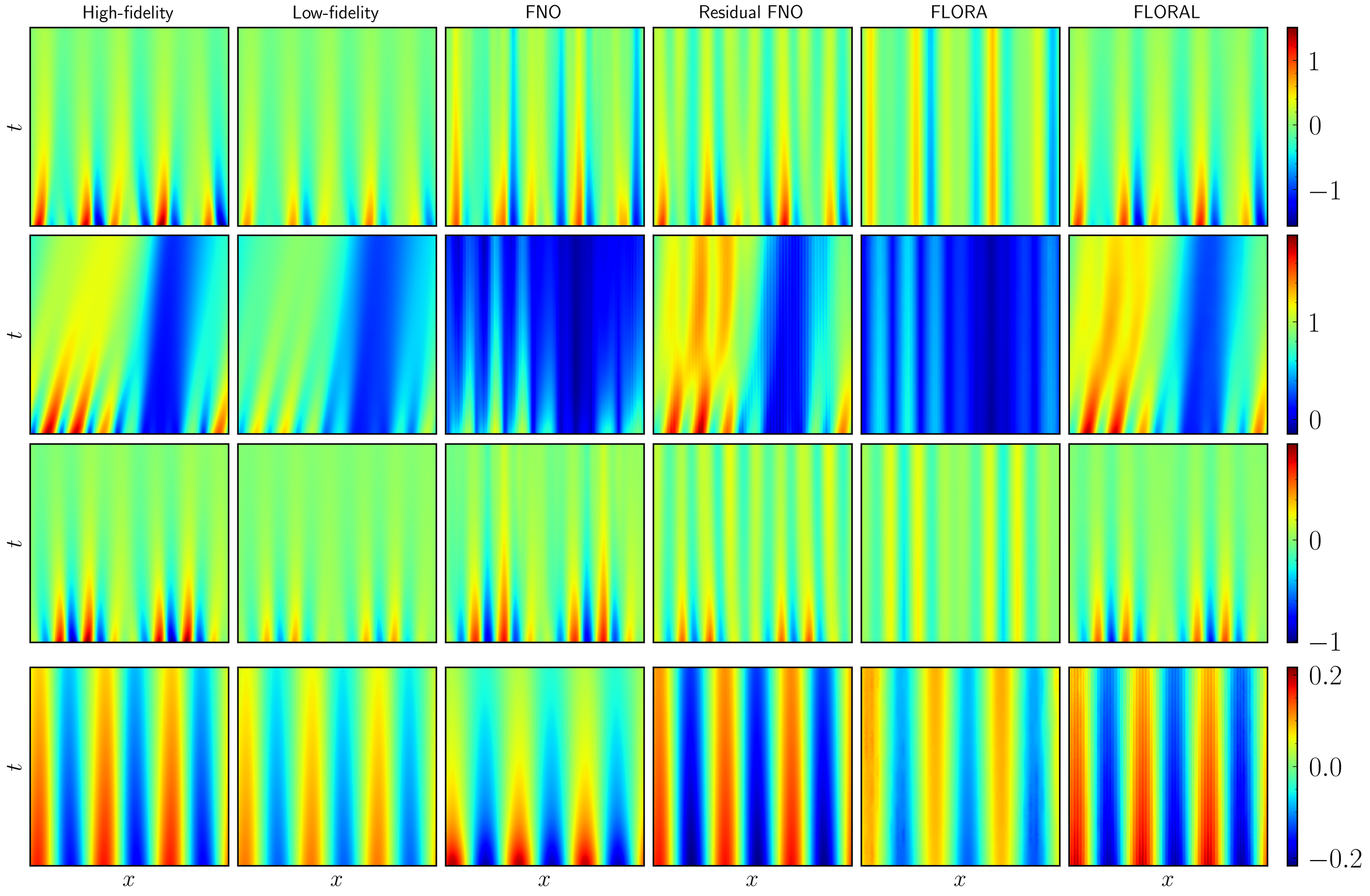}
    \caption{\em Predictive samples of the stochastic solution function for the 1D Burgers' benchmark at unseen input conditions.
    Each row corresponds to a distinct unseen input condition, with $2500$ realizations generated from the learned probabilistic operator, and mean solution function is shown.
    The model is trained using a total of $100$ training samples, corresponding to $50$ realizations of the solution function for each input condition. }
    \label{fig:appendix/burgers/field_samples_n_train_100_n_val_750}
\end{figure}
\begin{figure}
    \centering
    \includegraphics[width=0.99\linewidth]{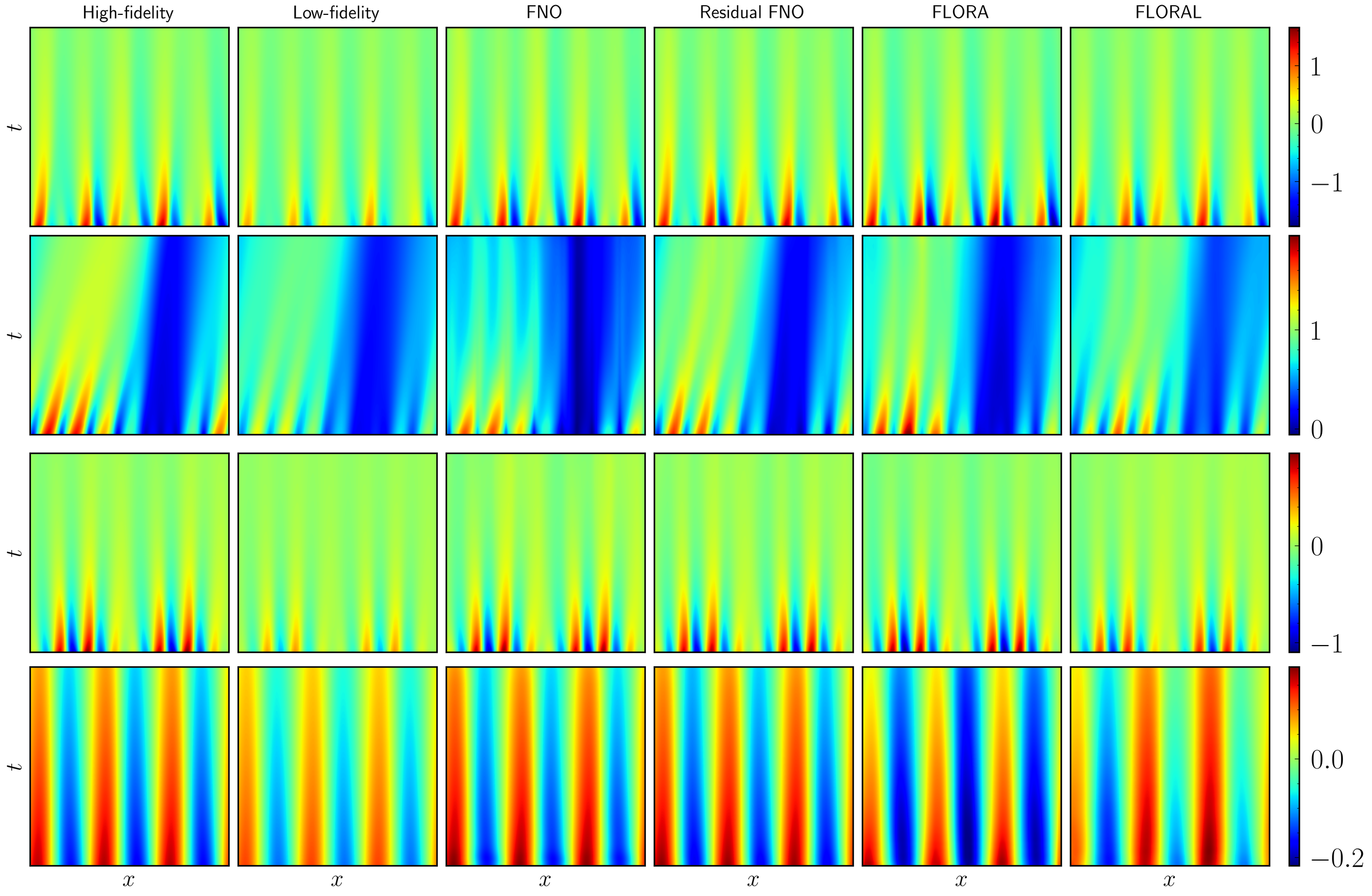}
    \caption{\em Predictive samples of the stochastic solution function for the 1D Burgers' benchmark at unseen input conditions (same as~\cref{fig:appendix/burgers/field_samples_n_train_100_n_val_750}).
    Each row corresponds to a distinct unseen input condition, with $2500$ realizations generated from the learned probabilistic operator, and mean solution function is shown.
    The model is trained using a total of $2000$ training samples, corresponding to $50$ realizations of the solution function for each input condition. }
    \label{fig:appendix/burgers/field_samples_n_train_2000_n_val_750}
\end{figure}
\clearpage
\section{Flow Through Porous Media with Uncertain Permeability field (Darcy)}\label{sec:appendix/darcy}
\begin{figure}[htbp]
    \centering
    \includegraphics[width=0.7\linewidth]{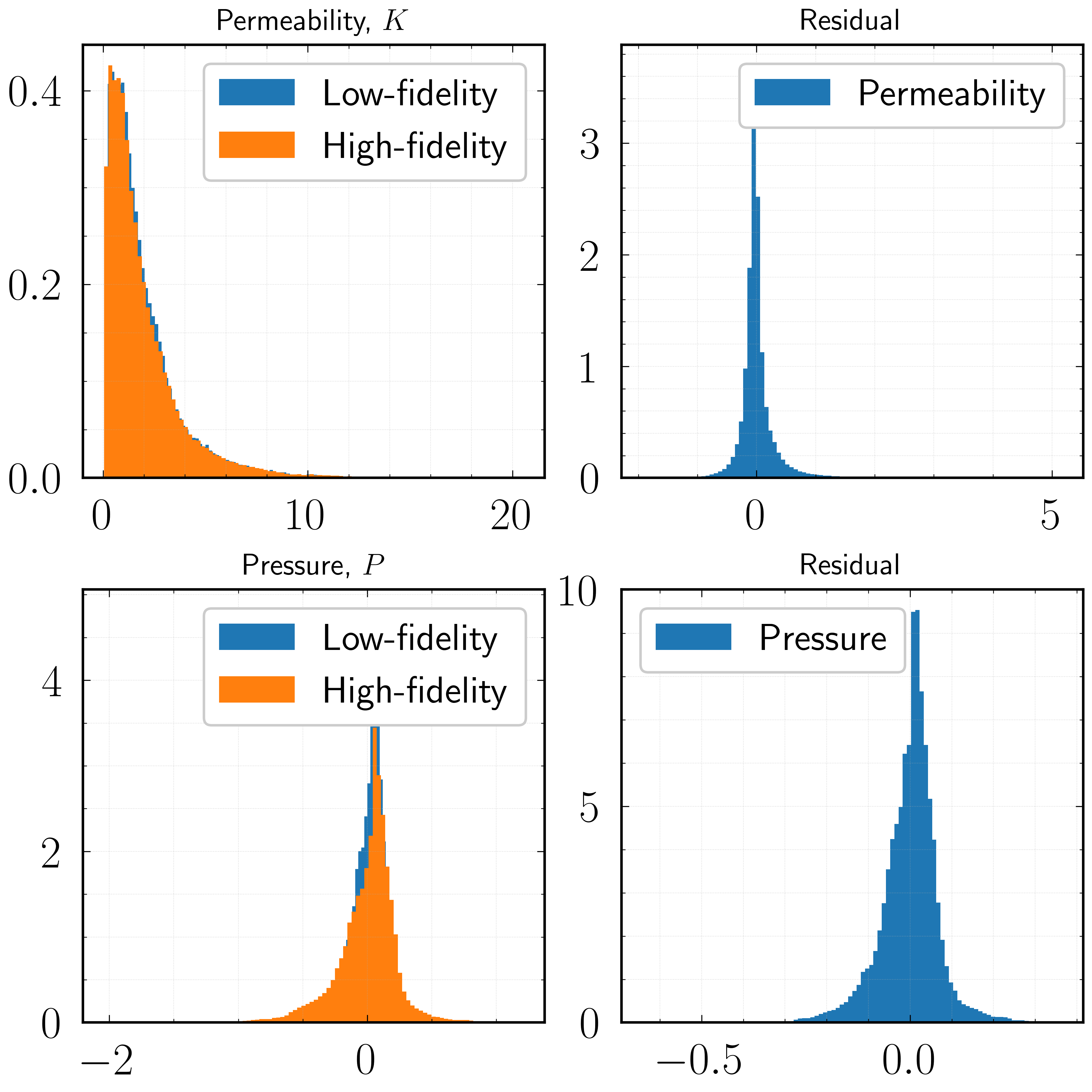}
    \caption{\em Marginal distributions for the flow through porous media. (\textbf{Top left}) Distribution of the low- and high-fidelity permeability fields. (\textbf{Top right}) Distribution of the corresponding permeability residual. (\textbf{Bottom left}) Distribution of the low- and high-fidelity pressure solution field. (\textbf{Bottom right}) Distribution of the corresponding pressure residual.}
    \label{fig:appendix/darcy/data_marginal}
\end{figure}
\begin{figure}[htbp]
    \centering
    \includegraphics[width=0.9\linewidth]{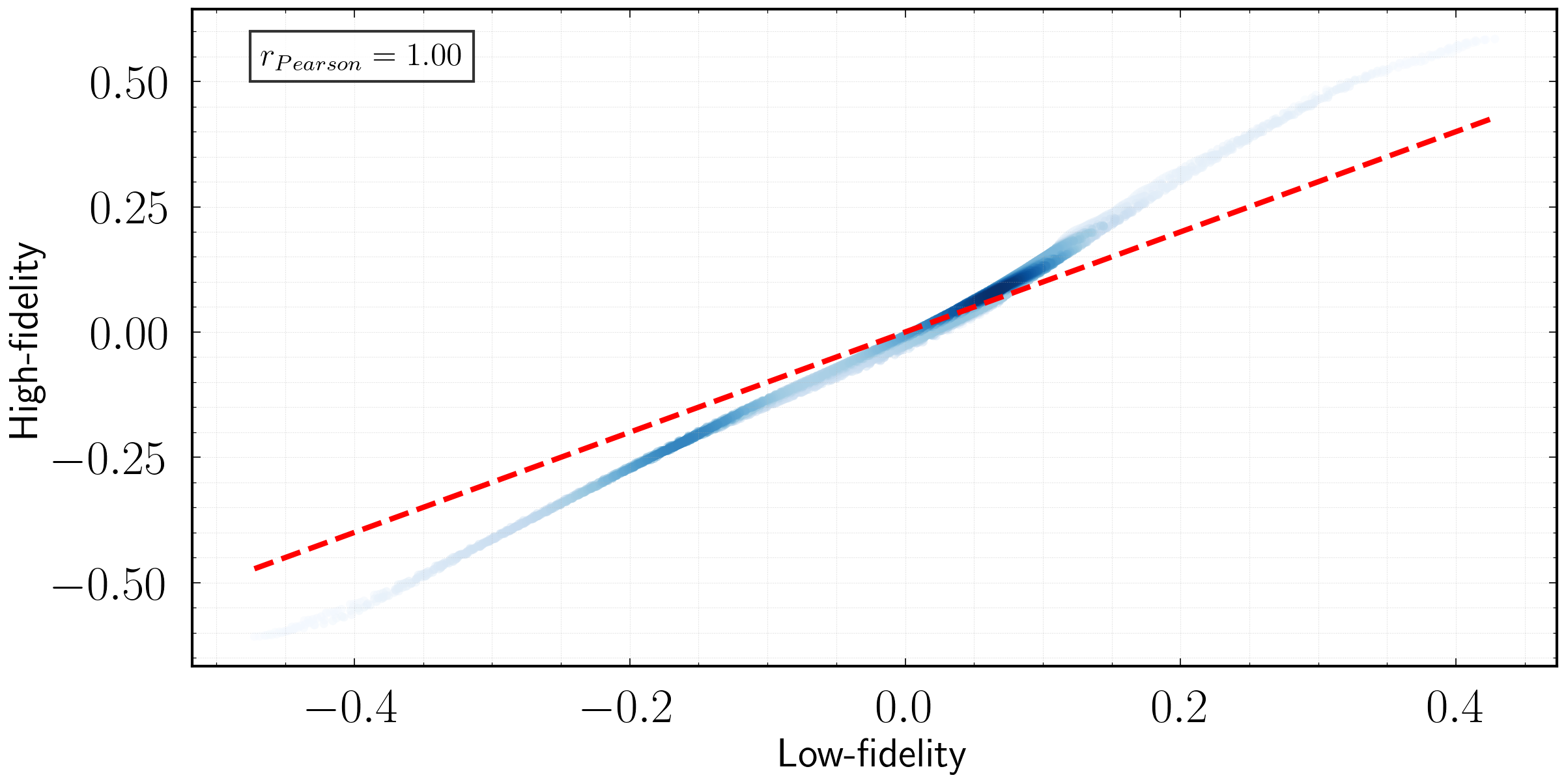}
    \caption{\em Joint distribution between the low- and high-fidelity pressure field for the flow through porous media.
    }
    \label{fig:appendix/darcy/LF_HF_Pressure_joint_plot}
\end{figure}
\begin{figure}[htbp]
    \centering
    \includegraphics[width=0.9\linewidth]{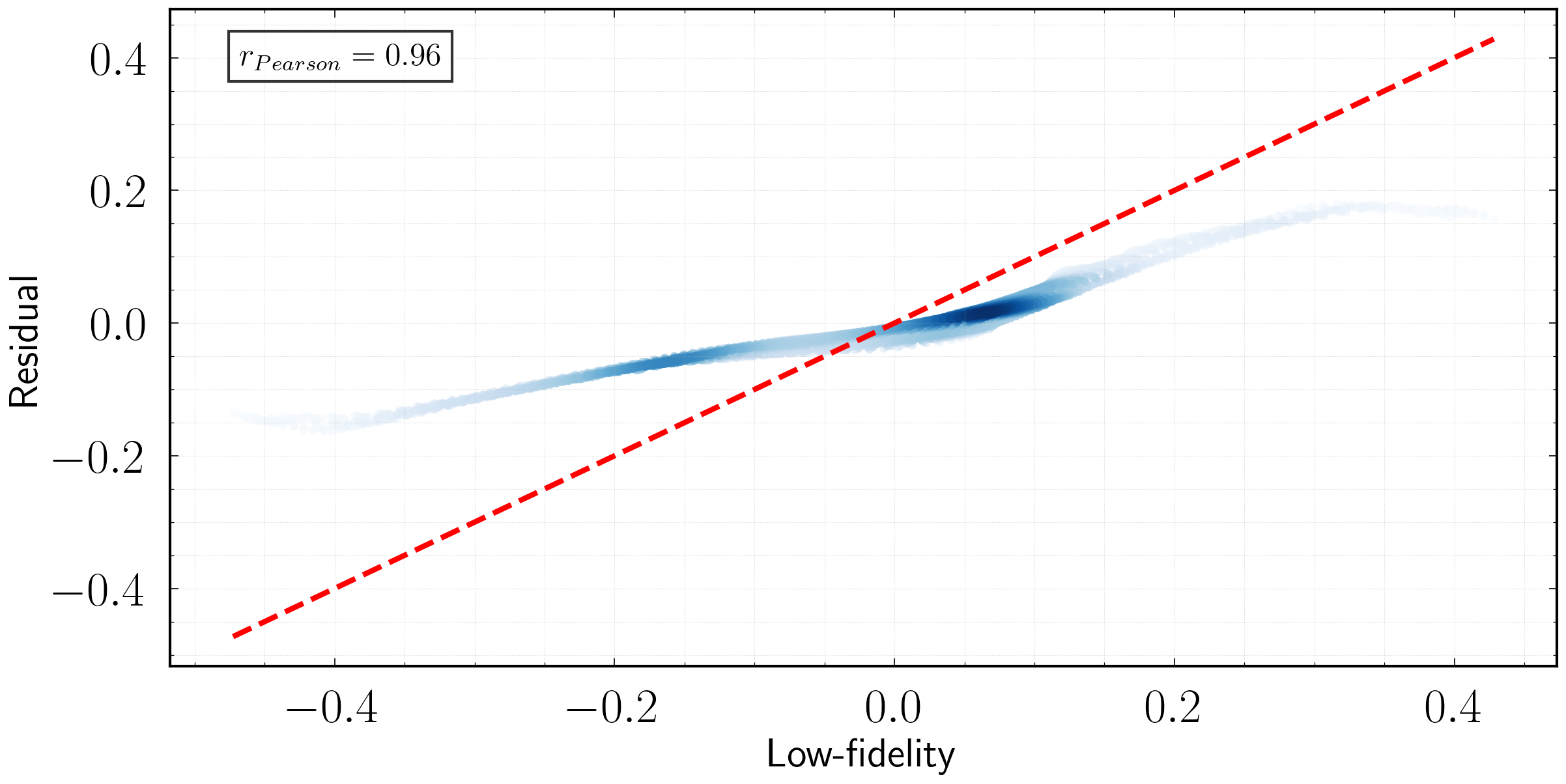}
    \caption{\em Joint distribution between the low-fidelity and residual pressure field for the  flow through porous media.
    }
    \label{fig:appendix/darcy/Residual_Pressure_joint_plot}
\end{figure}
%
%
%
\begin{figure}
    \centering
    \includegraphics[width=0.99\linewidth]{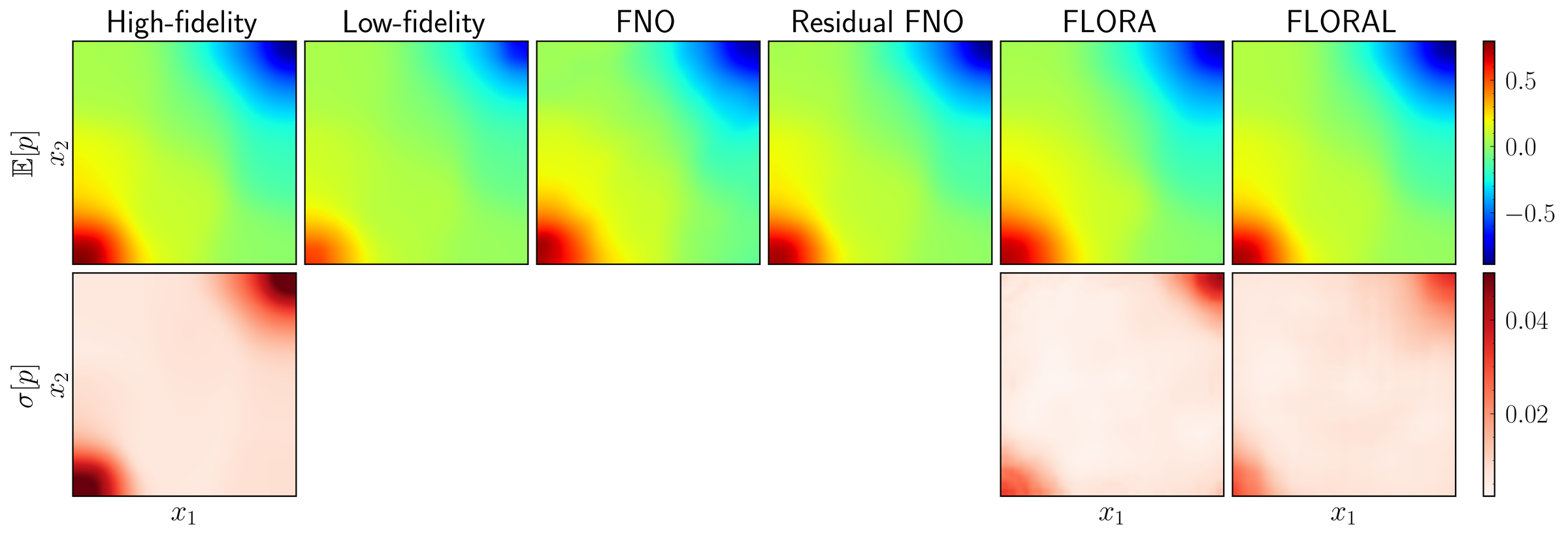}
    \caption{\em Predictive performance for approximating the solution function at an unseen input condition for the flow through porous media. 
    The model is trained using a total of $2000$ training samples, corresponding to $50$ realizations of the stochastic solution function for each input condition.
    (\textbf{Top}) Mean solution function.
    (\textbf{Bottom}) Standard deviation of the solution function.
    }
    \label{fig:appendix/darcy/mean_std_n_train_2000_n_val_750_train_res_full}
\end{figure}
\begin{figure}
    \centering
    \includegraphics[width=0.99\linewidth]{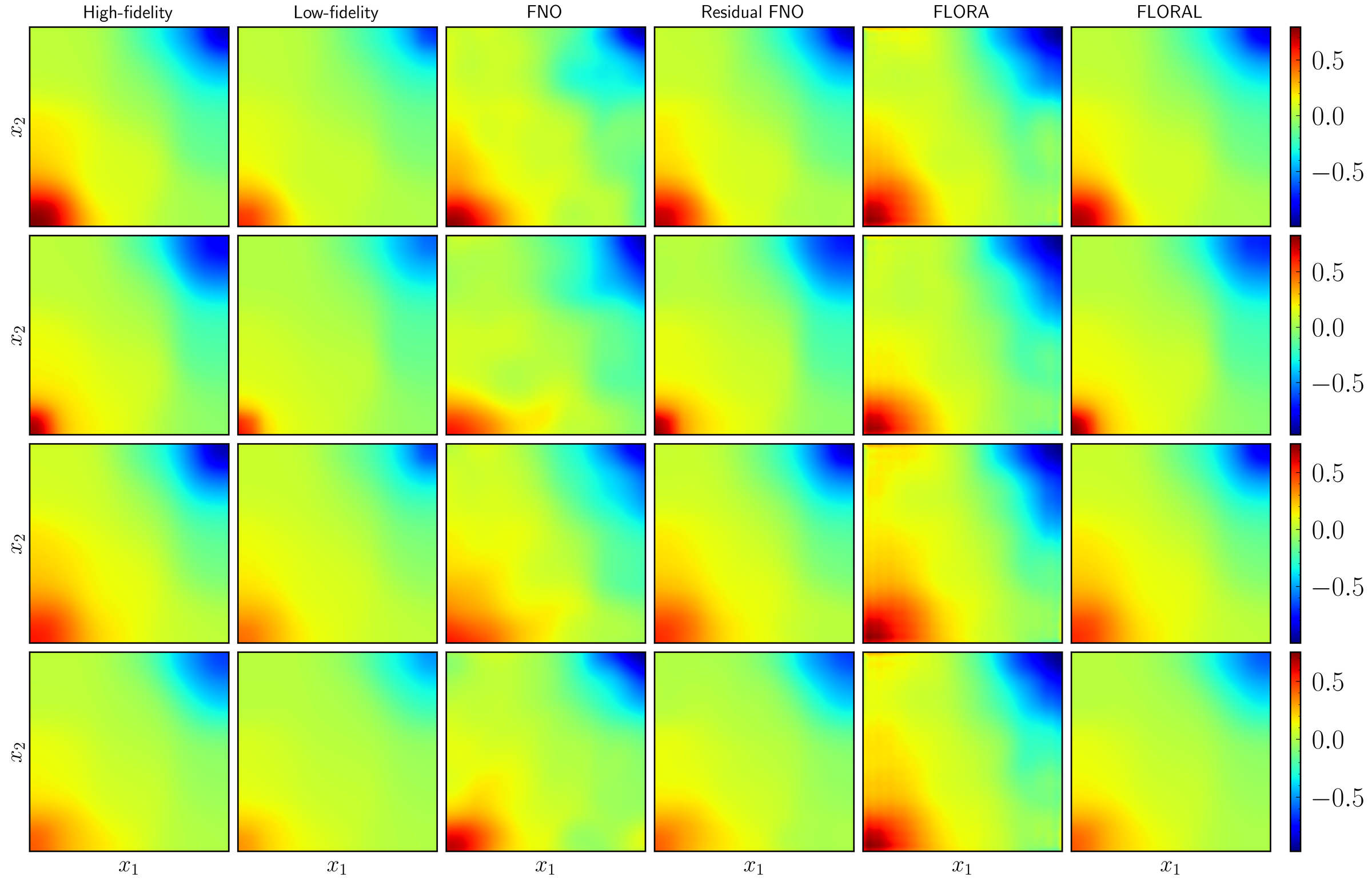}
    \caption{\em Predictive samples of the stochastic solution function for the flow through porous media at unseen input conditions.
    Each row corresponds to a distinct unseen input condition, with $2500$ realizations generated from the learned probabilistic operator, and mean solution function is shown.
    The model is trained using a total of $100$ training samples, corresponding to $50$ realizations of the solution function for each input condition. }
    \label{fig:appendix/darcy/field_samples_n_train_100_n_val_750}
\end{figure}
\begin{figure}
    \centering
    \includegraphics[width=0.99\linewidth]{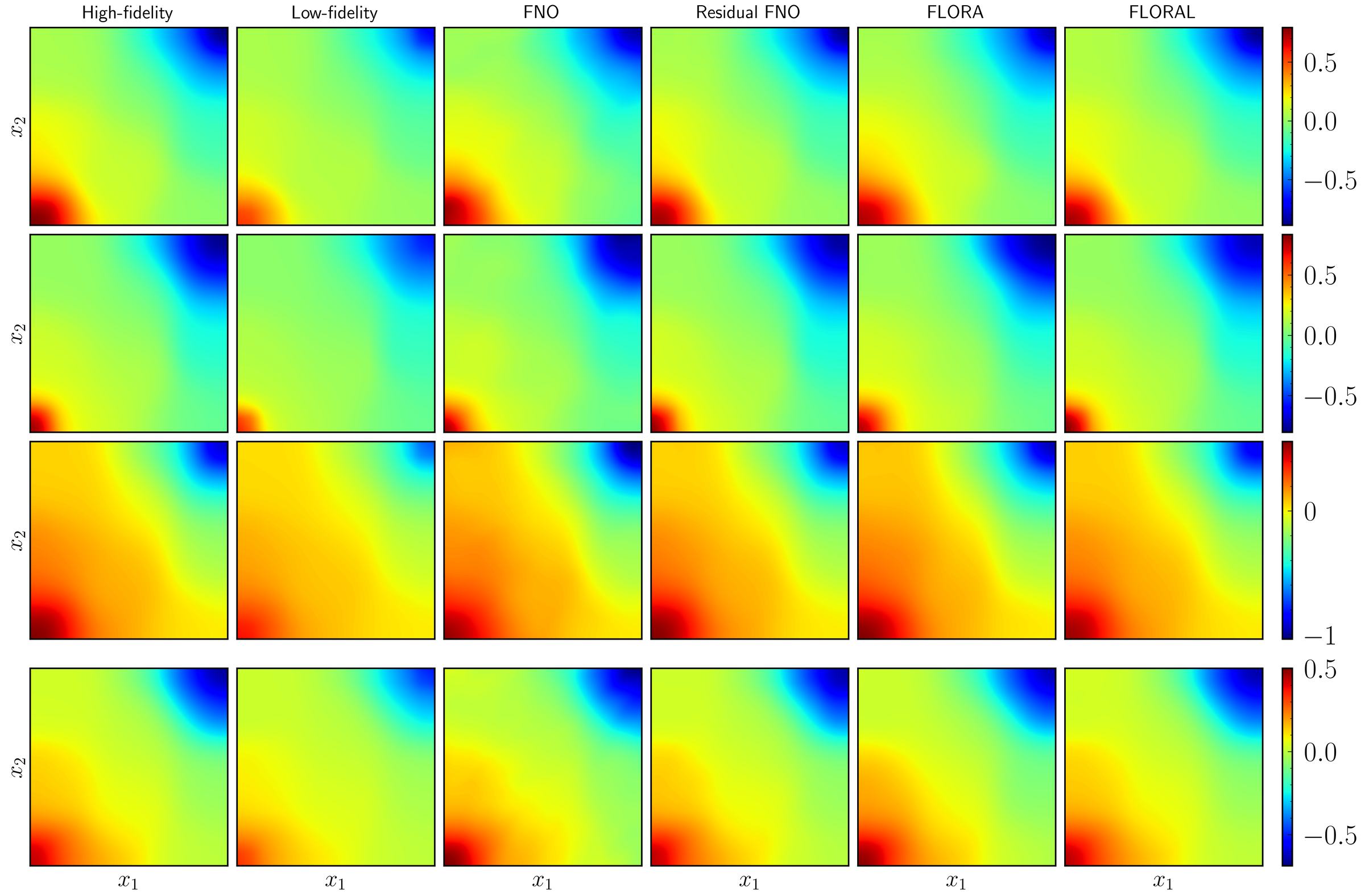}
    \caption{\em Predictive samples of the stochastic solution function for the flow through porous media at unseen input conditions (same as~\cref{fig:appendix/darcy/field_samples_n_train_100_n_val_750}).
    Each row corresponds to a distinct unseen input condition, with $2500$ realizations generated from the learned probabilistic operator, and mean solution function is shown.
    The model is trained using a total of $2000$ training samples, corresponding to $50$ realizations of the solution function for each input condition. }
    \label{fig:appendix/darcy/field_samples_n_train_2000_n_val_750}
\end{figure}
%


\end{document}